\pdfoutput=1
\documentclass[12pt]{iopart}
\usepackage{iopams}
\expandafter\let\csname equation*\endcsname\relax \expandafter\let\csname endequation*\endcsname\relax 
\usepackage{amsmath}
\usepackage[square,sort&compress, comma,numbers]{natbib}
\usepackage[capitalize]{cleveref} 

\usepackage{graphicx}
\usepackage{bm}
\usepackage{svg}
\usepackage{afterpage}
\usepackage{physics}
\usepackage{xr}
\usepackage[T1]{fontenc}
\usepackage[utf8]{inputenc}

\newcommand{\nyuphysics}{Center for Soft Matter Research, Department of Physics, New York University, New York 10003, USA}
\newcommand{\nyuchemistry}{Department of Chemistry, New York University, New York 10003, USA}
\newcommand{\nyusimons}{Simons Center for Computational Physical Chemistry, Department of Chemistry, New York University, New York 10003, USA}

\newcommand{\lpthe}{Sorbonne Universit\'{e}, CNRS UMR 7589, Laboratoire de Physique Th\'{e}orique et Hautes Energies, LPTHE, 4 place Jussieu, Couloir 13-14, 5ème \'{e}tage, 75252 Paris Cedex 05, France}

\begin{document}

\title[Hamiltonian flocks: Generalized TRS]{Hamiltonian flocks: Time-Reversal Symmetry and its consequences 
}

\author{Mathias Casiulis}

\address{\nyuphysics}
\address{\nyuchemistry}
\address{\nyusimons}

\author{Leticia F. Cugliandolo}
\address{\lpthe}

\begin{abstract}
The fluctuation-dissipation theorem is a hallmark of equilibrium systems that stem from their time-reversal symmetry.
In many non-equilibrium systems, in particular active ones, extensions and explicit violations of this theorem are used to assess their ``distance'' to equilibrium.
In Hamiltonian flocks, conservative yet non-Galilean models of polar liquids, previous work reported collective motion without the activity that usually underlies it.
In this paper, we show that this model obeys a generalized time-reversal symmetry that yields a fluctuation-dissipation theorem that mixes position and polarity degrees of freedom.
Due to the oddness of spin under time reversal, the system also obeys Onsager-Casimir reciprocity rather than standard Onsager relations.
The coupling also induces rich  spin orientation dynamics, including a non-trivial diffusion constant at long times.
Finally, we show that considering the naïve time-reversal operation rather than the generalized one that leaves the system invariant leads to a spurious entropy production rate, that could be wrongly interpreted as a distance to equilibrium.
Our findings suggest looking for possible extensions of time-reversal symmetry in active-looking systems, which may lead to yet unknown generalizations of the fluctuation-dissipation theorem.
\end{abstract}


\maketitle

\tableofcontents
\markboth{}{}

\section{Introduction}

Play a movie of Brownian particles backwards, and you will not be able to tell the difference with the original movie from any observable.
This is because their dynamics are invariant under time reversal -- a property shared with all systems in thermodynamic equilibrium.
This time-reversal symmetry (TRS) is associated to conservation laws obeyed by the dynamics, that encode the absence of entropy production, and are usually called the Fluctuation-Dissipation Theorem (FDT)~\cite{Kubo1966,Marconi2008,Cugliandolo2011}.
Conversely, in systems forced out of equilibrium, TRS is typically 
broken and there is non-zero entropy production, so that the standard FDT is violated,
and fluctuation theorems~\cite{Evans1993,Gallavotti1995,Jarzynski1997,Kurchan1998,Lebowitz1999,Crooks1999,Harada2005},
reviewed in~\cite{Sevick2008,Seifert2012,Murashita2021}, are proven.

Many works have focused on the time-reversal properties of active matter, in particular self-propelled systems that are able to transduce some form of local energy source into mechanical work, both in the theoretical~\cite{Fodor2016,Pigolotti2017,Mandal2017,Nardini2017,Shankar2018,Caprini2019,Dabelow2019,Borthne2020, Chiarantoni2020,Caprini2021,Caprini2021a,DalCengio2021,Markovich2021,Skinner2021a, Ro2022, Caprini2023,
Semeraro2023,Boffi2024,Agranov2025,Ferretti2025,Johnsrud2025} and experimental~\cite{Ro2022,Anand2024} realms.
In these systems, part of the interest for entropy production and FDT violations was sparked by the fact that irreversibility is related to the amount of work that they could produce, were one to design a smart energy extraction system~\cite{Ro2022,Anand2024}.
Yet, recent work has shown that the study of reversibility and entropy production in active systems is strongly affected by the dynamical observables that are considered -- and that in particular hidden degrees of freedom may in fact deeply change results on entropy production~\cite{Knight2025}, or even the amount of extractable work~\cite{Metzger2025}.
In experimental work, this limitation may be crucial, as it is unclear which dynamical degrees of freedom are necessary and sufficient to capture the qualitative behavior of a system which is, by definition, an open system extracting energy from the environment.

By contrast, Ref.~\cite{Bore2016} introduced a conservative model of a liquid which could display spontaneous collective motion thanks to a broken Galilean invariance.
This surprising feature gives rise to velocity-velocity correlations at finite low temperatures, leading to Hamiltonian flock states of moving droplets with active-like dynamics in finite size~\cite{Casiulis2019b,Casiulis2019c,MyThesis,Tasaki2020,Bhattacharya2025}, and endows the system with an effective ``swim pressure''~\cite{Chen2026}, akin to that of models of active swimmers~\cite{Takatori2014}.
This system is thus a unique one -- it displays flocking like active matter, yet it is a conservative, closed system in which all degrees of freedom can be tracked.
It thus behoves us to study time-reversal symmetry in this simple system, to highlight how collective motion can be reconciled with time-reversal symmetry, and show how the breaking of Galilean invariance alone affects the FDT.

Furthermore, the vast majority of numerical works on this model~\cite{Casiulis2019b,Casiulis2019c,MyThesis,Bissinger2023,Bhattacharya2025} have been performed in the microcanonical ensemble, using deterministic molecular dynamics.
Due to the non-Galilean nature of the system, switching to the canonical ensemble and to a Langevin-level description is not fully standard~\cite{Bore2016,Casiulis2019b,MyThesis}.
As a result, it is also interesting to study how to properly couple the system to a bath, and assess what regime recent work on overdamped dynamics~\cite{Chen2026} actually considered.
Furthermore, we study the behavior of noise-averaged single-particle dynamics within this Langevin formalism, and thus predict that angular dynamics develop rich phenomenology as the non-Galilean coupling and the average velocity of the system are varied.

The organization of the paper is the following. In Section~\ref{sec:model} we introduce the conservative and 
dissipative model and we display some representative trajectories for different values of the parameters. In 
Section~\ref{sec:TRS} we discuss the time-reveral symmetry and its consequences, including the fluctuation-dissipation
theorem and Onsager-Casimir relations which we check with analytic and numerical calculations. We discuss in 
detail several special features of the translational and rotational dynamics of these particles. We then discuss
our results in Sec.~\ref{sec:discussion}. Several Appendices complement the main body of the paper.
\ref{app:Langevin} explains the origin of the friction and noise contributions to the Langevin equations, 
\ref{app:EveryAction_Thing} presents the construction of the generating functional, its symmetries and the consequences, notably on entropy production under incomplete time reversal
\ref{app:DR} gives additional details on the parameter dependence of the angular diffusion constant,
\ref{app:MSD} discusses predictions for the mean-square displacement and the Einstein relations,
\ref{app:overdamped} touches on the overdamped limit of the Langevin equations in a model with a non-zero velocity imposed by the bath and, finally, the numerical method used to integrate the 
stochastic dynamics is specified in~\ref{app:NumericalMethods}.

\section{Model}
\label{sec:model}

\begin{figure}
    \centering
    \hspace{2.5cm}
    \includegraphics[width=0.8\columnwidth]{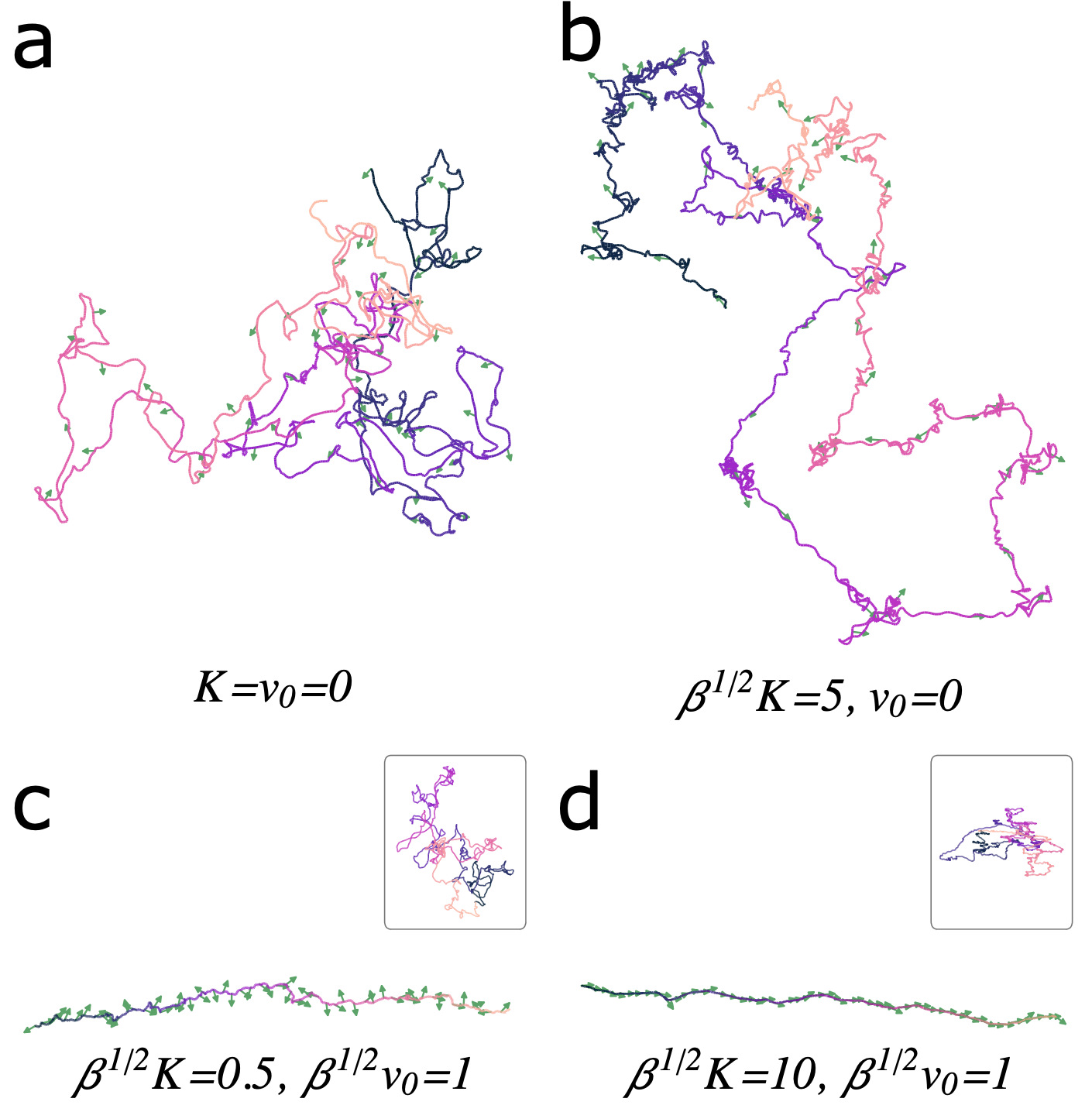}
    \caption{\textbf{Single-particle trajectories.}
    Example trajectories for single particles following the dynamics defined by Eqs.~\eqref{eq:Langevin_p} and~\eqref{eq:Langevin_omega} for $\gamma_t = \gamma_r = \gamma$.
    In the top row, $v_0 = 0$ with (a) $K = 0$ (Brownian particle) and (b) $\sqrt{\beta }K = 5$.
    In the bottom row, $\bm{v}_0 = v_0 \hat{\bm{e}}_x$ with $\sqrt{\beta}v_0 = 1$, and (c) $\sqrt{\beta}K = 0.5$ or (d) $\sqrt{\beta}K = 10$.
    Insets of the bottom row show the same trajectories in the frame moving at $\bm{v}_0$.
    Throughout panels, a gradient of colors represents time, flowing from black ($t=0$) to pink ($\gamma t= 50$), and a few snapshots of the spin are shown as green arrows.
    }
    \label{fig:underdamped_traj}
\end{figure}
We consider a Hamiltonian flock model.
For $N$ particles in $d =2$ dimensions, following the dimensionless rewriting of Refs.~\cite{Bore2016,Casiulis2019b,Casiulis2019c} to set masses and moments of inertia to unity, it is described by the Hamiltonian
\begin{align}
   \mathcal{H} &=\!\sum\limits_{i = 1}^{N} \left( \frac{\bm{p}_i^2}{2} + \frac{\omega_i^2}{2}  - K \bm{p}_i\! \cdot \!\bm{s}_i \right)
  + \frac{1}{2} \sum\limits_{k \neq i} \left[ U(r_{ik}) -  J(r_{ik}) \, \bm{s}_i\!\cdot\!\bm{s}_k \right] \, , \label{eq:Hamiltonian}
\end{align}
where each particle is characterized by its position $\bm{r}_i \in \mathbb{R}^2$, its polarization $\bm{s}_i = (\cos\theta_i, \sin\theta_i)$, and the associated canonical momenta $\bm{p}_i = \dot{\bm{r}}_i +K\bm{s}_i$ and $\omega_i = \dot{\theta}_i$, where $K$ is a spin-velocity coupling that breaks Galilean invariance.
These particles generically interact via a repulsive pair potential $U$ and a ferromagnetic pair interaction $J$.
This system is Hamiltonian, and can thus be described by standard statistical mechanics~\cite{Bore2016,Casiulis2019b,MyThesis}, within a canonical ensemble in which one needs to specify not just the inverse temperature, $\beta$, but also the velocity of the center of mass  of the system, $\bm{v}_0$, such that the probability of a configuration $\mathcal{C}$ in the canonical ensemble reads
\begin{align}
    \mathbb{P}[\mathcal{C}] = \frac{1}{Z(\beta, \bm{v}_0)} e^{- \beta(\mathcal{H}(\mathcal{C}) - \bm{v}_0\cdot \bm{P}\mathcal{(C)})} \label{eq:Canonical_Distribution}
\end{align}
with $\bm{P} = \sum_i \bm{p}_i$ the total momentum of the configuration~\cite{Diu1996}.
The total angular momentum in principle also needs to be added to the measure (in particular in systems with closed boundaries~\cite{Caravelli2026}) but that its statistical significance can be neglected assuming a much more extended bath~\cite{Diu1996,MyThesis}.

\subsection{Introducing damping and noise}

First, we propose to switch to a Langevin description of the dynamics, which will later let us derive noise-averaged properties of the system and check time-reversal properties in the canonical ensemble.
A challenge is here that the bath does not simply play the role of a thermostat, that imposes a temperature, but also that of a \textit{``tachostat''}, that imposes velocity $\bm{v}_0$.
To justify the right noise and friction terms to consider at the Langevin level, we resort to an explicit bath inspired by the Zwanzig-Mori construction~\cite{Zwanzig1961,Mori1965}.
In short, we introduce a super-system consisting of spin-velocity coupled particles, and additional degrees of freedom representing the bath, that are linearly coupled to positions and spin orientations, then write effective equations on the degrees of freedom of the system when the bath degrees of freedom are integrated out.
The full derivation, which is standard but cumbersome, is given in~\ref{app:DampingDerivation}.

For a single particle, we eventually find
\begin{align}
   \dot{\bm{p}} + \gamma_t \left(\bm{p} - K \bm{s} - \bm{v}_0 \right) - \sqrt{2 \gamma_t/\beta}\, \bm{\eta} &= \bm{0} \; \label{eq:Langevin_p} , \\
   \dot{\omega} - K \bm{p} \cdot \bm{s}_{\perp} + \gamma_r \omega - \sqrt{2 \gamma_r /\beta}\, \xi &= 0 \;  ,
   \label{eq:Langevin_omega}
\end{align}
with $\bm{s}_\perp = (-\sin\theta, \cos\theta)$ the spin vector rotated by a $\pi/2$ angle counterclockwise, $(\gamma_t, \gamma_r)$ the translational and rotational friction coefficients, and $(\eta_x$, $\eta_y$, $\xi$) independent sources of unit-variance Gaussian white noise.
These two equations of motion are completed by the definitions of the two canonical momenta,
\begin{align}
    \dot{\bm{r}} &= \bm{p} - K\bm{s} 
    \; , \label{eq:Langevin_r} \\
    \dot{\theta} &=  \omega\label{eq:Langevin_theta}
    \; .
\end{align}
While it is not the focus of this work, the calculation in~\ref{app:DampingDerivation} is performed in the presence of conservative interactions between particles such as those considered in Eq.~(\ref{eq:Hamiltonian}).
The only modification is that the $N$ copies of Eqs.~(\ref{eq:Langevin_p}) and~(\ref{eq:Langevin_omega}) thus obtained then have right-hand-sides given by the total forces and torques, respectively, acting on each particle.

A few trajectories generated with these single-particle stochastic dynamics are shown in Fig.~\ref{fig:underdamped_traj}.
For $K=0$ and $v_0=0$, there is simple unbiased Brownian motion; for $K\neq 0$ and $v_0=0$ the trajectories resemble those of an undirected persistent random walk; when $\bm{v_0} \neq 0$, this velocity imposes a direction of motion on the single particle movement.
In all these cases the motion of the particle is unbounded since there is no confining potential.

\begin{figure}
    \centering
    \hspace{2.5cm}
    \includegraphics[width=0.8\linewidth]{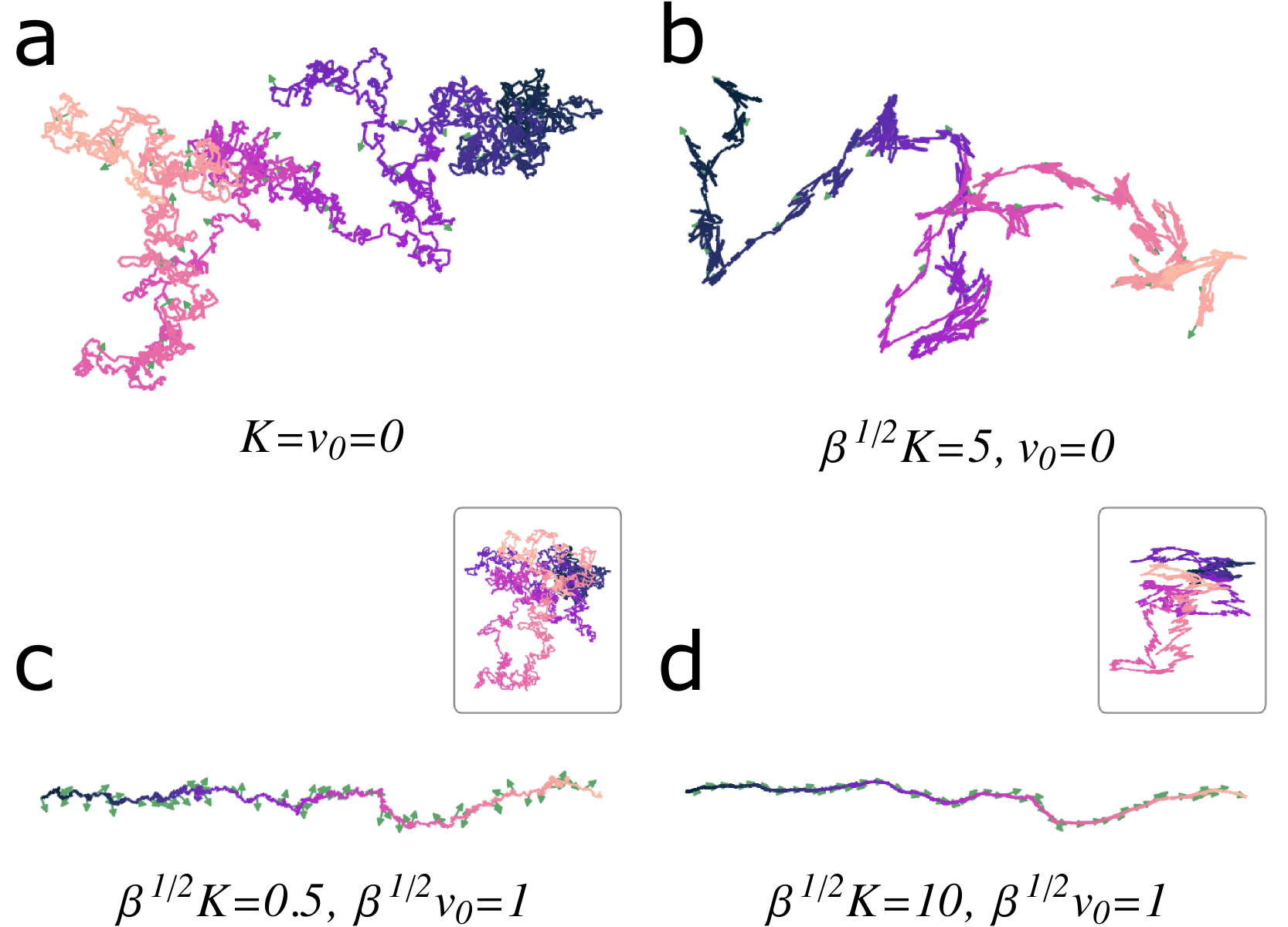}
    \caption{\textbf{Overdamped trajectories.}
    Example trajectories for single particles following the overdamped dynamics defined by Eqs.~\eqref{eq:Langevin_r_overdamped} and~\eqref{eq:Langevin_theta_overdamped} for $\gamma_t = \gamma_r = \gamma$.
    In the top row, $v_0 = 0$ with (a) $K = 0$ (Brownian particle) and (b) $\sqrt{\beta }K = 5$.
    In the bottom row, $\bm{v}_0 = v_0 \hat{\bm{e}}_x$ with $\sqrt{\beta}v_0 = 1$, and (c) $\sqrt{\beta}K = 0.5$ or (d) $\sqrt{\beta}K = 10$.
    Insets of the bottom row show the same trajectories in the frame moving at $\bm{v}_0$.
    Throughout panels, a gradient of colors represents time, flowing from black ($t=0$) to pink ($\gamma t= 50$), and a few snapshots of the spin are shown as green arrows.
    }
    \label{fig:overdamped_traj}
\end{figure}

\subsection{Overdamped and undamped limits}

There are two asymptotic regimes for damping, the overdamped limit ($\gamma_t \to \infty,\gamma_r \to \infty$) and the undamped and noiseless limit ($\gamma_t\to 0,\gamma_r \to 0$), which we briefly comment on here.
In the overdamped limit, one should heed the non-kinetic part of momentum to arrive to a consistent approximation.
It is thus easier to work with the Newtonian equations, obtained by combining Eqs.~\eqref{eq:Langevin_p} and~\eqref{eq:Langevin_r} on the one hand, and Eqs.~\eqref{eq:Langevin_omega} and~\eqref{eq:Langevin_theta} on the other, 
\begin{align}
    \ddot{\bm{r}} + K\dot{\bm{s}} + \gamma_t\left(\dot{\bm{r}} - \bm{v}_0\right) - \sqrt{2 \gamma_t/\beta} \, \bm{\eta} &= \bm{0}, \\
    \ddot{\theta} - K \dot{\bm{r}}\cdot\bm{s}_\perp + \gamma_r \dot{\theta} - \sqrt{2\gamma_r /\beta} \, \xi &= 0.
\end{align}
In these equations, one may then neglect all second-order derivatives, to obtain
\begin{align}
    \dot{\bm{r}} &= \bm{v}_0 - \frac{K}{\gamma_t} \dot{\theta}\bm{s}_\perp + \sqrt{2D_t} \bm{\eta} \label{eq:Langevin_r_overdamped} \\
    \dot{\theta} &= \frac{K}{\gamma_r} \dot{\bm{r}}\cdot \bm{s}_\perp + \sqrt{2 D_r} \, \xi \label{eq:Langevin_theta_overdamped}
\end{align}
where we introduced the $K=0$ values of the diffusion constants, $D_t = k_B T/\gamma_t$ and $D_r = k_BT / \gamma_r$.
These equations are essentially those considered in Ref.~\cite{Chen2026}, except that the authors did not consider the full canonical ensemble including a tachostat.
Note in particular that naïvely cancelling out $\dot{\bm{p}}$ instead of $\ddot{\bm{r}}$ to obtain overdamped dynamics would lead to a trivial equation on position, $\dot{\bm{r}} = \bm{v}_0 +\rm noise$ and thus completely miss the physics.

Injecting Eq.~\eqref{eq:Langevin_r_overdamped} into~\eqref{eq:Langevin_theta_overdamped} mixes sources of noise and will create a renormalized factor in front of $\dot{\theta}$, which already hints at rich angular dynamics even in the limit of large friction terms, which is confirmed qualitatively by the trajectories in Fig.~\ref{fig:overdamped_traj}, which mirrors the parameters of Fig.~\ref{fig:underdamped_traj}.
We shall study in detail the overdamped dynamics in~\ref{app:overdamped}.
Qualitatively for now, Fig.~\ref{fig:overdamped_traj}(b) shows persistent-looking polarities at $K > 0$ and $v_0 =0$, that affect intermediate-time spatial displacements.
Note however that the typical extent of the whole walk does not look very different from the free diffusing case, Fig.~\ref{fig:overdamped_traj}(a).
Furthermore, $\bm{v}_0$ still acts as a magnetic field on $\bm{s}$, so that this effect is added to progressive alignment of spins onto $\bm{v}_0$ as $v_0 >0$ grows, Fig.~\ref{fig:overdamped_traj}(c),(d).

\begin{figure}
    \centering
    \hspace{2.5cm}
    \includegraphics[width=0.8\linewidth]{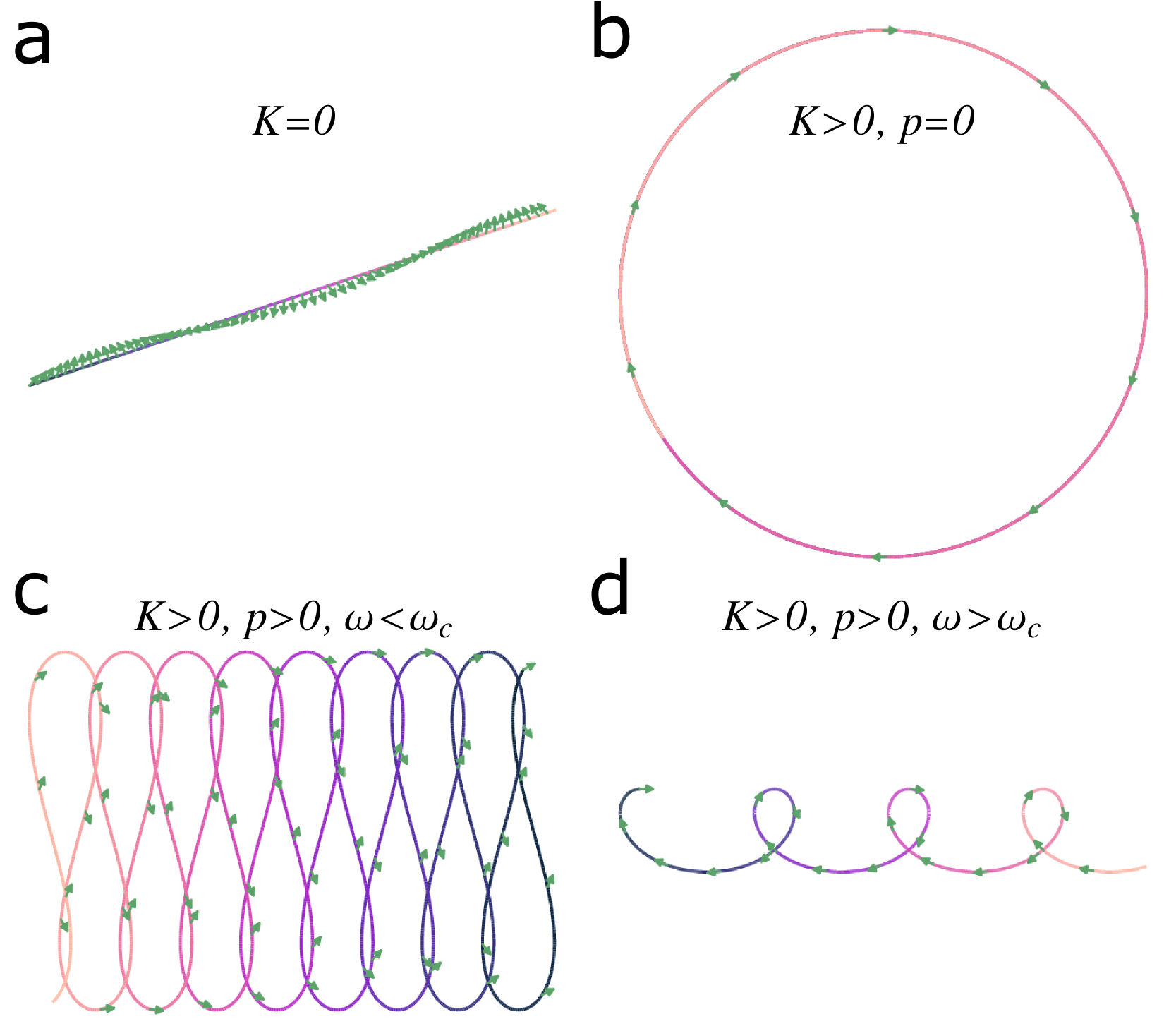}
    \caption{\textbf{Undamped and noiseless trajectories.}
    Conservative single-particle trajectories obtained by integrating Eqs.~\eqref{eq:undamped_p} and~\eqref{eq:undamped_omega}.
    (a) Galilean particle ($K = 0$) with finite $p$ and $\omega$.
    (b) Circular trajectory for $K > 0$, $p = 0$, $\omega >0$.
    (c) Stable-oscillation regime of the spin for $K > 0$, $p>0$, and $\omega < \omega_c$.
    (d) Winding regime of the spin for $K > 0$, $p>0$, and $\omega > \omega_c$.
    Throughout the figure, time flows from dark purple to light pink, and green arrows indicate a few points of measurement of the spin.
    }
    \label{fig:undamped_traj}
\end{figure}
In the other limit, $\gamma_t, \gamma_r \to 0$, one recovers fully conservative dynamics, as introduced in Refs.~\cite{Bore2016, Casiulis2019b, Casiulis2019c, Bhattacharya2025}, for which the momenta follow
\begin{align}
    \dot{\bm{p}} &= \bm{0}, \label{eq:undamped_p} \; ,
    \\
    \dot{\omega} &= K \bm{p}\cdot\bm{s}_\perp \; . 
    \label{eq:undamped_omega}
\end{align}
In this case, the relevant parameters are simply the value of $K$ and the amplitudes of the conserved momenta, $p$ and $\omega$ (since the system is translationally invariant for both $\bm{r}$ and $\theta$).
As already discussed in Ref.~\cite{Bore2016}, there are 4 regimes for the trajectories, illustrated in Fig.~\ref{fig:undamped_traj}.
For $K = 0$, spins and positions decouple so that the trajectories are simple straight lines with spins that rotate at a constant $\omega$, Fig.~\ref{fig:undamped_traj}(a).
For $K > 0$ but $p = 0$, the trajectories become circular, with a radius $R = K/\omega$, Fig.~\ref{fig:undamped_traj}(b).
For $K > 0$ and $p>0$, there are two regimes depending on $\omega$.
For $\omega < \omega_c = 2\sqrt{Kp}$, the spin follows the dynamics of a pendulum that oscillates around the velocity, so that the trajectories oscillate symmetrically around a straight line, Fig.~\ref{fig:undamped_traj}(c).
For $\omega > \omega_c$, $\theta$ winds continuously, like an undamped pendulum crossing the unstable equilibrium with the mass pointing upwards, and the trajectory becomes cycloidal, Fig.~\ref{fig:undamped_traj}(d).

We have summarized the rich variety of single particle trajectories which can be generated with this model (ignoring inter-particle interactions and any confining potential).
We shall now study the generic stochastic dynamics at finite damping, and discuss their time-reversal symmetry, its consequences both in terms of the fluctuation-dissipation theorem and of reciprocity between the responses of position and spin variables, and their noise-averaged behavior.

\section{Time-reversal symmetry}
\label{sec:TRS}

As the originality of the system is its kinetic energy, we henceforth focus on the single-particle dynamics of Eqs.~\eqref{eq:Langevin_p} and~\eqref{eq:Langevin_omega}.
To study the time-reversal symmetry of the system, we construct the Martin-Siggia-Rose-Janssen-De Dominicis (MSRJD) dynamical action~\cite{Martin1973,Janssen1976,DeDominicis1976,deDominicis1978,ArnoulxdePirey2022} corresponding to the equations of motion.
In short, this cumbersome but standard procedure consists in writing a path-integral representation of the dynamics, such that any observable $A(\mathcal{C})$ that depends on the configuration and is measured between times $0$ and ${\mathcal T}$ has a dynamical average
\begin{align}
    \left\langle A \right\rangle_{\text{dyn}} = \int \mathcal{D}\bm{X} \mathcal{D}\bm{Q} d\bm{X}_0 \, P_0(\bm{X}_0) \, A(\bm{X}) \; e^{-\mathcal{S}(\bm{X}, \bm{Q}|\bm{X}_0)}
\end{align}
where $\bm{X}$ is a short-hand notation for  $(\bm{r}, \theta)$, $\left\langle \cdots \right\rangle_{\text{dyn}}$ represents an average over both the distribution $P_0$ of initial conditions $\bm{X}_0$ and noise histories, and $\mathcal{S}$ is the MSRJD action.
The latter depends on auxiliary fields, also called response fields~\cite{ArnoulxdePirey2022}, that we group up under the notation $\bm{Q}$ which, like $\bm{X}$, is a short-hand for $(\bm{q}, \lambda)$.
After some cumbersome but overall standard algebra detailed in~\ref{app:MSRJD}, we find $\mathcal{S} = \mathcal{S}_{det} + \mathcal{S}_{diss}$ with
\begin{align}
    \mathcal{S}_{det} &= i \int dt \, \left[ \bm{q}\cdot \dot{\bm{p}}  + \lambda \left( \dot{\omega} - K \bm{p} \cdot \bm{s}_{\perp} \right)\right] - \ln P_0(\bm{X}_0)
    \; ,\\
    \mathcal{S}_{diss}&= i \int dt \ \left[  \gamma_t \bm{q}\cdot  \left(\bm{v} - \bm{v}_0\right) + \gamma_r \lambda\omega \right] + \int dt \, \left(\gamma_r\lambda^2 + \gamma_t\bm{q}^2 \right)
    \;,
\end{align}
where the action is split into a deterministic part and a dissipative part, the latter containing all terms that couple to the bath.
We do not explicitly discuss the discretization scheme used in the path integral, in spite of the central role that it plays in some models, see \textit{e.g.} Refs.~\cite{Aron2010,Aron2016,ArnoulxdePirey2022}.
The reason, which is further discussed in~\ref{app:EveryAction_Thing}, is that we only consider underdamped dynamics in the procedure that defines the action -- in this situation, no spurious ``It\={o}-Stratonovich'' term appears at leading order during the discretization procedure.

One may then study the effect of time reversal on the dynamical action.
For simplicity, we consider a symmetrized time interval $[-{\mathcal T};{\mathcal T}]$ with ${\mathcal T}$ some arbitrary time.
Introducing the short-hand notation $\bm{X}_{-t} = \bm{X}(-t)$, we find that the following extended time-reversal operation,
\begin{align}
\begin{split}
   \mathbb{T}\!: \;\; &  \bm{r}_t \to \bm{r}_{-t} \; , \\
     & \theta_t \to \theta_{-t} + \pi \; , \\
   &  i\bm{q}_t \to i \bm{q}_{-t} + \beta \left( d_t \bm{r}_{-t} + \bm{v}_0\right)
    \; , \\
    & i \lambda_t \to i \lambda_{-t} + \beta d_t \theta_{-t} \; ,
    \end{split}
\end{align}
leaves the dynamical action unchanged, under the hypothesis that $P_0$ is taken as the canonical equilibrium distribution of Eq.~\eqref{eq:Canonical_Distribution}. 
We note that for the equilibrium distribution of position variables to be normalizable one has to consider that the particle moves in a large but finite surface, or else one has to add a possibly smooth and very flat confining potential which does not modify the dynamic properties far from the boundaries (see~\ref{app:TRS}).
Furthermore, the time-reversal operation flips both velocities and spins, as expected from the Hamiltonian, Eq.~\eqref{eq:Hamiltonian}.

The existence of this TRS validates the choice of dissipative terms introduced by a Zwanzig-Mori construction.
Other choices could have been tempting, for instance a damping of the form $\gamma_p \bm{p}$, which could be introduced by the argument that friction affects momentum conservation.
However, this variant of the damping would not just violate the generalized TRS we just established, it would also simply \textit{not admit} a time-reversal symmetry at all (see~\ref{app:TRS}).

Finally, this generalized TRS leaves both $\mathcal{S}_{det}$ and $\mathcal{S}_{diss}$ invariant independently.
This implies that the bath is not necessary for the time-reversal symmetry to hold: deterministic dynamics, in spite of the breaking of Galilean invariance, are also symmetric under this operation, provided the initial conditions are drawn from canonical equilibrium.
Any other choice of initial conditions leads to relaxation dynamics that are explicitly not time-reversal symmetric, and  in that case neither the FDT nor the reciprocity relations we discuss in the following apply.

\subsection{Fluctuation-Dissipation Theorem}

Having identified the TRS, we can now derive the corresponding FDT.
The theorem consists in equations that relate linear response functions $R_{\bm{a},\bm{b}}(t,t')$ to 
two-time correlation functions $C_{\bm{a},\bm{b}}(t,t')$ or mean-square displacements $\Delta^2_{\bm{a},\bm{b}}(t,t')$. 
The response functions are defined as
\begin{align}
    R_{\bm{a},\bm{b}}(t,t') &= \left. \frac{\delta \left\langle \bm{a}(t) \right\rangle_{\bm{b}} }{\delta \bm{f}_{\bm{b}}(t')}\right|_{\bm{f}_{\bm b} = 0},
\end{align}
where $\bm{a}$ and  $\bm{b}$ are either a position $\bm{r}$ or a spin $\bm{s}$, and $\bm{f}_{\bm{b}}$ is a force-like object entering the Hamiltonian through a linear coupling with $\bm{b}$.
In this paper, we consider three such kinds of couplings,
\begin{itemize}
    \item standard mechanical forces $\bm{f}_{\rm ext}$ if $\bm{b} = \bm{r}$, that enter the Hamiltonian~\eqref{eq:Hamiltonian} through an extra term $-\bm{f}_{\rm ext}\cdot\bm{r}$;
    \item magnetic fields $\bm{h}_{\rm ext}$ if $\bm{b} = \bm{s}$, that enter the Hamiltonian through an extra term $-\bm{h}_{\rm ext} \cdot \bm{s}$;
    \item torques $\Gamma_{\rm ext}$ if $\bm{b} = \theta$, that enter the Hamiltonian through an extra term $-\Gamma_{\rm ext} \theta$ (leading to a constant forcing of $\omega$).
\end{itemize}
The connected correlation functions are defined as
\begin{align}
    C^c_{\bm{a},\bm{b}}(t,t') = \left\langle \bm{a}(t)\cdot\bm{b}(t')\right\rangle_{\text{dyn}} - \left\langle \bm{a}(t) \right\rangle_{\text{dyn}}\cdot\left\langle \bm{b}(t') \right\rangle_{\text{dyn}}
\end{align}
and the self mean-square displacement, the only one we will use, as 
\begin{align}
    \Delta a^2 (t,t') = 
    \left\langle (\bm{a}(t)-\bm{a}(t'))^2 \right\rangle_{\text{dyn}}.
    \label{eq:def-Delta2}
\end{align}
After some simple algebra that follows from the definition of the time-reversal operator (see App.~\ref{app:FDT}), we find
\begin{align}
    R_{\bm{r},\bm{r}}(\tau) &= 
    - \beta d_\tau C^c_{\bm{r},\bm{r}}(\tau) \; , \label{eq:Rrr}\\
    R_{\bm{s},\bm{s}}(\tau)  &= - \beta d_\tau C^c_{\bm{s},\bm{s}}(\tau)
    \; , \label{eq:Rss}\\
    R_{\bm{s},\bm{r}}(\tau) &= - \beta d_\tau C^c_{\bm{s},\bm{r}}(\tau)
    \; , \label{eq:Rsr}\\
    R_{\bm{r},\bm{s}}(\tau) &= - \beta d_\tau C^c_{\bm{r},\bm{s}}(\tau)
    \; , \label{eq:Rrs}
\end{align}
where we introduced $\tau = t - t'$, used time-translational invariance, and assumed $\tau >0$ to use the causality of $R_{{\bm a},{\bm b}}$ (which cancels out response before the excitation).
The average velocity $\bm{v}_0$ is such that $\langle\bm{r}(t) - \bm{r}(0)\rangle = \bm{v}_0 t$, so that the ``connected'' version of the correlation here maps to the correlation of positions as seen in the co-moving frame at $\bm{v}_0$.
Similar Galilean boosts have been considered before~\cite{Gomez-Solano2009} to study the FDT in non-equilibrium steady states obtained by driving particles at a constant speed.
These equations are the standard FDT for Galilean systems (including the ones that consider cross-terms between variables~\cite{Crisanti2012}).
However, the oddness of spins under the generalized time reversal operation affects the link between the cross-responses Eqs.~\eqref{eq:Rsr} and~\eqref{eq:Rrs}.
Indeed, applying the generalized time reversal operation on the definitions of correlation and response (see~\ref{app:FDT}), we find
\begin{align}
    R_{\bm{s},\bm{r}} = -\beta\,\partial_\tau C^c_{\bm{s},\bm{r}} = \beta\,\partial_\tau C^c_{\bm{r},\bm{s}} = -R_{\bm{r},\bm{s}} \; .
\end{align}
These relations differ from the standard Onsager reciprocity~\cite{Onsager1931,Onsager1931a} by a sign: the cross-terms obey Onsager-Casimir~\cite{Casimir1945} rather than Onsager reciprocity.

\begin{figure}
    \centering
    \includegraphics[width=0.48\columnwidth]{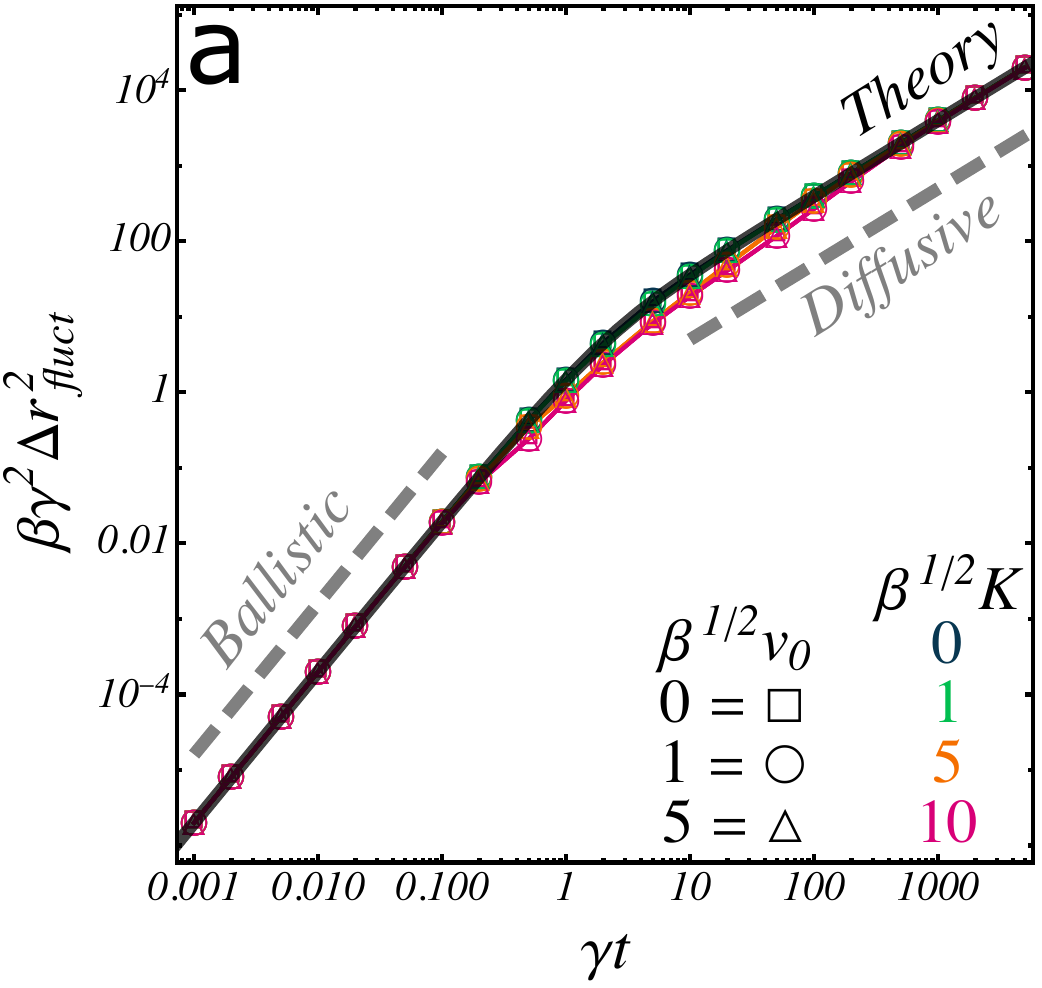}
    \includegraphics[width=0.48\columnwidth]{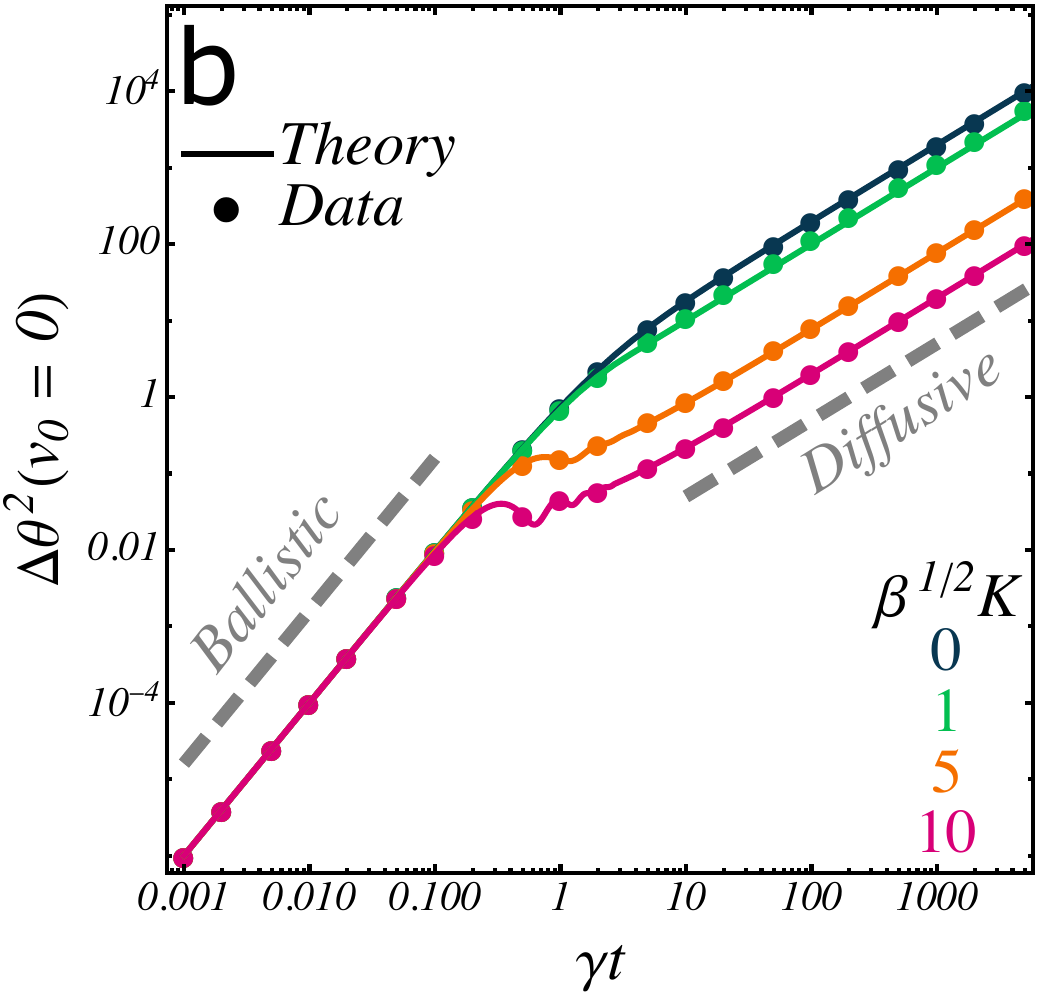}
    \caption{\textbf{Einstein-Smoluchowski-Sutherland Relations.}
    We numerically integrate Eqs.~\eqref{eq:Langevin_p},~\eqref{eq:Langevin_omega},~\eqref{eq:Langevin_r},~\eqref{eq:Langevin_theta} and compute the mean-square displacements (\ref{eq:def-Delta2}) 
    across $N$ trajectories then compute averages over noise and equilibrium initial conditions.
    (a) MSD in the frame co-moving at $\bm{v}_0$, $\Delta r^2_{\text{fluct}}$, against time, both dimensionless.
    Colored symbols present data across values of $v_0$ (symbols) and $K$ (colors).
    All curves fall very close to a Brownian prediction (black line), Eq.~(\ref{eq:Delta2r-Brownian}), with a ballistic regime at short times and a diffusive one at long times (gray dashed guide line).
    (b) MSAD against dimensionless time for $v_0 = 0$ and across $K$ values (colors).
    Symbols indicate numerical data, and solid lines theoretical predictions from Eq.~\eqref{eq:msad_v0=0}.
    We here set $\gamma_r = \gamma_t = \gamma$ and average over $N=10^6$ trajectories.}
    \label{fig:ESSR}
\end{figure}
Importantly, unbounded diffusive models are out of equilibrium and the FDT linking linear response and correlation functions is violated~\cite{Cugliandolo1994,Pottier2003}, due to the fact that there is no proper equilibrium measure for the  position variables in an infinite space. 
For such problems, and to avoid including a confining potential, it is convenient to test the relation between the linear response and the mean-square displacement $\Delta r^2 (\tau)= \left\langle \left[\bm{r}(\tau) - \bm{r}(0)\right]^2\right\rangle$, via
\begin{equation}
R_{\bm{r},\bm{r}} = \frac{1}{2k_BT} \partial_{\tau}\Delta r^2(\tau)
\; , 
\label{eq:FDT-Delta2}
\end{equation}
which is valid even under unbounded diffusion in simple one particle cases~\cite{Cugliandolo1994}.
Angular dynamics do not have that problem as angles are naturally defined on a finite-measure domain.
It is however useful to introduce the equivalent of Eq.~\eqref{eq:FDT-Delta2} for angular dynamics (see App.~\ref{app:MSAD},
\begin{align}
    R_{\theta,\theta} = \frac{1}{2 k_B T}\partial_\tau \Delta\theta^2(\tau) 
    \; .
    \label{eq:FDT-Delta2-angle}
\end{align}

We now discuss the concrete effects of our predictions.
After some standard manipulations (see~\ref{app:MSD}), one can link the long-time mean-square displacements of single particles in the co-moving frame with velocity $\bm{v}_0$
\begin{align}
    \Delta r_{\text{fluct}}^2 &\equiv \left\langle (\bm{r}(t) - \bm{v}_0t - \bm{r}_0 )^2 \right\rangle \underset{t\to\infty}{\sim} \frac{k_BT}{\gamma_t} t 
    \label{eq:ESSR}
    \; , 
\end{align}
to the linear response to an external force.
Equation~(\ref{eq:ESSR}) is 
the Einstein-Smoluchowski-Sutherland relation (ESSR) which defines the long-time translational diffusion constant 
\begin{equation}
D_T = k_BT/\gamma_t
\end{equation}
like in standard free diffusion, regardless of the values of $K$ and $\bm{v}_0$.
The fact that the relation (\ref{eq:FDT-Delta2}) is verified in Fig.~\ref{fig:ESSR}(a), 
where we show that all MSD curves essentially collapse on the same line, which is well predicted by usual Brownian Motion theory, 
\begin{equation}
 \Delta r^2_{\rm BM} = 2d D_T [ t - (1-e^{-\gamma_t t})/\gamma_t]
 \; . 
 \label{eq:Delta2r-Brownian}
\end{equation}

Following active matter standards, one may also introduce a mean-aquare angular displacement (MSAD).
Assuming that, at long times, the MSAD is diffusive with
\begin{align}
    \Delta \theta^2 \equiv \left\langle (\theta(t) - \theta_0)^2 \right\rangle \underset{t\to\infty}{\sim} 2D_R t 
\end{align}
the relevant FDT, Eq.~\eqref{eq:FDT-Delta2-angle}, then implies that the rotational diffusion constant is linked to the linear response function through the time-integrated response $\chi_{\theta,\theta}$,
\begin{align}
    D_R = \frac{1}{\beta }\lim_{t\to\infty} \frac{1}{t} \int\limits_{0}^t d\tau \; 
    R_{\theta,\theta} (\tau) 
     \equiv \frac{1}{\beta} \lim_{t\to\infty} \frac{1}{t} \; \chi_{\theta,\theta}(t)\; . 
    \label{eq:LongTimeResponse}
\end{align}
However, the parameter dependence of $D_R$ is subtler than that of $D_T$.
Interestingly, this is the converse situation compared to results in the microcanonical ensemble, where angular fluctuations were used as a simple thermometer and momentum fluctuations were far less simple~\cite{Bore2016,Casiulis2019b}.

First, we consider the case $v_0 = 0$, and study the response of a spin to a constant external torque $\Gamma_{\text{ext}}$.
In the steady state, and averaging over the noise in the presence of the torque, the equation of motion~\eqref{eq:Langevin_omega} yields
\begin{align}
    \gamma_r \Omega =  K\langle \bm{p} \cdot \bm{s}_{\perp}\rangle_{\Gamma_{\text{ext}}} + \Gamma_{\text{ext}}
\end{align}
where $\Omega = \langle\omega\rangle_{\Gamma_{\text{ext}}}$.
The instantaneous momentum $\bm{p}(t)$ can be expressed by integrating Eq.~\eqref{eq:Langevin_p}.
We set $\bm{p}_0=\bm{0}$.
Projecting the solution onto $\bm{s}_\perp$ and averaging over noise,
\begin{align}
    \langle\bm{p}(t)\cdot\bm{s}_\perp(t)\rangle_{\rm ext} = K\gamma_t\int_0^t d\tau 
    \, e^{-\gamma_t\tau}\,\langle\bm{s}(t-\tau)\cdot\bm{s}_\perp(t)\rangle_{\Gamma_{\rm{ext}}} \;,
    \label{eq:p_dot_sperp}
\end{align}
where we used the fact that $\langle \bm{\eta}(t') \cdot \bm{s}_\perp(t)\rangle = 0$.
Then, we decompose the angular dynamics into $\theta(t) = \theta(0) + \Omega t + \delta\theta(t)$ where $\delta \theta$ represents stochastic fluctuations around the long-time constant rotation rate, 
we rewrite
\begin{align}
    \langle\bm{s}(t-\tau)\cdot\bm{s}_\perp(t)\rangle_{\Gamma_{\text{ext}}} = -\sin(\Omega \tau) \left\langle \cos(\Delta\theta(t,\tau)) \right\rangle_{\Gamma_{\text{ext}}}
\end{align}
where we only used trigonometry and the oddness of $\sin$, and defined $\Delta\theta(t,\tau) =\delta\theta(t) - \delta\theta(t-\tau)$.
In the limit $\Gamma_{\text{ext}}\to 0$, which implies $\Omega \to 0$, this expression can be approximated by
\begin{align}
    \langle\bm{s}(t-\tau)\cdot\bm{s}_\perp(t)\rangle_{\Gamma_{\text{ext}}} \approx -\Omega \tau \left\langle \cos(\Delta\theta(t,\tau)) \right\rangle
\end{align}
where the average on the right-hand side is approximated by that of torqueless dynamics.
Finally, we assume that $\Delta \theta$ is well approximated by Gaussian statistics, so that $\langle \cos \Delta\theta\rangle \approx e^{-\langle\Delta\theta^2\rangle/2}$.
Assuming, furthermore, that the value $\tau \propto \gamma_t^{-1}$ that dominates the integral lies in the terminal diffusive regime for angles, $\langle \cos \Delta\theta\rangle \approx e^{-D_R \tau}$, we find
\begin{align}
\langle\bm{p}\cdot\bm{s}_\perp\rangle
  &\approx -K\gamma_t\,\Omega \int_0^\infty e^{-\gamma_t\tau}\,\tau\,
     e^{-D_R\tau}\,d\tau
  = -\frac{K\gamma_t\,\Omega}{(\gamma_t + D_R)^2} \;,
\end{align}
where we also used the limit $t \to \infty$ in the upper bound of the integral.
Altogether, the steady-state linear response equation becomes
\begin{align}
    \gamma_{\text{eff}}\,\Omega = \Gamma_{\text{ext}}
\end{align}
with 
\begin{align}
    \gamma_{\text{eff}} = \gamma_r + \frac{K^2 \gamma_t}{(\gamma_t + D_R)^2}.
\end{align}
Using Eq.~\eqref{eq:LongTimeResponse} this yields an implicit equation on $D_R$,
\begin{align}
    D_R = \frac{k_BT}{\gamma_r +K^2\gamma_t/(\gamma_t + D_R)^2}.
\end{align}
In particular, for $D_R \ll \gamma_t$, which is consistent with the Gaussian approximation (as it implies that $\bm{p}$ can be adiabatically eliminated from the angular dynamics), 
\begin{align}
    \gamma_{\text{eff}} & \approx \gamma_r + K^2/\gamma_t \; ,\\
    D_R & \approx \frac{k_BT}{\gamma_r +K^2/\gamma_t} \; .
\end{align}
There is thus an effective (still postive) friction on the angular dynamics generated by the non-Galilean coupling.
In particular, notice that for $K=0$, one recovers the standard ESSR $D_R = k_BT/\gamma_r$ as expected.
If one naïvely tried to check the latter, it would be easy to be misled into concluding that this system violates the ESSR on angles and instead displays an effective temperature, an effect reported in both active~\cite{Fily2012} and glassy~\cite{Cugliandolo2011} dynamics.
Furthermore, looking at the trajectories in Figs.~\ref{fig:underdamped_traj}$(b)$ and~\ref{fig:overdamped_traj}$(b)$ compared to their $K=0$ equivalents in Figs.~\ref{fig:underdamped_traj}$(a)$ and~\ref{fig:overdamped_traj}$(a)$, the introduction of $K$ and the associated lower angular mobility lead 
to visibly more persistent walks at finite times: trajectories with finite $K$ look more like broken lines, while the ones with $K = 0$ have more curvature.
One may naïvely think that this lower curvature maps to a larger persistence length, or even superdiffusion -- which is clearly not the case looking at Fig.~\ref{fig:ESSR}$(a)$.
This is a reminder that the model displays unusual features for an equilibrium model.

\begin{figure}
    \centering
    \includegraphics[width=0.5\linewidth]{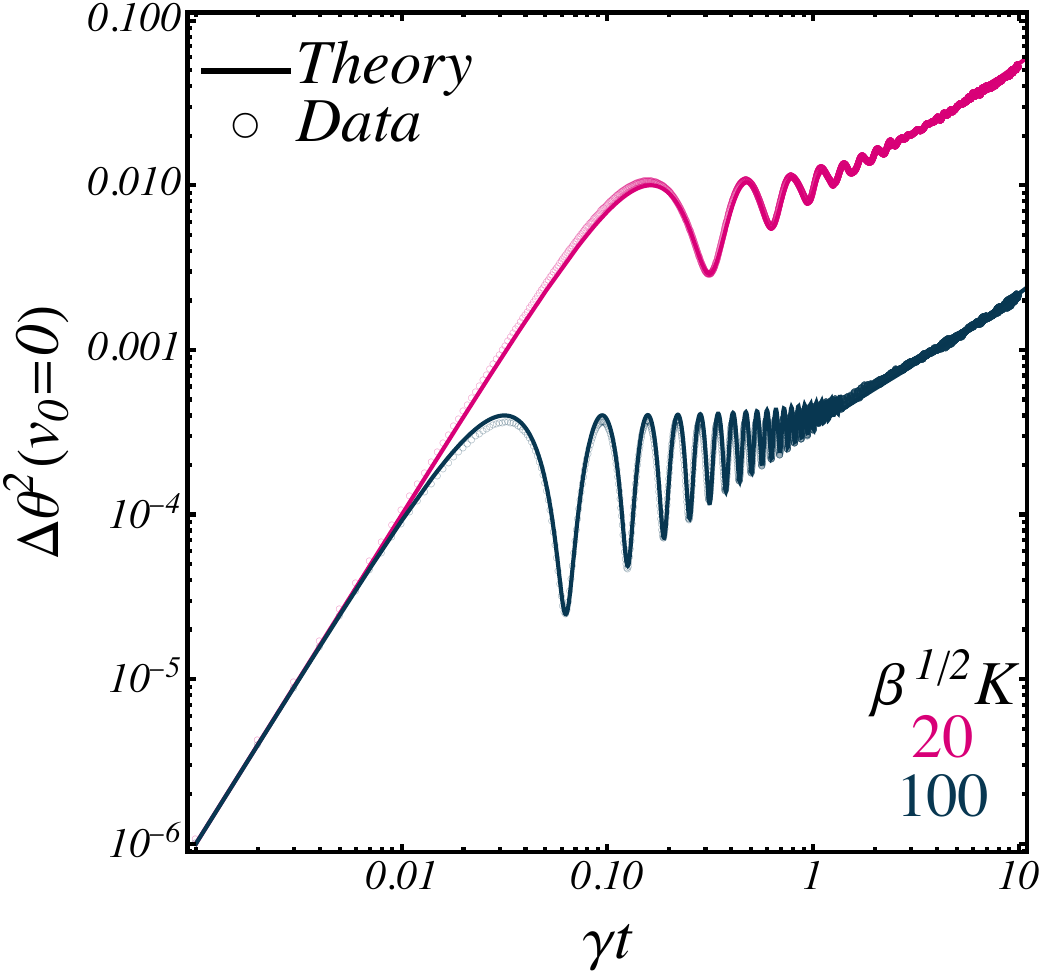}
    \caption{\textbf{Oscillatory behavior.}
    MSAD against dimensionless time for $v_0 = 0$, analogous to Fig.~\ref{fig:ESSR}(b), but for $\sqrt{\beta}K = 20$ (red) and $\sqrt{\beta}K = 100$ (dark blue).
    Open symbols indicate numerical measurements from integrating the full 
    Langevin dynamics and averaging over $N = 10^6$ trajectories, solid lines indicate the 
    analytical prediction from Eq.~\eqref{eq:msad_v0=0}.
    }
    \label{fig:Oscillations}
\end{figure}

A full prediction of MSAD across time ranges, which terminates with diffusion constant $D_R$, can be derived similarly  (see~\ref{app:MSAD_v0=0}), and reads, in the simpler case $\gamma_r = \gamma_t = \gamma$,
\begin{align}
  \Delta\theta^2(t)
  = 2 D_R\left[
    t - \cos(2\phi)/\gamma
    + e^{-\gamma t}\cos(Kt + 2\phi)/\gamma
  \right] ,
  \label{eq:msad_v0=0}
\end{align}
with $\phi = \arctan(K/\gamma)$.
It is shown to agree remarkably well with numerical data in Fig.~\ref{fig:ESSR}(b).
In particular, this expression captures non-trivial oscillatory dynamics at intermediate times for $K >0$.
These oscillations are further illustrated at higher $K$, for which they are faster and more noticeable, in Fig.~\ref{fig:Oscillations}.

\begin{figure}
    \centering
    \includegraphics[height=0.47\columnwidth]{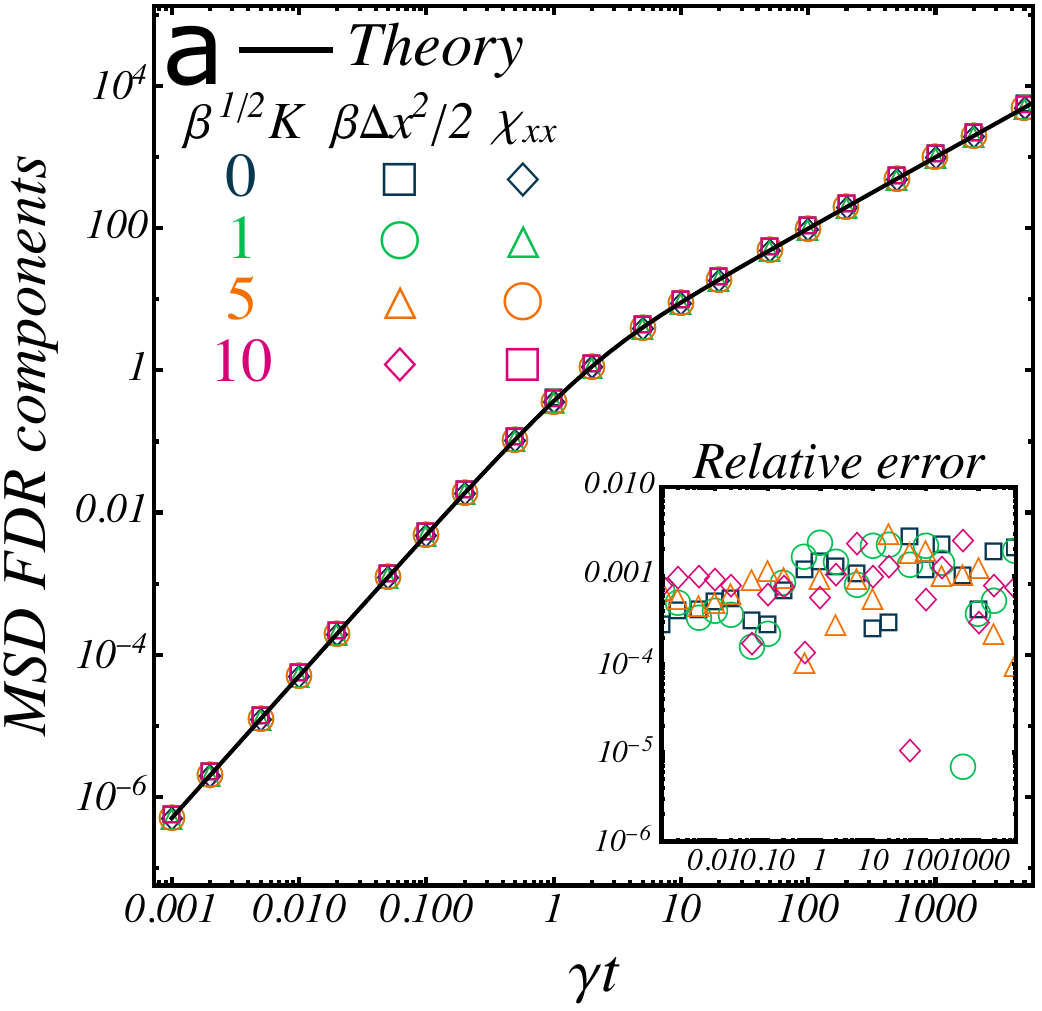}
    \includegraphics[height=0.47\columnwidth]{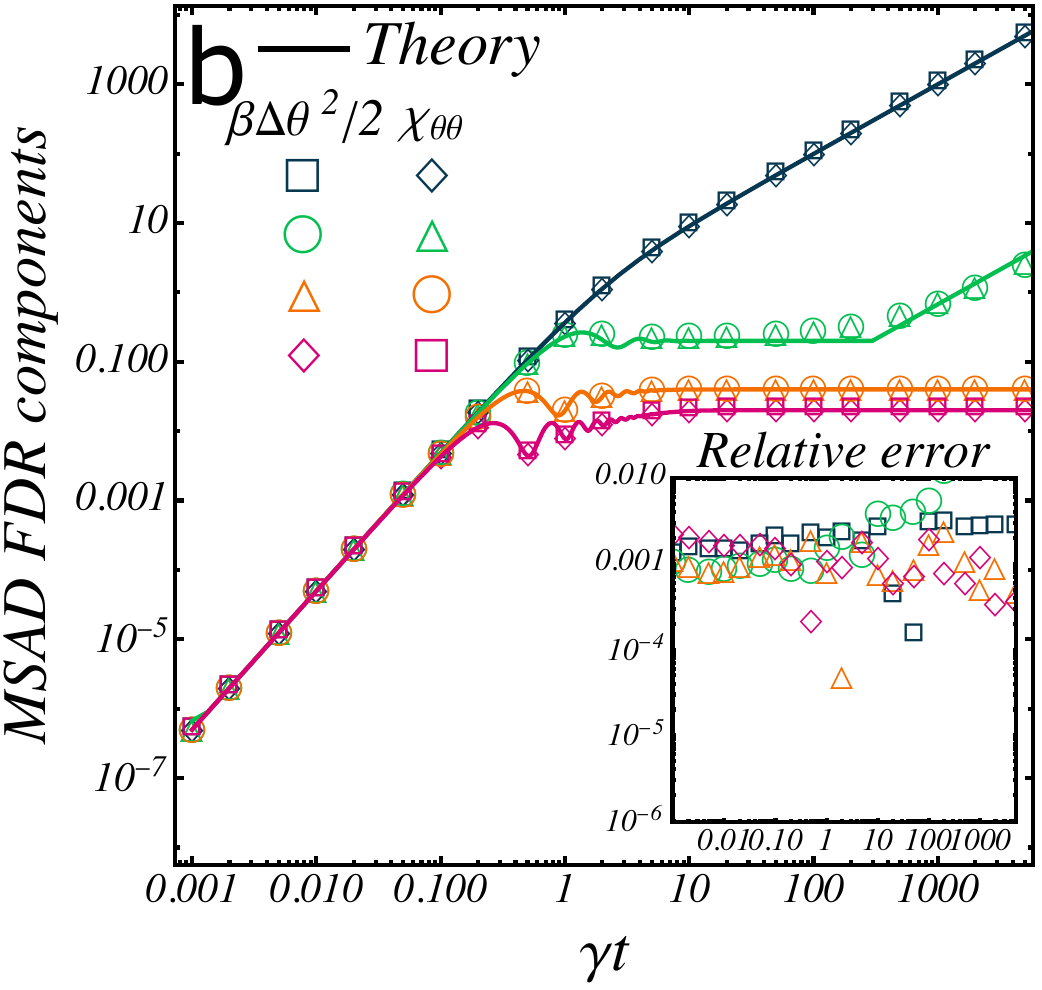}
    \caption{\textbf{Diagonal Fluctuation-Dissipation Equations.}
    (a) Position-position FDT, Eq.~\eqref{eq:Rrr}.
    We plot data for free particle fluctuations $\beta \Delta x_{\text{fluc}}^2/2$, and for the linear response to a weak force along $x$ constantly applied, $(\langle x\rangle_{f_x} - \langle x\rangle_{0})/f_x$ (noted $\chi_{xx}$) across $K$ values, coded by color.
    A black line shows the prediction assuming Brownian behavior, Eq.~\eqref{eq:Delta2r-Brownian}, and the inset shows the relative error between correlation and response data.
    (b) Analogue figure for the angular FDT, Eq.~\eqref{eq:FDT-Delta2-angle}, with fluctuations $\beta \Delta\theta^2/2$, and the response to a constant torque $\Gamma$, $(\langle \theta\rangle_{\Gamma} - \langle \theta\rangle_{0})/\Gamma$.
    Solid lines show analytical predictions for $\Delta\theta^2$ for $Kv_0 >0$ using Eq.~\eqref{eq:msad_composite_maintext}.
    Throughout the figure, $\beta^{1/2} v_0 = 5$ and averages were obtained using $N = 10^6$ trajectories.
    Response curves are obtained using the common random number method, see~\ref{app:CRN}.}
    \label{fig:diagonal_FDT}
\end{figure}

When varying $K$ and $v_0$, we check that the FDT between linear response and displacement consistently holds, by comparing the linear response due to an external force or torque to spontaneous fluctuations, see Fig.~\ref{fig:diagonal_FDT}.

For $v_0 \neq 0$, the angular dynamics develop more intricate behavior.
This is explained by the fact that, rewriting the equations of motion in the co-moving frame moving at $\bm{v}_0$, $K v_0$ plays the role of an effective field on the spins.
Thus, when $Kv_0$ becomes large, the spin dynamics turn into a Kramers problem in a confining potential around the tachostat's velocity.
The short-time regime can be captured by a linearization of the dynamics around the state $(\bm{v}=\bm{v}_0,\dot{\bm{v}}=\bm{0}, \bm{s}=\bm{v}_0/v_0, \omega =0)$.
A full derivation of this linearized theory, given in~\ref{app:MSAD_v0>0}, yields the short-time linearized behavior
\begin{align}
  \Delta\theta^2_{\rm lin}(\tau)
  = \frac{2k_BT}{Kv_0}\left[
    1 - \sum_{i=1}^{3}
    \frac{(s_i + \gamma)^2 + K^2}{(s_i - s_j)(s_i - s_k)}\,
    e^{s_i\tau}\right]
  \label{eq:msad_lin_maintext}
\end{align}
where the $s_i$'s are the roots of the cubic polynomial 
\begin{align}
  s^3 + 2\gamma\,s^2 + (\gamma^2 + K^2 + Kv_0)\,s + Kv_0\gamma = 0 \;.
  \label{eq:char_poly_maintext}
\end{align}

Intuitively, at very short time scales ($\gamma_rt \ll 1$), inertia dominates and 
$\Delta\theta^2  \sim k_B T \, t^2$. 
The prefactor $k_B T$ simply follows from equipartition: since initial conditions are drawn from the equilibrium distribution, Eq.~\eqref{eq:Canonical_Distribution}, $\left\langle \omega^2 \right\rangle \sim k_B T$, which sets the ballistic angular speed at short times.
At time scales $\gamma_r t \gg 1$ but much shorter than the Kramers escape time $\tau_K$, the MSAD is constant, given by the ratio of temperature to the typical torque $\Delta\theta^2  \sim k_BT/(Kv_0)$.
As $t \to \infty$, the angle escapes the potential due to $Kv_0$ and the dynamics are, again, diffusive,
\begin{align}
    \Delta\theta_{\rm Kramers}^2 (\tau) \underset{\tau\to \infty}{\sim} 2 D_R(v_0) \tau
    \; . \label{eq:msad_Kramers_diffusion}
\end{align}
The long-time dynamics of a Brownian particle in a periodic potential are well studied~\cite{Festa1978,Risken1989} and (after a standard calculation detailed in~\ref{app:DRwithv0}) yields an analytical expression for the terminal diffusion constant
\begin{align}
    D_R(v_0) &= \frac{D_R(v_0=0)}{I_0( K v_0 / k_BT)^2}
    \; ,
\end{align}
where $I_0$ is the modified Bessel function.

The crossover between the confined plateau and the final diffusive regime is difficult to predict analytically so that we propose to model MSAD by simply taking the maximum between Eqs.~\eqref{eq:msad_lin_maintext} and~\eqref{eq:msad_Kramers_diffusion},
\begin{align}
  \Delta\theta^2_{\rm theory}(\tau) = \max\bigl(
    \Delta\theta^2_{\rm lin}(\tau),\;
    \Delta\theta^2_{\rm Kramers}(\tau)
  \bigr) \;.
  \label{eq:msad_composite_maintext}
\end{align}
This model, and its asymptotic ballistic-confined-diffusive description, is quantitatively verified in Fig.~\ref{fig:diagonal_FDT}(b).
For instance the $\beta K^2 = 1$ curve (green line and symbols) shows good agreement across regimes.

\begin{figure}[t!]
    \centering
    \includegraphics[height=0.33\linewidth]{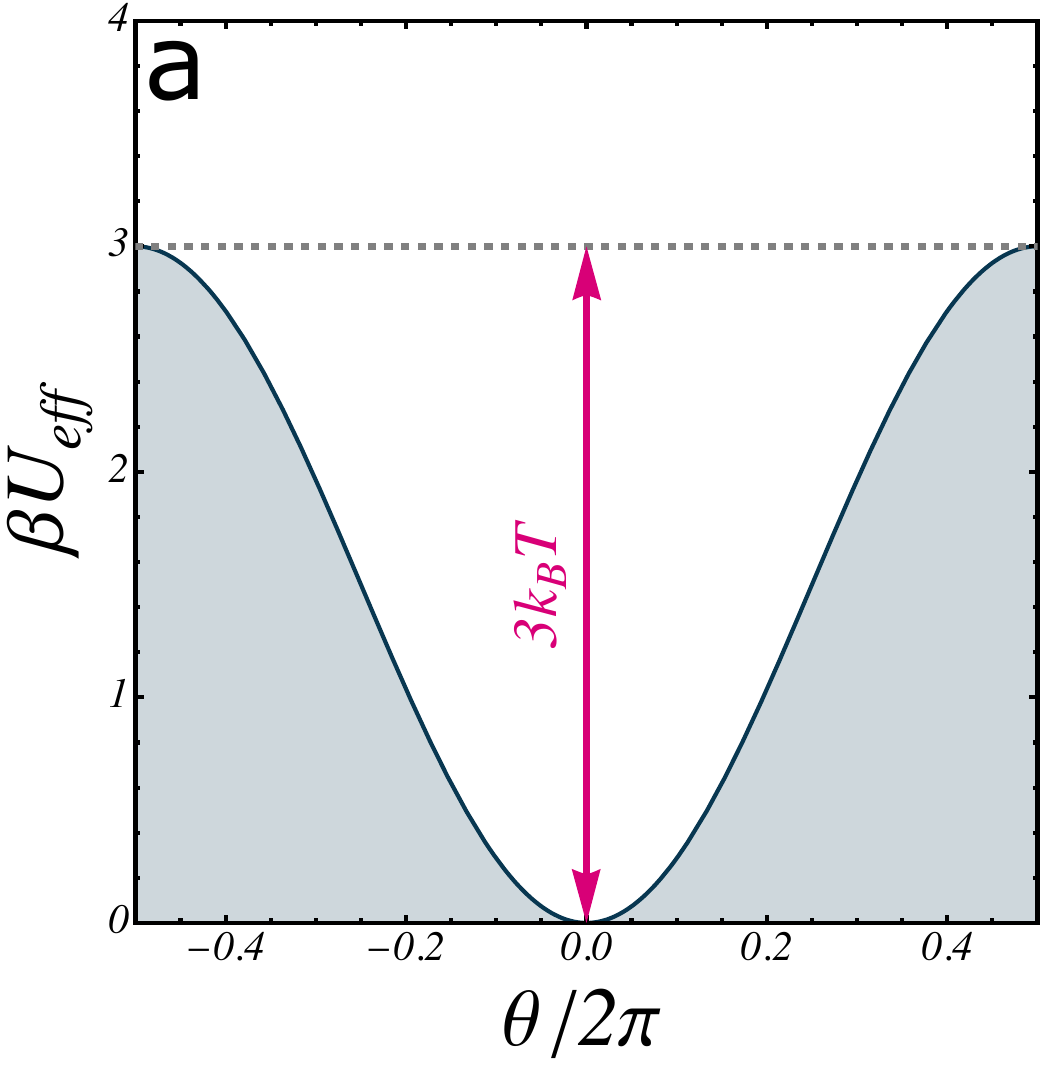}
    \includegraphics[height=0.33\linewidth]{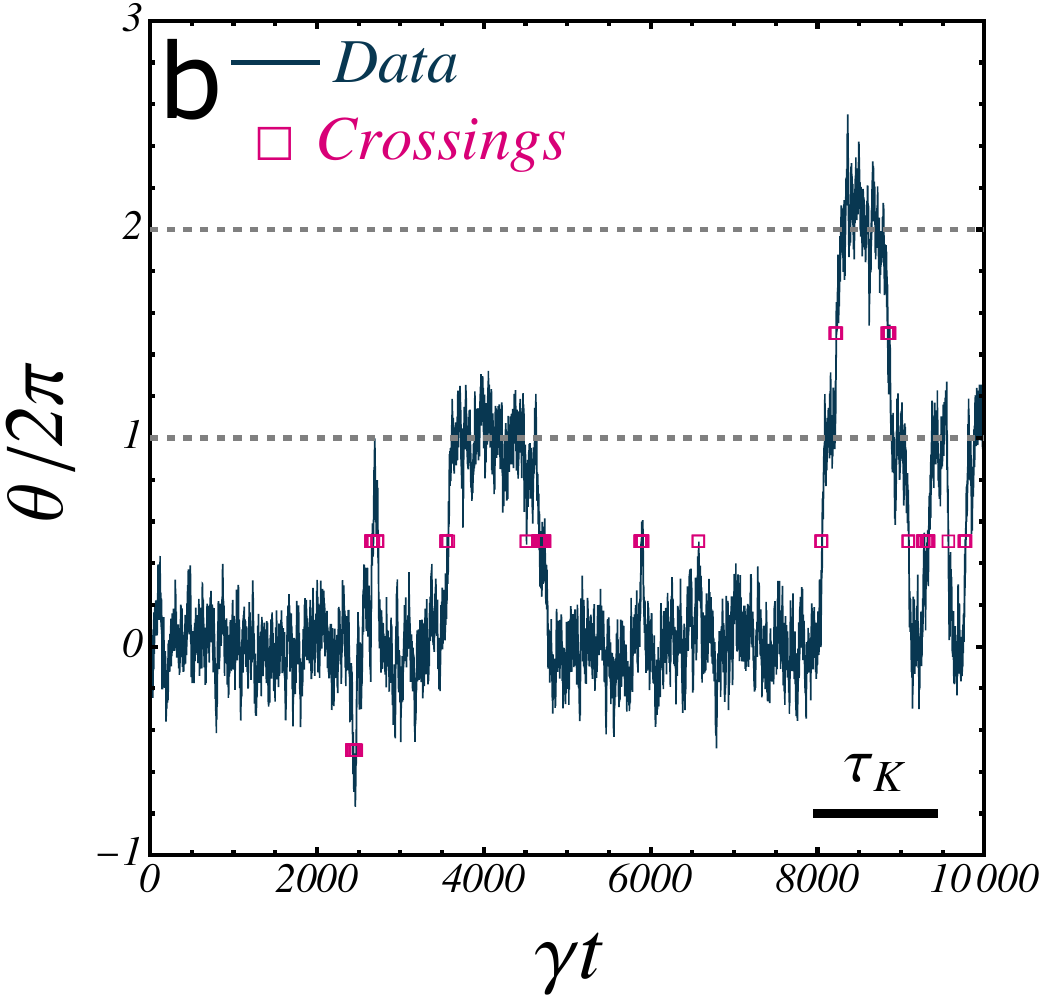}\\
    \includegraphics[width=0.82\linewidth]{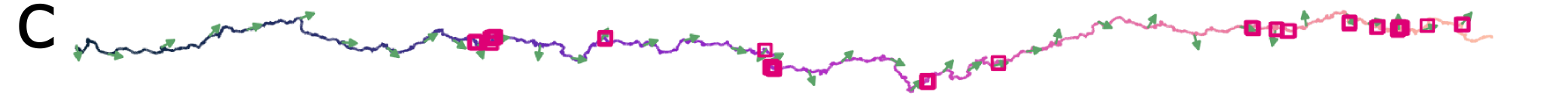} \\
    \includegraphics[height=0.33\linewidth]{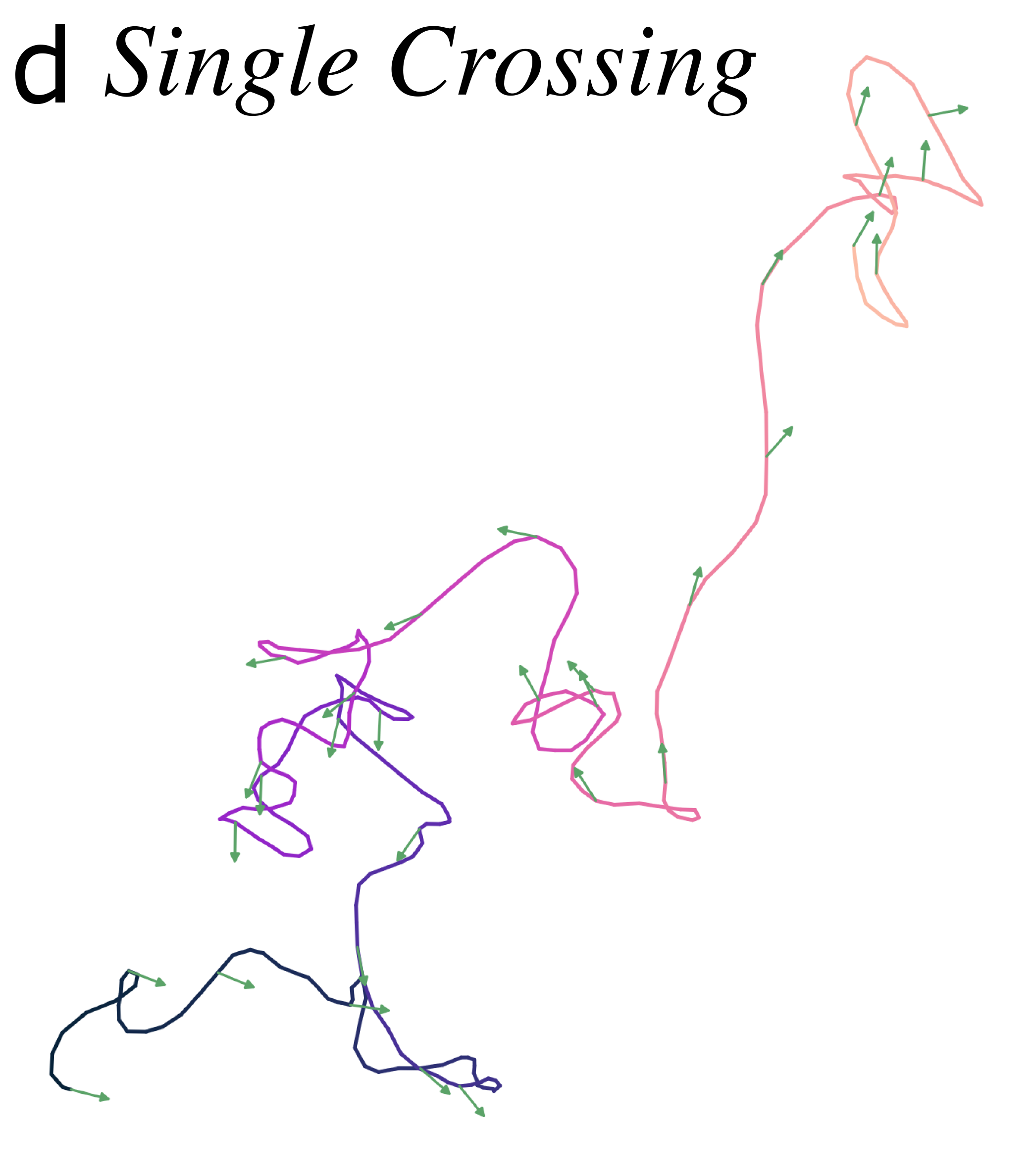}
    \includegraphics[height=0.33\linewidth]{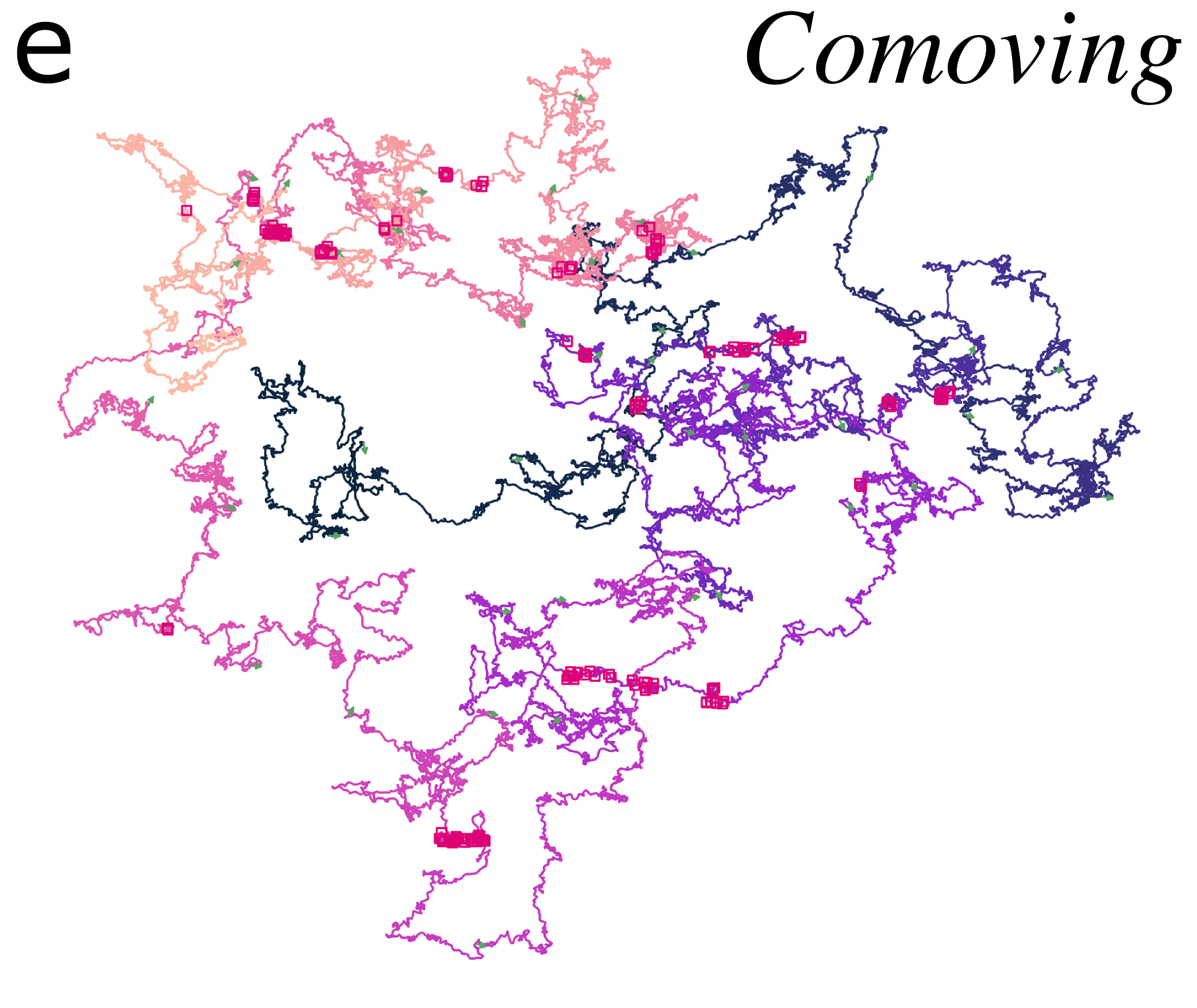}
    \caption{\textbf{Confined angular dynamics as a Kramers problem.}
    (a) Effective potential felt by the spin in the special case we choose to illustrate Kramers dynamics, $\beta K v_0 = 3/2$ (dark blue).
    A gray dashed line highlights the height of the maximum and a magenta arrow its distance from the minimum, $2 K v_0 = 3k_B T$.
    (b) Example trajectory of the angle for these parameters (dark blue).
    Empty red squares highlight barrier crossings, and dashed gray lines indicate multiples of $2\pi$, where periodic copies of the minimum lie.
    A scale bar (bottom right) indicates the estimate of the Kramers crossing time obtained from the terminal diffusion constant $D_R(v_0)$.
    Here, $\gamma\tau_K \approx 1.4\times10^3$.
    (c) Full trajectory in the lab frame.
    Green arrows show a few values of the spin, magenta squares show barrier crossings, and time flows from black to pink.
    (d) Zoom on a single crossing, for $\gamma t \in [9550; 9580]$.
    (e) Full trajectory in the co-moving frame, with the same plotting conventions as panel (c).
    }
    \label{fig:kramers}
\end{figure}

Before discussing reciprocity relations, we briefly focus on the Kramers escape problem view of the system to validate it.
To do so, we focus on one set of parameters, $\sqrt{\beta} v_0 = 0.3$ and $\sqrt{\beta} K = 5$, so that $\beta K v_0 = 3/2$.
This particular choice means that the effective potential acting on the spin $U_{\rm eff}(\theta) = - K v_0 \cos\theta$, sketched in Fig.~\ref{fig:kramers}(a), has a barrier with height $3 k_B T$, high enough that the spin is confined for an appreciable time but low enough that crossings are observed in a reasonable simulation time.
In Fig.~\ref{fig:kramers}(b), we show a simulated trajectory for the orientation $\theta$ of the spin of a single particle, and highlight every time the angle crosses the maximum of the effective potential (magenta symbols).
The crossing time is a good match for the estimate of the Kramers time found using only the final diffusion constant, via $\Delta\theta^2 \sim 2 D_R(v_0) t \sim (2\pi)^2 t/\tau_K$, or more explicitly
\begin{align}
    \tau_K \approx \frac{2 \pi^2}{D_R(v_0)}
    \; .
\end{align}
We show the corresponding trajectory in the lab frame in Fig.~\ref{fig:kramers}(c), highlighting the crossings with magenta symbols again, as well as a zoom on a single crossing event in Fig.~\ref{fig:kramers}(d).
In the latter, the spin starts and ends aligned with the average velocity along $\hat{\bm{e}}_x$ but undergoes a full clockwise rotation of $2\pi$.
We also show for completeness the trajectory in the co-moving frame in Fig.~\ref{fig:kramers}(e).

\subsection{Reciprocity relations}

\begin{figure}
    \centering
    \includegraphics[height=0.47\columnwidth]{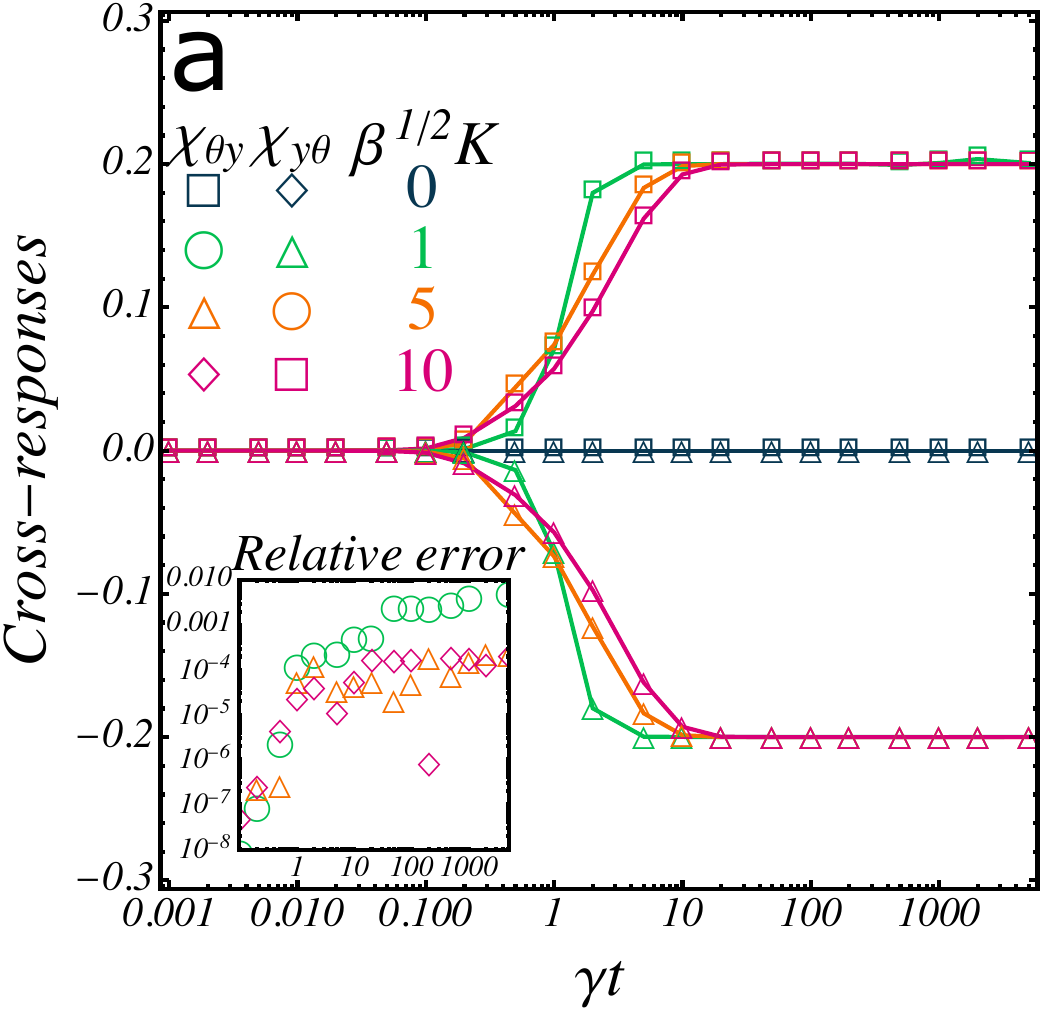}
    \includegraphics[height=0.47\columnwidth]{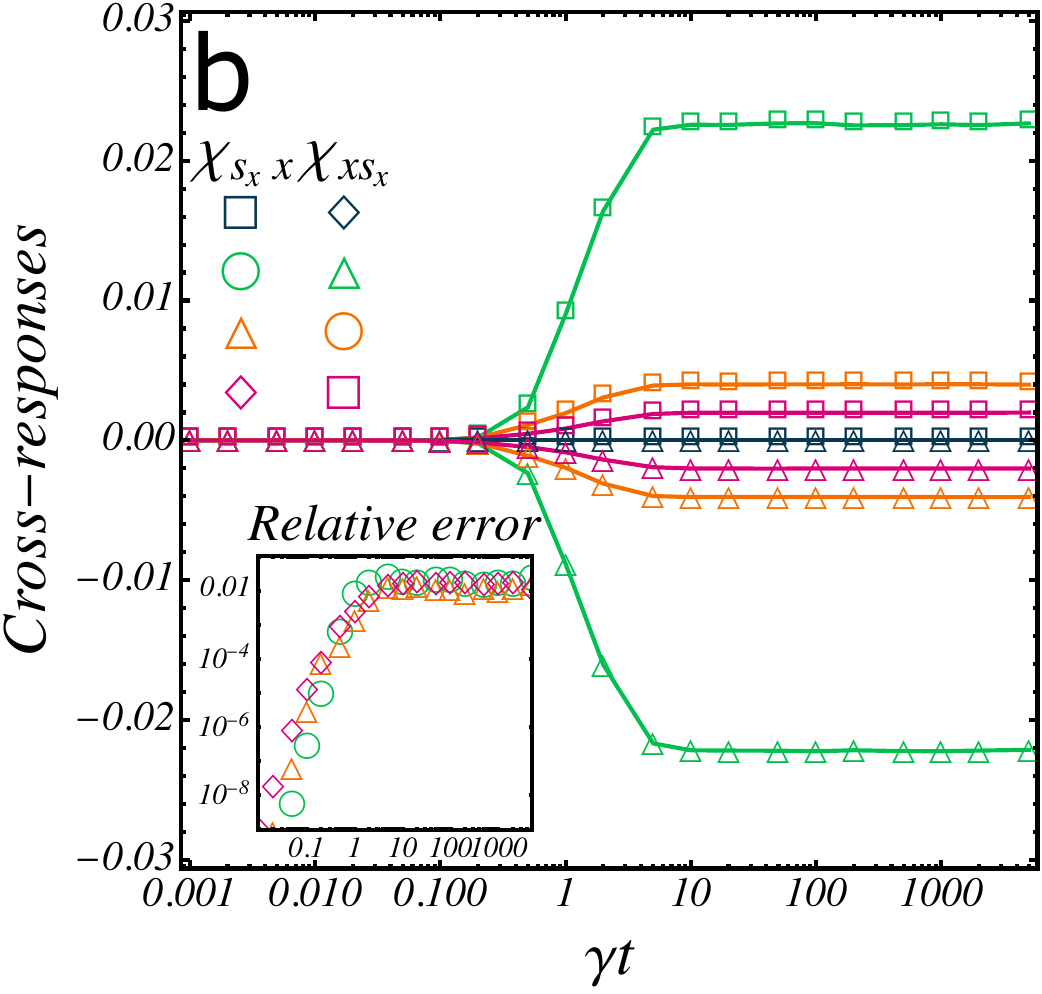}
    \caption{\textbf{Onsager-Casimir Reciprocity.}
    (a) Integrated response of $\theta$ to a force along $y$, $\chi_{\theta y}$ and 
    integrated response of $y$ to a positive torque, $\chi_{y \theta}$, across a few values of $K$.
    (b) Response of $s_x = \cos\theta$ to a force along $x$, $\chi_{s_xx}$, 
    and response of $x$ to a magnetic field along $x$, $\chi_{xs_x}$.
    In the insets, we show the relative error on the responses being opposite.
    Across this figure, $\beta^{1/2}v_0 = 5$.}
    \label{fig:Onsager_Casimir}
\end{figure}

Another non-trivial feature of the FDT is the non-standard sign linking cross-responses.
This sign leads to a violation of the standard Onsager~\cite{Onsager1931,Onsager1931a} reciprocity relations, which state that the linear susceptibility $\chi_{xy}$ of variable $x$ to the force $f_y$ that acts linearly on $y$ starting at time $0$, defined as
\begin{align}
    \chi_{xy}(t) \equiv \frac{\left\langle x(t) \right\rangle_{f_y} - \left\langle x(t)\right\rangle_0}{f_y} = \int\limits_{0}^t dt' \; R_{xy} (t,t') 
    \; ,
\end{align}
verifies $\chi_{xy} = \chi_{yx}$.
Because the spin is analogous to a magnetic moment with respect to time reversal, the reciprocity relations are here replaced by the analog of Onsager-Casimir~\cite{Casimir1945,Benenti2017} relations, $\chi_{rs} = - \chi_{sr}$.
We check that this relation holds in Fig.~\ref{fig:Onsager_Casimir}, both when considering constant torques (Fig.~\ref{fig:Onsager_Casimir}(a)) and constant vector fields (Fig.~\ref{fig:Onsager_Casimir}(b)) as spin excitations.

\subsection{Entropy Production under wrong time reversal}

Going further, the study of the path-integral representation of the dynamics lets us study the 
entropy production of the system under various hypotheses.
To do so, we introduce the Onsager-Machlup dynamical action $\mathcal{A}_{OM}$ as the noise-averaged version of the MSRJD action~\cite{Onsager1953}.
$\mathcal{A}_{\text{OM}}$ yields the probability of observing the system in state $\bm{X}_{\cal T}$ at time $t=\cal T$ given the initial condition $\bm{X}_0$ at time $t=0$ through
\begin{align}
    \mathbb{P}[\bm{X}_{\cal T} | \bm{X}_0] \propto e^{\mathcal{A}_{OM}[\bm{X}_{\cal T}| \bm{X}_0]}
    \; .
\end{align}
The ratio of the probability of a trajectory and its time-reversed version is then given by,
\begin{align}
    \frac{\mathbb{P}[\bm{X}_0 | \bm{X}_{\cal T}]}{\mathbb{P}[\bm{X}_{\cal T} | \bm{X}_0]} = e^{ \mathbb{T}\mathcal{A}_{OM}[\bm{X}_{\cal T}|\bm{X}_0] - \mathcal{A}_{OM}[\bm{X}_{\cal T}|\bm{X}_0]} \equiv e^{-\Delta S[\bm{X}_{\cal T}|\bm{X}_0]}
    \; ,
\end{align}
where $\Delta S$ is the entropy production for time $\cal T$~\cite{Seifert2005,Seifert2012}.

The calculation, which is standard but cumbersome, is reproduced in~\ref{app:EP}.
Since $\mathbb{T}$ is a true symmetry of the action, it is guaranteed that any many-body system described by Hamiltonian~\eqref{eq:Hamiltonian} and initialized in the canonical ensemble through distribution~\eqref{eq:Canonical_Distribution} will have $\Delta S = 0$.

However, it is common in the context of active matter to measure entropy productions using only partial information about the system, see \textit{e.g.}~\cite{Ro2022, Anand2024}, which was recently shown to have subtle effects depending on the symmetries of the underlying self-propulsion dynamics~\cite{Knight2025}.
This raises a natural question: if one were to observe an experiment following dynamics like those of Hamiltonian flocks but not know the underlying dynamics, would entropy production be measured?

To answer this, we first introduce a ``naïve'' time-reversal operator, $\widehat{\mathbb{T}}$, that reverses velocities but \textit{leaves spins and $\bm{v}_0$ unchanged}.
For this ``wrong'' operator, we show (see~\ref{app:SpuriousEP}) that, even when $\bm{v}_0 = \bm{0}$ (which ensures that the equilibrium distribution is the standard Boltzmann one whence free space is limited by a boundary of confining potential) there is a constant non-negative entropy production rate (EPR),
\begin{align}
 \sigma_{\widehat{\mathbb{T}}}\equiv \frac{d}{dt}\Delta S= \frac{2K^2}{\gamma},
\end{align}
where we set $\gamma_t = \gamma_r = \gamma$ for simplicity.
In other words, if one did not have fine knowledge of the dynamics that underlie the system's motion, they would be likely to measure a spurious entropy production.
We show in ~\ref{app:SpuriousEP} that the many-body EPR is extensive and has an additional positive term that depends on ferromagnetic ordering.

\subsection{Probe particle}

An even more treacherous situation is that of a standard particle with $K = 0$, \textit{e.g.} a polar colloidal particle treated as a probe, placed in a ``bath'' of our particles.
In such a situation, assume that the total Hamiltonian of the fluid with spin-velocity coupling and of the extra particle decomposes into 
\begin{align}
    \mathcal{H} = \mathcal{H}_{HF} + \mathcal{H}_C + \mathcal{H}_{\text{int}}
\end{align}
where the first term $\mathcal{H}_{HF}$ is the Hamiltonian of Eq.~(\ref{eq:Hamiltonian}), $\mathcal{H}_C$ is the (standard) kinetic energy of the colloid, and $\mathcal{H}_{\text{int}}$ is an interaction term between the colloid and the background fluid, which we assume is written as
\begin{align}
    \mathcal{H}_{\text{int}} = \sum\limits_{i=1}^N \left[ U_{\text{int}}(\bm{r}_C-\bm{r}_i) - J_{\text{int}}(\bm{r}_C - \bm{r}_i) \bm{s}_C\cdot \bm{s}_i \right]
\end{align}
where $\bm{r}_C$ and $\bm{s}_C$ represent the position and the polarity of the colloid, and $U_{\text{int}}$, $J_{\text{int}}$ represent the strengths of attracto-repulsive and alignment interactions, respectively.

There are two very different situations in this system.
The first one is that in which $J = 0$, so that the colloid is only coupled to translational degrees of freedom of the bath.
In this case, the overall system of $N+1$ particles obeys the generalized time-reversal symmetry we established for the Hamiltonian flock particles, completed by the standard operation for the colloid, namely that both its position and polarity are even under time reversal.
As a result, if one only had access to $\bm{r}_C$ and $\bm{s}_C$, it would follow the standard TRS, and its position, like those of the particles of the background fluid, would simply diffuse at long times.

In contrast, if $J \neq 0$, the only way to leave the overall system unchanged under a generalized time-reversal operation is to also flip $\bm{s}_C$.
As a result, as soon as $J_{\text{int}} \neq 0$, the usual time-reversal operation would produce a spurious entropy production even for the colloid; and Onsager-Casimir reciprocity rather than Onsager reciprocity would be observed.
Intuitively, this is due to the fact that the spin of the colloid couples to the spins of the fluid, which themselves couple to the velocities of the fluid, which finally couple to the velocity of the colloid through position variables -- meaning that the colloid itself develops an effective bath-mediated spin-velocity coupling.
In particular, this effective coupling would create an effective angular mobility, which could also be confused for an effective temperature due to a non-equilibrium effect as we discussed earlier.
In short, a standard colloid placed in a bath of particles described by Eq.~\ref{eq:Hamiltonian}, if only its own degrees of freedom were observed, would look like it follows non-equilibrium dynamics due to the generalized time-reversal symmetry followed by the bath, and the only way to evade this pitfall is to know that the polarity of the colloid itself should be flipped.

\section{Discussion}
\label{sec:discussion}
We have derived FDT equations that apply to particles following non-Galilean dynamics and thus obeying an extended time-reversal symmetry.
We have shown that these FDT equations, in spite of the introduction of a tachostat in non-Galilean dynamics, are very similar to the standard ones, differing only by the need to work in the co-moving frame. 
From this result, we showed that a standard Einstein-Smoluchowski-Sutherland relation is recovered 
for the position MSDs, with $D_T = k_BT/\gamma_t$, but that angular dynamics are much richer.
For $K > 0$, the long-time behavior of MSADs takes the form $D_R = k_BT/\gamma_{\rm eff}$,
with an effective friction that stems from the spin-velocity coupling and approaches the simple $\gamma_r$ for $K\to 0$. 
For $Kv_0 \neq 0 $ the angular dynamics additionally display Kramers-type confined behavior.
We also showed that the oddness of the spins to time-reversal symmetry leads to Onsager-Casimir reciprocity relations.
This effect, analogous to the Hall conductivity~\cite{LandauStatPhysI} naïvely appears to break standard Onsager relations.
Furthermore, observing the system naïvely, one may conclude that it produces entropy as it appears to display fluxes, as noted in previous works~\cite{Bore2016, Casiulis2019b,Casiulis2019c}.
However, we show that this is a consequence of considering the standard time-reversal symmetry operation instead of the generalized one obeyed by the system.

Beyond the specific case of this system, our results constitute a warning sign for studies of self-propelled systems.
In observations of systems that develop spontaneous velocities, it is tempting to measure entropy production, or even the breaking of the FDT, using standard definitions of the FDT.
We highlight that in polar active systems, that typically break Galilean invariance, these quantities may not use the right conservative baseline for the FDT.
Furthermore, such measurements typically rely on partial observations of the degrees of freedom of the system -- as highlighted in recent work~\cite{Dieball2025,Knight2025}, the symmetry of the non-measured degrees of freedom has deep repercussions on measured entropy productions.
In particular, the effects of Galilean invariance (or lack thereof) are still ill-understood, and may play an important role in getting correct estimates of extractable work in active matter.

In this context, the interest of understanding equilibrium non-Galilean systems is two-fold.
First, if, inspired by this work, one finds that a system that looked non-equilibrium is convincingly captured by a conservative non-Galilean theory instead, the equilibrium structure of the theory gives access to the whole toolbox of equilibrium statistical mechanics, from the definition of a Gibbs measure~\cite{Bore2016,Casiulis2019b,MyThesis} to the use of numerical tools designed for conservative systems~\cite{Bhattacharya2025}, or to the use of FDT and reciprocity as we have just shown.
Second, and perhaps more interestingly, Hamiltonian flocks have blurred the line between equilibrium systems and active ones~\cite{Bore2016,Casiulis2019,Casiulis2019b,Casiulis2019c,MyThesis,Bhattacharya2025,Chen2026}.
As a result, inspired by recent work~\cite{Shi2026} that managed to map systems with non-reciprocal interactions onto Hamiltonian systems with additional auxiliary variables and constraints, one can wonder whether some classes of active systems could likewise be mapped onto higher-dimensional Hamiltonian, yet non-Galilean, systems.
Doing so, one may be able to use numerical and analytical tools from equilibrium statistical physics and shed light on the respective effects of the non-equilibrium and non-Galilean natures of a system.

\section*{Acknowledgments}
M.C. would like to thank Satyam Anand for motivating the publication of these results, some of which were originally derived during COVID -- and then left behind with other bad memories.
M.C. would also like to thank the many people whose comments over the years helped guide the direction of this work, whether they still remember it or not, in particular (and in alphabetical order)
Éric Bertin, Anna Frishman, Jean-Noël Fuchs, Silke Henkes, Yariv Kafri, Jorge Kurchan, Sunghan Ro, Julien Tailleur, Frédéric von Wijland.
M.C. acknowledges the Israel Science Foundation under grant No. 1866/16 and the Simons Center for Computational Physical Chemistry for financial support.
This work was supported in part through the NYU IT High Performance Computing resources, services, and staff expertise.
L.F.C. acknowledges  funding from ANR-20-CE30-0031 THEMA and NYU for hospitality. We both thank 
Olivier Dauchot and Marco Tarzia for numerous discussions on this very same model. 

\appendix

\section[\hspace{2cm} Justifying the Langevin dynamics]{Justifying the Langevin dynamics \label{app:DampingDerivation}}
\label{app:Langevin}

In this section we justify the possible shapes of the noise and damping using explicit derivations of the Langevin equation, using a Zwanzig-Mori type construction~\cite{Zwanzig1961,Mori1965}.
Typically, the idea is to model the bath by a collection of independent oscillators, like in Ref.~\cite{Ford1965}.
We consider the fully deterministic Hamiltonian
\begin{align}
    \mathcal{H} = \mathcal{H}_{particles} + \mathcal{H}_{bath} + \mathcal{H}_{coupling} + \mathcal{H}_{counter-terms},
\end{align}
where $\mathcal{H}_{particles}$ contains the terms that describe the system of $N$ particles that we want to write a Langevin equation for.
Here, it reads
\begin{align}
    \mathcal{H}_{particles} &= \sum\limits_{i=1}^{N}\left[ \frac{\bm{p}_i^2}{2} + \frac{\omega_i^2}{2} - K \bm{p}_i\cdot\bm{s}_i + \sum\limits_{k\neq i}\Big( U(r_{ik}) - J(r_{ik}) \bm{s}_i\cdot \bm{s}_k\Big)\right]
    \; .
     \label{eq:Hparticles}
\end{align}
Each of these particles is coupled to $M$ independent oscillators (each set of oscillators is also independent from the others).
There is some freedom in the choice of description of these oscillators, we here present the most natural choices: a linear coupling to the angular degree of freedom of spins, and a linear coupling to the spin vector itself, showing that both yield the same Langevin dynamics in the limit of a broad spectrum of oscillator frequencies.

\subsection[\hspace{1.75cm} Linear coupling to the angle]{Linear coupling to angles}

First, we consider oscillators with positions $\bm{q}_{a,i}$, momenta $\bm{\pi}_{a,i}$, and a polarisation degree of freedom $\phi_{a,i}$ associated to momenta $\mu_{a,i}$.
They are coupled to the particles in the simplest possible way, \textit{i.e.} via linear interactions on positions and angles:
\begin{align}
    \mathcal{H}_{coupling} &= \sum\limits_{i=1}^{N}\sum\limits_{a = 1}^M \left[ c_{a,i} \bm{q}_{a,i}\cdot\bm{r}_i + d_{a,i} \theta_i \phi_{a,i} \right]
    \; ,
     \label{eq:Hcoupling}
\end{align}
with $c_{a,i}$ and $d_{a,i}$ the position and spin coupling strengths. We have chosen to couple the angular variables $\theta_i$ to the  oscillators.
This is a choice that simplifies the calculations and the Langevin equations that we will derive.
In particular, the angular Langevin equation will have additive noise and not multiplicative one, as the one found in other problems with translational and rotational degrees of freedom, such as magnetic systems~\cite{Bertotti2009} or Brownian dipoles~\cite{Cugliandolo2015}.
We comment on other choices at the end of this section.

The Hamiltonian describing the oscillators of the bath is chosen as the simplest possible harmonic energy function,
\begin{align}
    \mathcal{H}_{oscillators} &= \sum\limits_{i = 1}^{N}\sum\limits_{a = 1}^{M}\left[ \frac{\bm{\pi}_{a,i}^2}{2} + \frac{\mu_{a,i}^2}{2} + \frac{\Omega_{a,i}^2}{2} \bm{q}_{a,i}^2  + \frac{\nu_{a,i}^2}{2} \phi_{a,i}^2 \right]
    \; ,
     \label{eq:Hoscillators}
\end{align}
with $\nu_{a,i}$ and $\Omega_{a,i}$ the natural frequencies of the oscillations of the bath polarisations and positions, respectively.
Finally, one needs to add counter terms to the full Hamiltonian so that, if one assumes that the whole system is put into contact with a super-bath at inverse temperature $\beta$, when the bath degrees of freedom are integrated over in the full partition function $Z_{tot}$, one recovers (up to a constant factor) the partition function $Z_{particles}$ without any extra factor.
Here,
\begin{align}
    Z_{tot} &= \sum\limits_{\mathcal{C}} e^{- \beta \mathcal{H}} 
         = \sum\limits_{\mathcal{C}_{particles}} e^{-\beta \left(\mathcal{H}_{particles} + \mathcal{H}_{counter-terms} \right)} \sum\limits_{\mathcal{C}_{bath}} e^{- \beta \left( \mathcal{H}_{bath} + \mathcal{H}_{coupling} \right)},
\end{align}
so that the counter terms are defined by
\begin{align}
    \mathcal{H}_{counter-terms} &\equiv k_B T \ln  \sum\limits_{\mathcal{C}_{bath}} e^{- \beta \left( \mathcal{H}_{bath} + \mathcal{H}_{coupling} \right)}  
    = k_BT \ln Z_{bath}
    \; . 
\end{align}

The partition function of the bath is given by 
\begin{align}
    Z_{bath} &\equiv \int 
    \prod\limits_{i=1}^{N}\prod\limits_{a = 1}^{M} d^2 \bm{q}_{a,i}  d^2\bm{\pi}_{a,i} d\phi_{a,i} d\mu_{a,i}
    \ e^{-\beta \left( \frac{\bm{\pi}_{a,i}^2}{2} + \frac{\mu_{a,i}^2}{2} + \frac{\Omega_{a,i}^2}{2} \bm{q}_{a,i}^2 + \frac{\nu_{a,i}^2}{2} \phi_{a,i}^2 + c_{a,i} \bm{q}_{a,i}\cdot\bm{r}_i + d_{a,i} \theta_i \phi_{a,i} \right) } \; . 
\end{align}
The integrals over momenta (both linear and angular) can be computed and yield a constant that one can include into an irrelevant normalization factor.
The remaining integrals can be factorised, since the oscillators of the bath are not coupled to each other, 
\begin{align}
    Z_{bath} &\propto \prod\limits_{i = 1}^N\prod\limits_{a = 1}^{M}  \left[ \int d^2\bm{q} e^{-\beta \left( \frac{\Omega_{a,i}^2}{2} \bm{q}^2 +\bm{q}\cdot c_{a,i}  \bm{r}_i  \right) }  \int d\phi   \ e^{- \beta \left( \frac{\nu_{a,i}^2}{2} \phi^2 + \phi d_{a,i}  \theta_i  \right)} \right]
    \; .
\end{align}
The integral over positions is simply a Gaussian integral:
\begin{align}
    \int d^2\bm{q} \ e^{-\beta \left[ \frac{\Omega_{a,i}^2}{2} \bm{q}^2 + c_{a,i} \bm{q}\cdot \bm{r}_i  \right] } &= \frac{2 \pi}{\beta \Omega_{a,i}^2} \ e^{\beta \frac{c_{a,i}^2}{2 \Omega_{a,i}^2} \bm{r}_i^2 }
\end{align}
and, likewise, the integral over $\phi$'s is also Gaussian. 
If one assumes that $\phi_a$ takes values  in $\left(- \infty, \infty \right)$ (which does \textit{not} represent a realistic bath of particles with unit spins each oriented with angle $\phi_{a,i}$, but \textit{could} represent some coarse-grained field due to very small particles in the bath that impose an effective field $\phi_{a,i}$), one has
\begin{align}
    \int d\phi \ e^{- \beta \left( \frac{\nu_{a,i}^2}{2} \phi^2 +  d_{a,i} \phi \theta_i  \right)} 
    &= \sqrt{\frac{2 \pi}{\beta \nu_{a,i}^2}} \ e^{\beta \frac{d_{a,i}^2}{2 \nu_{a,i}^2} \theta_i^2 }
    \; .
\end{align}
Putting these results together
\begin{align}
    Z_{bath} &\propto \prod\limits_{a = 1}^M e^{\beta \frac{c_{a,i}^2}{2 \Omega_{a,i}^2} \bm{r}_i^2  + \beta \frac{d_{a,i}^2}{2 \nu_{a,i}^2} \theta_i^2 }
    \; , 
\end{align}
with a sub-exponential prefactor. As a result, the counter-terms needed to avoid a drift of the positions and angles in the effective equations on the tagged particles read, at leading order:
\begin{align}
    \mathcal{H}_{counter-terms} &\equiv k_B T \ln Z_{bath} 
    = \sum\limits_{i = 1}^N\sum\limits_{a = 1}^{M}\left[\frac{c_{a,i}^2}{2 \Omega_{a,i}^2} \bm{r}_i^2  +  \frac{d_{a,i}^2}{2 \nu_{a,i}^2}\theta_i^2\right]
    \; . 
\label{eq:Hcounter-term}
\end{align}

We now have all terms that contribute to the total Hamiltonian, they are given in Eqs.~\eqref{eq:Hparticles}, \eqref{eq:Hcoupling}, \eqref{eq:Hoscillators}, \eqref{eq:Hcounter-term}.
The associated Hamiltonian equations of motion are, for the bath
\begin{align}
\begin{split}
    \dot{\bm{q}}_{a,i}   &= \bm{\pi}_{a,i} \; , \\
    \dot{\bm{\pi}}_{a,i} &= - \Omega_{a,i}^2 \bm{q}_{a,i} - c_{a,i} \bm{r}_i \;  ,\\
    \dot{\phi}_{a,i} &= \mu_{a,i} \; , \\
    \dot{\mu}_{a,i} &= - \nu_{a,i}^2 \phi_{a,i} - d_{a,i} \theta_i \; ,
    \end{split}
\end{align}
and they can be rewritten as two Lagrangian equations of motion,
\begin{align}
\begin{split}
    \ddot{\bm{q}}_{a,i} + \Omega_{a,i}^2 \bm{q}_{a,i} &=  - c_{a,i} \bm{r}_i \;   ,\\
    \ddot{\phi}_{a,i} + \nu_{a,i}^2 \phi_{a,i} &= - d_{a,i}  \theta_i \; .
    \end{split}
\end{align}
They can both be solved using the initial conditions on both variables and their time-derivative, yielding
\begin{align}
\begin{split}
    \bm{q}_{a,i}(t) &= \bm{q}_{a,i}(0) \cos \Omega_{a,i} t + \frac{\dot{\bm{q}}_{a,i}(0)}{\Omega_{a,i}} \sin \Omega_{a,i} t - \frac{c_{a,i}}{\Omega_{a,i}} \int\limits_{0}^t dt' \sin \Omega_{a,i} \left(t - t'\right) \bm{r}_i(t')
    \; , \\
    \phi_{a,i}(t) &= \phi_{a,i}(0) \cos \nu_{a,i} t + \frac{\dot{\phi}_{a,i}(0)}{\nu_{a,i}} \sin \nu_{a,i} t - \frac{d_{a,i}}{\nu_{a,i}} \int\limits_{0}^t dt' \sin \nu_{a,i} \left(t - t'\right) \theta_i(t')
    \; .
    \end{split}
\end{align}
It is convenient to rewrite the last integrals using an integration by parts, for instance,
\begin{align}
    & \int\limits_{0}^t dt' \sin \Omega_{a,i} \left(t - t'\right) \bm{r}_i(t') = \left[ \frac{\cos \Omega_a (t - t')}{\Omega_{a,i}} \bm{r}_i(t')\right]_{0}^{t} - \int dt' \  \frac{\cos \Omega_{a,i} (t - t')}{\Omega_{a,i}} \ \dot{\bm{r}}_i(t') \nonumber \\
    & \qquad\quad = \frac{\bm{r}_i (t)}{\Omega_{a,i}} - \frac{\cos \Omega_{a,i} t}{\Omega_{a,i}} \bm{r}_i(0) - \int dt' \  \frac{\cos \Omega_{a,i} (t - t')}{\Omega_{a,i}} \ \dot{\bm{r}}_i(t') 
    \; .
\end{align}
Using the same trick on the other integral, the time evolution of the degrees of freedom of the bath reads
\begin{align}
    \bm{q}_{a,i}(t) &= \cos \Omega_{a,i} t \left(\bm{q}_{a,i}(0) +  \frac{c_{a,i} \bm{r}_i (0)}{\Omega_{a,i}^2} \right) + \frac{\dot{\bm{q}}_{a,i}(0)}{\Omega_{a,i}} \sin \Omega_{a,i} t 
    \nonumber\\
    & \quad + \frac{c_{a,i}}{\Omega_{a,i}^2}  \left[\bm{r}_i (t) + \int\limits_{0}^t dt' \cos \Omega_{a,i} \left(t - t'\right) \dot{\bm{r}}_i(t') \right] 
    \; , \\
    \phi_a(t) &=  \cos \nu_{a,i} t \left( \phi_{a,i}(0) + \frac{d_{a,i} \theta_i (0)}{\nu_{a,i}^2} \right) + \frac{\dot{\phi}_{a,i}(0)}{\nu_{a,i}} \sin \nu_{a,i} t 
    \nonumber\\
    & \quad + \frac{d_{a,i}}{\nu_{a,i}^2} \left[\theta_i (t) + \int\limits_{0}^t dt' \cos \nu_{a,i} \left(t - t'\right) \dot{\theta}_i(t') \right]
    \; .
\end{align}

The Hamiltonian equations of motion for the tagged particles, on the other hand, read
\begin{align}
\begin{split}
    \dot{\bm{r}}_i &= \bm{p}_i - K \bm{s}_i \; , 
    \\
    \dot{\bm{p}}_i &= - \sum\limits_{a = 1}^{M} \left[c_{a,i} \bm{q}_{a,i} + \frac{c_{a,i}^2}{\Omega_{a,i}^2} \bm{r}_i \right] + \sum\limits_{k (\neq i)} \left(\frac{\partial J (r_{ik})}{\partial \bm{r}_i} \cos\theta_{ik} - \frac{\partial U (r_{ik})}{\partial \bm{r}_i} \right)
    \; , \\
    \dot{\theta}_i &= \omega_i
    \; , \\
    \dot{\omega}_i &= -\sum\limits_{a = 1}^{M} \left[d_{a,i} \phi_{a,i} + \frac{d_{a,i}^2}{\nu_{a,i}^2} \theta_i \right] + K \bm{p}_i \cdot \bm{s}_{\perp, i} + \sum\limits_{k (\neq i)} 
    J(r_{ik})\sin\theta_{ik}
    \;,
   \end{split}
\end{align}
where $\bm{s}_{\perp,i} = (-\sin\theta_i, \cos\theta_i) = \partial_{\theta_i}\bm{s}_i$ is introduced to make equations more palatable.
In these equations, one can notice the need to add the counter-term in the Hamiltonian: terms arising from $\mathcal{H}_{counter-terms}$ cancel out drift terms coming from the solutions for $\bm{q}_a(t)$ and $\phi_a(t)$.
We then use the explicit expressions of $\bm{q}_a(t)$ and $\phi_a(t)$ to write
\begin{align}
    \dot{\bm{r}}_i = & \; \bm{p}_i - K \bm{s}_i
    \; , 
    \nonumber\\
    \dot{\bm{p}}_i = &  - \sum\limits_{a = 1}^{M} c_{a,i}  \left[\cos \Omega_{a,i} t \left(\bm{q}_{a,i}(0) + \frac{c_{a,i} \bm{r}_i (0)}{\Omega_{a,i}^2} \right) + \frac{\dot{\bm{q}}_{a,i}(0)}{\Omega_{a,i}}     \sin \Omega_{a,i} t \right.  
    \nonumber\\
    & \qquad\qquad\;\;   \left. + \frac{c_{a,i}}{\Omega_{a,i}^2}  \int\limits_{0}^t dt' \cos \Omega_{a,i} \left(t - t'\right) \dot{\bm{r}}_i(t')\right]  
     + \sum\limits_{k (\neq i)} \left(\frac{\partial J (r_{ik})}{\partial \bm{r}_i} \cos\theta_{ik} - \frac{\partial U (r_{ik})}{\partial \bm{r}_i} \right)
    \; , 
    \nonumber\\
    \dot{\theta}_i = & \; \omega_i 
    \; , \\
    \dot{\omega}_i = & \; -\sum\limits_{a = 1}^{M} d_{a,i} \left[ \cos \nu_{a,i} t \left( \phi_{a,i}(0) +\frac{d_{a,i} \theta_i (0)}{\nu_{a,i}^2} \right) + \frac{\dot{\phi}_{a,i}(0)}{\nu_{a,i}} \sin \nu_{a,i} t 
    \right. 
    \nonumber\\
    & 
    \left. \qquad\qquad \;\; + \frac{d_{a,i}}{\nu_{a,i}^2} \int\limits_{0}^t dt' \cos \nu_{a,i} \left(t - t'\right) \dot{\theta}_i(t') \right] 
    + K \bm{p}_i \cdot \bm{s}_{\perp, i} + \sum\limits_{k (\neq i)} J(r_{ik})\sin\theta_{ik} 
    \; . 
    \nonumber
\end{align}
The terms coming from the coupling to the bath are of two kinds: one set of terms simply yield instantaneous forces that do not depend on the history of the trajectories of the tagged particles, only on initial conditions,
\begin{align}
\begin{split}
    \bm{\eta}_i (t) &\equiv  - \sum\limits_{a = 1}^{M} c_{a,i}  \left[\cos \Omega_{a,i} t \left(\bm{q}_{a,i}(0) + \frac{c_{a,i} \bm{r}_i (0)}{\Omega_{a,i}^2} \right) + \frac{\dot{\bm{q}}_{a,i}(0)}{\Omega_{a,i}} \sin \Omega_{a,i} t \right] 
    \; , \\
    \xi_i(t) &\equiv -\sum\limits_{a = 1}^{M} d_{a,i} \left[ \cos \nu_{a,i} t \left( \phi_{a,i}(0) + \frac{d_{a,i} \theta_i (0)}{\nu_{a,i}^2} \right) + \frac{\dot{\phi}_{a,i}(0)}{\nu_{a,i}} \sin \nu_{a,i} t\right]
    \; ,
    \end{split}
\end{align}
and another set of terms explicitly depend on the history of the velocities of the particles, in which we define
\begin{align}
\begin{split}
    \Gamma_{t,i}(t - t')&\equiv \sum\limits_{a = 1}^{M} \frac{c_{a,i}^2}{\Omega_{a,i}^2} \cos \Omega_{a,i} \left(t - t'\right), \\
    \Gamma_{r,i}(t - t') &\equiv \sum\limits_{a = 1}^{M} \frac{d_{a,i}^2}{\nu_{a,i}^2} \cos \nu_{a,i} \left(t - t'\right).
    \end{split}
\end{align}
Using these definitions, the equations of motion of the tagged particle then read
\begin{align}
\begin{split}
    \dot{\bm{r}}_i &= \bm{p}_i - K \bm{s}_i
    \; , \\
    \dot{\bm{p}}_i &= \sum\limits_{k (\neq i)} \left(\frac{\partial J (r_{ik})}{\partial \bm{r}_i} \cos\theta_{ik} - \frac{\partial U (r_{ik})}{\partial \bm{r}_i} \right) + \bm{\eta}_i(t) 
    - \int_0^t dt' \ \Gamma_{t,i}(t - t') \dot{\bm{r}}_i (t') \; , \\
    \dot{\theta}_i &= \omega_i \; , \\
    \dot{\omega}_i &= K \bm{p}_i \cdot \bm{s}_{\perp, i} + \sum\limits_{k (\neq i)} J(r_{ik})\sin\theta_{ik} + \xi_i(t) - 
    \int_0^t dt'  \ \Gamma_{r,i}(t - t') \dot{\theta}_i (t')
    \; .
    \end{split}
\end{align}
These are still deterministic equations, but they depend on the precise initial conditions and dynamics of the collection of oscillators in the bath.
One can check that, if one assumes that the degrees of freedom of the oscillators are initially drawn from an equilibrium distribution at inverse temperature $\beta$,
\begin{align}
    & \mathbb{P}(\left\{\bm{q}_{a,i}(0),\bm{\pi}_{a,i}(0),\phi_{a,i}(0), \mu_{a,i}(0) \right\}
    \! ; \{\bm{r}_i(0), \theta_i(0)\}) 
    \nonumber\\
    & \qquad\qquad\qquad\qquad \equiv \frac{1}{\mathcal{N}} 
    e^{-\beta \left(\mathcal{H}_{oscillators} + \mathcal{H}_{counter-terms} + \mathcal{H}_{coupling} \right)}
    \; ,
\end{align}
the mean value of the force terms above, averaged over initial conditions, have Gaussian statistics and read
\begin{align}
\begin{split}
   &  \left\langle \bm{\eta}_i(t) \right\rangle_0 = \bm{0}
   \; , 
    \qquad\qquad
        \left\langle \bm{\eta}_i(t) \cdot  \bm{\eta}_j(t')   \right\rangle_0 = \beta^{-1} \delta_{ij} \Gamma_{t,i}(t-t')
        \; , \\
   &  \, \left\langle \xi_i(t) \right\rangle_0 = 0
   \; , 
     \qquad\qquad
    \left\langle \xi_i(t)\xi_j(t')   \right\rangle_0 = \beta^{-1} \delta_{ij} \Gamma_{r,i}(t-t')
    \; .
    \end{split}
\end{align}
Finally, in the limit of oscillators with a very broad distribution of frequencies, the kernels tend to Dirac $\delta$ functions, 
\begin{align}
    & \Gamma_{t,i} (t-t') \to 2 \gamma_t \delta(t -t')
    \; , \qquad\qquad
    \Gamma_{r,i} (t-t') \to 2 \gamma_r \delta(t -t')
    \; ,
\end{align}
and the evolution equations simplify to 
\begin{align}
\begin{split}
    \dot{\bm{r}}_i &= \bm{p}_i - K \bm{s}_i 
    \; , \\
    \dot{\bm{p}}_i &= \sum\limits_{k (\neq i)} \left(\frac{\partial J (r_{ik})}{\partial \bm{r}_i} \cos\theta_{ik} - \frac{\partial U (r_{ik})}{\partial \bm{r}_i} \right) + \bm{\eta}_i(t) - \gamma_{t} \dot{\bm{r}}_i  
    \; , \\
    \dot{\theta}_i &= \omega_i
    \; , \\
    \dot{\omega}_i &= K \bm{p}_i \cdot \bm{s}_{\perp, i} + \sum\limits_{k (\neq i)} J(r_{ik})\sin\theta_{ik} + \xi_i(t) - \gamma_{r}  \dot{\theta}_i
    \; .
\end{split}
\end{align}
The integral of the memory kernel yields only $\gamma_{r,t}$, without the factor of $2$, as the Dirac delta selects a boundary value.
These equations have the form proposed for a Langevin equation with damping acting on the velocities, for the special choice $\bm{v}_0 = \bm{0}.$ 
The simple additive noise structure in the angular equation is due to the choice of coupling to the bath variables that we made.

In order to recover the proposed $\bm{v}_0$ dependence, that corresponds to a velocity imposed by the (Galilean invariant) bath, one should add to the Hamiltonian a term that reads
\begin{align}
    \mathcal{H}_{friction} \equiv - \sum\limits_{i = 1}^N \bm{r}_i \cdot \int\limits_{0}^{t} dt' \ \Gamma_{t,i}(t-t') \bm{v}_0
    \; ,
\end{align}
which simply removes the work of the friction force between the bath and the tagged particles, when the latter are moving at $\bm{v}_0$.
In other words, it is an energetic way of stating that the bath sets the velocity of the tagged particles to $\bm{v}_0$ (in simple cases, it should mean that the bath itself is moving at $\bm{v}_0$).

In the main text, we consider the simplest possible case, that of a memoryless damping, so that 
\begin{align}
    \Gamma_{t,i} (t-t') = 2 \gamma_t \delta(t-t')
    \; , 
    \qquad\qquad
    \Gamma_{r,i} (t-t') = 2 \gamma_r \delta(t-t')
    \; . 
\end{align}

\subsection[\hspace{1.75cm} Linear coupling to the spin]{Linear coupling to the spins}

Another natural choice of environment would be one that does not couple linearly to the angle, but linearly to the spin.
In that case, one would consider unit vector variables $\bm{\sigma}_{a,i} = (\cos\vartheta, \sin\vartheta)$ in the environment such that the coupling would be $- d_{a,i} \sigma_{a,i}\cdot \bm{s}_i$ for each spin-environment pair.
Here, the spin part of the damping and noise are standard, so the construction is similar to that of past works on Langevin-level equations for magnetic moments expressed in their vector form~\cite{Brown1963,Aron2014,DosSantos2023}.

In short, the spin part of the coupling Hamiltonian now reads
\begin{align}
    \mathcal{H}_{\text{coupling}}^{\text{spin}} = -\sum_{i=1}^{N}\sum_{a=1}^{M} d_{a,i}\,\bm{\sigma}_{a,i}\!\cdot\!\bm{s}_i \;,
    \label{eq:Hcoupling_vec}
\end{align}
while the part of the Hamiltonian that controls the oscillators coupled to the spins reads
\begin{align}
    \mathcal{H}_{\text{oscillators}}^{\text{spin}} = \sum_{i=1}^{N}\sum_{a=1}^{M}\left[ \frac{\bm{\mu}_{a,i}^2}{2} + \frac{\nu_{a,i}^2}{2}\,\bm{\sigma}_{a,i}^2\right].
    \label{eq:Hoscillators_vec}
\end{align}
The $\bm{\sigma}$ part of the partition function is then given by
\begin{align}
    Z_{\text{bath}}^{\text{spin}}
    &\propto \prod_{i,a} \int d^2\bm{\sigma} \; 
    e^{-\beta\left(\frac{\nu_{a,i}^2}{2}\bm{\sigma}^2 -d_{a,i}\,\bm{\sigma}\cdot\bm{s}_i\right)} = \prod_{i,a} \frac{2\pi}{\beta\nu_{a,i}^2}\; e^{\,\beta\frac{d_{a,i}^2}{2\nu_{a,i}^2}\,|\bm{s}_i|^2} = \rm cst
    \; ,
\end{align}
where we use the unit-vector nature of the spins.
Since this partition function is a constant, the corresponding counter terms are constant,
\begin{align}
    \mathcal{H}_{\text{counter-terms}}^{\text{spin}} = \sum_{i,a}\frac{d_{a,i}^2}{2\nu_{a,i}^2}
    \; .
    \label{eq:Hcounterterm_vec}
\end{align}

Then, analogously to the vector variables connected to oscillator positions in the previous derivation, 
the equations of motion for the new bath vector variables are
\begin{align}
    \ddot{\bm{\sigma}}_{a,i} + \nu_{a,i}^2\,\bm{\sigma}_{a,i} = d_{a,i}\,\bm{s}_i(t)
    \; .
    \label{eq:bath_eom_vec}
\end{align}
The solutions are given by
\begin{align}
    \bm{\sigma}_{a,i}(t)
    &= \bm{\sigma}_{a,i}(0)\cos\nu_{a,i} t + \frac{\dot{\bm{\sigma}}_{a,i}(0)}{\nu_{a,i}}\sin\nu_{a,i} t + \frac{d_{a,i}}{\nu_{a,i}} \int_0^t\!dt'\;\sin\nu_{a,i}(t-t')\,\bm{s}_i(t')
    \; ,
    \label{eq:sigma_sol_raw}
\end{align}
and the integral term can be integrated by parts to yield 
\begin{align}
    \bm{\sigma}_{a,i}(t)
    &= \left(\bm{\sigma}_{a,i}(0) - \frac{d_{a,i}\,\bm{s}_i(0)}{\nu_{a,i}^2}\right) \, \cos\nu_{a,i} t 
    + \frac{\dot{\bm{\sigma}}_{a,i}(0)}{\nu_{a,i}}\sin\nu_{a,i} t
    \nonumber\\
    &\quad
    + \frac{d_{a,i}}{\nu_{a,i}^2}\,\bm{s}_i(t)
    - \frac{d_{a,i}}{\nu_{a,i}^2}
    \int_0^t\!dt'\;\cos\nu_{a,i}(t\!-\!t')\,\dot{\bm{s}}_i(t')
    \; .
    \label{eq:sigma_sol}
\end{align}
Since the counter-terms for the spins are constant, the equations of motions on the rotational velocities is simply given by
\begin{align}
    \dot{\omega}_i = K\,\bm{p}_i\cdot\!\bm{s}_{\perp,i} + \sum_{k(\neq i)} J(r_{ik})\sin\theta_{ik} + \sum_{a=1}^{M} d_{a,i}\,\bm{\sigma}_{a,i}(t)\!\cdot\!\bm{s}_{\perp,i}(t).
    \label{eq:omega_eom_vec}
\end{align}
We now insert~\eqref{eq:sigma_sol} into~\eqref{eq:omega_eom_vec} and project onto $\bm{s}_{\perp,i}(t)$.
The instantaneous term proportional to $\bm{s}_i(t)$ in~\eqref{eq:sigma_sol} yields
\begin{align}
    \sum_a \frac{d_{a,i}^2}{\nu_{a,i}^2}\,
    \bm{s}_i(t)\!\cdot\!\bm{s}_{\perp,i}(t) = 0.
\end{align}
This is the intuitive reason why no spin counter-term is needed: the unit-length constraint $|\bm{s}|=1$ ensures that the ``self-interaction'' term that the counter-term would cancel vanishes automatically.

Finally, combining all spin-related equations, and defining
\begin{align}
    \bm{\xi}_i(t) &\equiv \sum_{a=1}^{M} d_{a,i}\left[ \left(\bm{\sigma}_{a,i}(0) - \frac{d_{a,i}\,\bm{s}_i(0)}{\nu_{a,i}^2}\right) \cos\nu_{a,i} t + \frac{\dot{\bm{\sigma}}_{a,i}(0)}{\nu_{a,i}}\sin\nu_{a,i} t \right]\label{eq:vec_noise} \\
    \Gamma_{r,i}(t-t') &\equiv \sum_{a=1}^{M} \frac{d_{a,i}^2}{\nu_{a,i}^2}\,\cos\nu_{a,i}(t-t'),
    \label{eq:kernel_vec}
\end{align}
the full equation of motion for spins can be rewritten as
\begin{align}
    \dot\omega_i = &
    \; K\,\bm{p}_i\!\cdot\!\bm{s}_{\perp,i} + \sum_{k\neq i}J(r_{ik})\sin\theta_{ik} + \bm{\xi}_i(t)\cdot\bm{s}_{\perp,i}(t) 
    \nonumber\\
    & 
    - \int_0^t\!dt'\; \Gamma_{r,i}(t-t') \cos[\theta_i(t)-\theta_i(t')] \, \dot\theta_i(t')
    \; .
\end{align}
The correlations and noise are slightly more complicated than when considering the angular coupling for a generic $\Gamma$.
In particular, the initial-condition-averaged correlations of the projected noise term verify
\begin{align}
    \left\langle \xi_{\perp,i}(t)\xi_{\perp,j}(t') \right\rangle_0
    &= \frac{1}{\beta}\delta_{ij}\Gamma_{r,i}(t-t') \bm{s}_{\perp,i}(t)\cdot\bm{s}_{\perp,i}(t')\nonumber\\
    &= \frac{1}{\beta}\,\delta_{ij}\,\Gamma_{r,i}(t-t') \cos\left[\theta_i(t) - \theta_i(t')\right].
    \label{eq:projected_noise_cov}
\end{align}
It is no longer a function of $t-t'$ alone, and instead depends on the angular trajectory explicitly reflecting the multiplicative character of the noise.
However, in the limit of white noise, $\Gamma_{r,i}(t-t') \to 2\gamma_r\,\delta(t-t')$, the cosine factor has limit $1$, so that the angular equation of motion becomes
\begin{align}
    \dot\omega_i &= K\,\bm{p}_i\!\cdot\!\bm{s}_{\perp,i} + \sum_{k\neq i}J(r_{ik})\sin\theta_{ik} + \bm{\xi}_i(t)\!\cdot\!\bm{s}_{\perp,i}(t) - \gamma_r\omega_i.
    \label{eq:Langevin_angle_vec}
\end{align}

Finally, one may rewrite the equation of motion purely in terms of the spin vector for a single particle, which yields
\begin{align}
	\ddot{\bm{s}} + |\dot{\bm{s}}|^2 \bm{s} - K (\dot{\bm{r}} \cdot \bm{s}_{\perp})\bm{s}_{\perp} + \gamma_r \dot{\bm{s}} - \sqrt{2 \gamma_r k_B T} \xi \bm{s}_{\perp} &= \bm{\Gamma}_{\rm ext}
	\;,
    \label{eq:Langevin-s}
\end{align}
where $\bm{\Gamma}_{\rm ext}$ is the deterministic external torque acting on the spin, and both the second term and the multiplicative noise term proportional to the Gaussian scalar white noise $\xi$ ensure that the spin and does not change its modulus.

\section[\hspace{2cm} From dynamics to FDT and Entropy Production]{From dynamics to FDT then Entropy Production\label{app:EveryAction_Thing}}

In this Appendix, we go through the full construction of the Martin-Siggia-Rose-Janssen-De Dominicis (MSRJD) dynamical action~\cite{Martin1973,Janssen1976,DeDominicis1976,deDominicis1978,ArnoulxdePirey2022}, the derivation of the right generalized time-reversal operation, the derivation of the corresponding fluctuation-dissipation theorem, then the switch to the Onsager-Machlup~\cite{Onsager1953} action, and the derivation of results on entropy productions.
All calculations within this section are cumbersome but nowadays rather standard~\cite{ArnoulxdePirey2022}, since the dynamics at hand do not contain explicit colored or multiplicative noise.

\subsection[\hspace{1.75cm} MSRJD action construction]{MSRJD action construction\label{app:MSRJD}}

For any observable $A(\bm{r},\bm{p},\theta,\omega)$ that depends on all  variables that characterize the particle, 
 the mean value over realisations of the noise and the initial conditions can be written as a path integral
\begin{align}
   & 
    \left\langle A(\bm{r}, \bm{p},\theta, \omega) \right\rangle = 
    \Big\langle \int \mathcal{D}\left[\bm{r},\bm{p}\right] \mathcal{D}\left[\theta,\omega\right]  d\bm{r}_0 d\bm{p}_0 d\theta_0 d\omega_0 
    \ P_0(\bm{r}_0, \bm{p}_0, \theta_0, \omega_0) 
    \  A(\bm{r}, \bm{p},\theta, \omega)  
    \nonumber\\
    &  \qquad \qquad \qquad \qquad \qquad
    \times
     \prod\limits_{t} \delta\left(\dot{\bm{r}} - \bm{p} + K \bm{s} \right) \delta\left(\dot{\theta} - \omega \right)
  \nonumber \\
    & \qquad \qquad \qquad \qquad \qquad
    \times  
    \delta\left( \dot{\bm{p}} + \gamma_t \left(\bm{p} - K \bm{s} - \bm{v}_0 \right) - \sqrt{2 \gamma_t k_B T}\bm{\eta} \right) 
     \nonumber \\
    & \qquad \qquad \qquad \qquad \qquad
    \times  
    \delta\left( \dot{\omega} - K \bm{p} \cdot \bm{s}_{\perp} +  \gamma_r \omega - \sqrt{2 \gamma_r k_B T} \xi \right)
    \Big{\rangle}
\; .
\end{align}
$P_0(\bm{r}_0, \bm{p}_0, \theta_0, \omega_0)$ is the distribution of the particle's phase space variables 
in the considered ensemble of initial conditions.
A reasonable choice would be the canonical equilibrium distribution of these coordinates, at the temperature and velocity imposed 
by the bath, that is
\begin{align}
    P_0(\bm{r}_0, \bm{p}_0, \theta_0, \omega_0) &= Z^{-1} \; 
    e^{- \beta \left( \frac{p_0^2}{2} + \frac{\omega_0^2}{2} - K \bm{p}_0\cdot \bm{s}_0 - \bm{v_0}\cdot \bm{p_0}\right)}
\end{align}
with $\beta = 1/(k_B T)$. The symbols $\mathcal{D}\left[\bm{r},\bm{p}\right] $ and $\mathcal{D}\left[\theta,\omega\right] $
represent the path integral elements, {\it e.g.} $\mathcal{D} \bm{r} = \prod_{k=1}^N d\bm{r}(t_k)$ with the discretized time 
$t_k = k\Delta t$ and $k=0, \dots, N$, $t_0$ the initial time at which $\bm{r}_0$, {\it etc.} are measured, and $\Delta t$ an infinitesimal increment.  We grouped them in this way for notational convenience. Here and in the following we use a  
continuous time notation that should be interpreted as the 
limit of a proper time-discretization that we do not need to specify in this Section.
Indeed, since our equations are underdamped, with second-order time derivatives, there are no discretization issues to worry about, see, e.g.~\cite{Aron2010}.
Therefore, products over time are simply 
denoted $\prod\limits_t$ and include the initial time.
For compactness, we will henceforth use the notation $\bm{X}_0 = \left(\bm{r}_0, \bm{p}_0, \theta_0, \omega_0\right)$ for the 
initial conditions and the definitions
\begin{align}
    & \delta_{\bm{p}} \equiv \prod\limits_t \delta\left(\dot{\bm{r}} - \bm{p} + K \bm{s} \right), \qquad\qquad
    \delta_{\omega} \equiv \prod\limits_t \delta\left(\dot{\theta} - \omega \right).
\end{align}

One can then rewrite the mean value $\langle A \rangle $ using the Fourier representation of the $\delta$ at each (discrete) 
time,
\begin{align}
\begin{split}
    \delta\left( \dot{\bm{p}} + \gamma_t \left(\bm{v} - \bm{v}_0\right) - \sqrt{2 \gamma_t k_B T}\bm{\eta} \right)  &= 
    \int d{\bm q} \; e^{-i \bm{q}\cdot \left(\dot{\bm{p}} + \gamma_t \left(  \bm{v} - \bm{v}_0 \right) - \sqrt{2 \gamma_t k_B T}\bm{\eta}\right)} \; , \\
    \delta\left( \dot{\omega} - K \bm{p} \cdot \bm{s}_{\perp} +  \gamma_r \omega - \sqrt{2 \gamma_r k_B T} \xi \right) &= 
    \int d\lambda \;  e^{-i \lambda \left(\dot{\omega} - K \bm{p} \cdot \bm{s}_{\perp} +  \gamma_r \omega - \sqrt{2 \gamma_r k_B T} \xi \right)}
    \; .
    \end{split}
\end{align}
The product over times can then be used to turn these integrals into functional integrals, and one can exponentiate the product to yield time-integrals in the exponentials, leading to
\begin{align}
    & \left\langle A(\bm{r}, \bm{p},\theta, \omega) \right\rangle = 
    \left\langle \int \mathcal{D}\left[\bm{r},\bm{p},\bm{q}\right] \mathcal{D}\left[\theta,\omega,\lambda\right] d\bm{X}_0 \
     \ P_0(\bm{X}_0) \ A(\bm{r}, \bm{p},\theta, \omega)  \ \delta_{\bm{p}}\delta_{\omega}  \right. 
    \nonumber\\
    & \qquad\qquad\qquad \times \left. e^{-i \int dt \bm{q}\cdot \left(\dot{\bm{p}} + \gamma_t \left(\bm{v} - \bm{v}_0\right) - \sqrt{2 \gamma_t k_B T}\bm{\eta}\right) -i \int dt \lambda \left(\dot{\omega} - K \bm{p} \cdot \bm{s}_{\perp} +  \gamma_r \omega - \sqrt{2 \gamma_r k_B T} \xi \right)} \right\rangle.
\end{align}
We now isolate the stochastic terms in the mean value and take the noise average within the integral to write
\begin{align}
    \left\langle A(\bm{r}, \bm{p},\theta, \omega) \right\rangle = &
    \int \mathcal{D}\left[\bm{r},\bm{p},\bm{q}\right] \mathcal{D}
    \left[\theta,\omega,\lambda\right] d\bm{X}_0 \ P_0(\bm{X}_0) \ A(\bm{r}, \bm{p},\theta, \omega) \ \delta_{\bm{p}}\delta_{\omega} 
     \nonumber \\
    & \qquad 
    \times  
    e^{-i \int dt \ \left[ \bm{q}\cdot\left( \dot{\bm{p}} + \gamma_t \left(\bm{v} - \bm{v}_0\right) \right) + \lambda \left( \dot{\omega} - K \bm{p} \cdot \bm{s}_{\perp} +  \gamma_r \omega \right)\right] }  \nonumber\\ 
    & \qquad
    \times \left\langle  e^{i \sqrt{2\gamma_t k_B T} \int dt \ \bm{q}\cdot  \bm{\eta} + i \sqrt{2 \gamma_r k_B T} \int dt \ \lambda\xi } \right\rangle.
\end{align}
Next, one usually uses a property of averages over Gaussian variables: if an average is performed over a Gaussian random variable $x$ with zero mean and variance $\sigma^2$, one can write that
\begin{align}
    \left\langle e^{- a x} \right\rangle &= \frac{1}{\sqrt{2\pi\sigma^2}} \int dx \ e^{- a x - x^2/2\sigma^2} = e^{a^2 \sigma^2/2}
    \; .
\end{align}
Here, $\xi$, $\eta_x$ and $\eta_y$ are independent Gaussian variables, so that one can split the mean value of the exponential above into three parts:
\begin{align}
    \left\langle  e^{i \sqrt{2 \gamma_t k_B T} \int dt \ \bm{q}\cdot\bm{\eta} + i \sqrt{2 \gamma_r k_B T} \int dt \ \lambda \xi}\right\rangle &= \left\langle e^{ i \sqrt{2 \gamma_r k_B T} \int dt \ \lambda \xi }\right\rangle_{\xi} \times \left\langle e^{i \sqrt{2 \gamma_t k_B T} \int dt \ q_x \eta_x }\right\rangle_{\eta_x} 
     \nonumber \\
    & \qquad \qquad \times \left\langle e^{i \sqrt{2\gamma_t k_B T} \int dt \ q_y \eta_y} \right\rangle_{\eta_y}
    \; . 
\end{align}
Furthermore, the noises are all zero mean and $\delta$-correlated in time, so that, for instance,
\begin{align}
    \left\langle e^{ i \sqrt{2 \gamma_r k_B T } \int dt \ \lambda \xi }\right\rangle_{\xi} & = \prod\limits_t \left\langle e^{ i \sqrt{2 \gamma_r k_B T } \ \lambda (t) \xi (t)} \right\rangle_{\xi(t)} =
    \prod\limits_t e^{- \gamma_r k_B T \ \lambda^2(t)} 
    \nonumber\\
    & = e^{- \gamma_r k_B T \int dt \ \lambda^2(t)}
    \; .
\end{align}
All in all, one obtains
\begin{align}
    \left\langle A(\bm{r}, \bm{p},\theta, \omega) \right\rangle = 
    &\int \mathcal{D}\left[\bm{r},\bm{p},\bm{q}\right] \mathcal{D}\left[\theta,\omega,\lambda\right] 
     d\bm{X}_0 \ P_0(\bm{X}_0) \ A(\bm{r}, \bm{p},\theta, \omega)  \ \delta_{\bm{p}}\delta_{\omega} \nonumber\\ 
    &\qquad \times 
    e^{-i \int dt \ \left[ \bm{q}\cdot\left( \dot{\bm{p}} + \gamma_t \left(\bm{v} - \bm{v}_0\right) \right) + \lambda \left( \dot{\omega} - K \bm{p} \cdot \bm{s}_{\perp} +  \gamma_r \omega \right)\right] - k_B T \int dt \ \left(\gamma_r\lambda^2 + \gamma_t \bm{q}^2 \right)}.
\end{align}
Finally, we introduce the dynamical action $\mathcal{S}$, such that
\begin{align}
    &\left\langle A(\bm{r}, \bm{p},\theta, \omega) \right\rangle =
    \int \mathcal{D}\left[\bm{r},\bm{p},\bm{q}\right] \mathcal{D}\left[\theta,\omega,\lambda\right] 
    d\bm{X}_0 \ A(\bm{r}, \bm{p},\theta, \omega) \ \delta_{\bm{p}}\delta_{\omega} \ 
     e^{-\mathcal{S}\left[\bm{r},\bm{p},\bm{q}, \theta,\omega,\lambda, \bm{X}_0 \right]} 
    \; .
\end{align}
By definition, this action reads
\begin{align}
    \mathcal{S} \; \equiv \; & \; i \int dt \, \left[ \bm{q}\cdot\left( \dot{\bm{p}} + \gamma_t \left(\bm{v} - \bm{v}_0\right) \right) + \lambda \left( \dot{\omega} - K \bm{p} \cdot \bm{s}_{\perp} +  \gamma_r \omega \right)\right] 
    \nonumber\\
    & + 
     k_B T \int dt \, \left(\gamma_r\lambda^2 + \gamma_t\bm{q}^2 \right) - \ln P_0(\bm{X}_0)
    \; .
\end{align}
It is convenient to split this action into a term that contains all the dependency on the nature of the bath, $\mathcal{S}_{diss}$, and a term that only contains the deterministic part of the dynamics, $\mathcal{S}_{det}$. They are defined by
\begin{align}
      \mathcal{S}_{det} &= i \int dt \, \left[ \bm{q}\cdot \dot{\bm{p}}  + \lambda \left( \dot{\omega} - K \bm{p} \cdot \bm{s}_{\perp} \right)\right] - \ln P_0(\bm{X}_0)
    \; ,\\
    \mathcal{S}_{diss}&= i  \int dt \ \left[  \gamma_t\bm{q}\cdot  \left(\bm{v} - \bm{v}_0\right) + \gamma_r\lambda\omega \right]  + k_B T \int dt \, \left(\gamma_r\lambda^2 + \gamma_t\bm{q}^2 \right)
    \; .
\end{align}
One notices that by setting $\gamma_{t,r}=0$ the action and generating functional for the isolated model are recovered.

At this step of the calculation, one can rewrite every term in the Lagrangian formalism, 
using the $\delta_{\bm{p}}$ and $\delta_{\omega}$ that we have carried all along the calculation, 
\begin{align}
    &\left\langle A[\bm{r},\theta] \right\rangle =\int \mathcal{D}\left[\bm{r},\bm{q}\right] \mathcal{D}\left[\theta,\lambda\right] 
    d\bm{X}_0 \ A\left[\bm{r}, \theta \right]  \ e^{-\mathcal{S}\left[\bm{r},\bm{q}, \theta,\lambda, \bm{X}_0 \right]}
    \; ,
\end{align}
with an abuse of notation we still call $\bm{X}_0$ the initial values $\bm{r}_0$ and $\theta_0$, and the two 
contributions to the action 
$\mathcal{S} = \mathcal{S}_{det} + \mathcal{S}_{diss}$ now read
\begin{align}
    \mathcal{S}_{det} &=  i \int dt \; [ \bm{q}\cdot ( \ddot{\bm{r}} + K \dot{\bm{s}})  + \lambda ( \ddot{\theta} 
    - K \dot{\bm{r}} \cdot \bm{s}_{\perp} )] - \ln P_0(\bm{X}_0)
    \; , 
    \label{eq:Sdet}
    \\
    \mathcal{S}_{diss}&= 
     i \int dt \; [  \gamma_t\bm{q}\cdot  (\dot{\bm{r}} - \bm{v}_0) + 
    \gamma_r\lambda\dot{\theta} ]  +  k_B T \int dt \, (\gamma_r\lambda^2 + \gamma_t\bm{q}^2 )
    \;  .
       \label{eq:Sdiss}
\end{align}
Using this action, one can define a dynamical generating function $Z_d$, through
\begin{align}
    Z_d\left[\bm{J}_{\bm{r}}, \bm{J}_{\bm{q}}, J_{\theta}, J_{\lambda} \right] &\equiv \int \mathcal{D}\left[\bm{r},\bm{q}\right] \mathcal{D}\left[\theta,\lambda\right] d\bm{X}_0 \ e^{-\mathcal{S}\left[\bm{r},\bm{q}, \theta,\lambda, \bm{X}_0 \right]} 
    \ e^{\int dt \,\left[ \bm{J}_{\bm{r}}\cdot \bm{r} +   \bm{J}_{\bm{q}}\cdot\bm{q} + J_{\theta}\theta +  J_{\lambda}\lambda \right]}
    \; ,
    \label{eq:Zd}
\end{align}
where the $J$'s are sources for each of the real- and reciprocal-space variables.
By definition, $Z_d[0,0,0,0] = \left\langle 1 \right\rangle = 1$. In the rest of this Section we will 
mostly work with the generating functional \eqref{eq:Zd} and the deterministic and dissipative actions in 
Eqs.~\eqref{eq:Sdet}-\eqref{eq:Sdiss}.

The generating functional  in Eq.~\eqref{eq:Zd}  is, as usual, useful, because derivatives of $Z_d$ with respect to the sources, evaluated at zero sources, yield moments of the various variables.
For instance, the 2-point correlation function of the particle's position reads
\begin{align}
    C_{\bm{r},\bm{r}}(t,t') \equiv \left\langle \bm{r}(t)\bm{r}(t')\right\rangle &= \left.\frac{\delta^2 Z_d\left[\bm{J}_{\bm{r}}, \bm{J}_{\bm{q}}, J_{\theta}, J_{\lambda}\right] }{\delta \bm{J}_{\bm{r}}(t) \delta \bm{J}_{\bm{r}}(t')}\right|_{\bm{J}_{\bm{r}}= \bm{J}_{\bm{q}}= J_{\theta}= J_{\lambda} = 0} \; , 
\end{align}
with an obvious short-hand notation for the variation with respect to a vector.
One can also seek the linear response of the particle's position in response to an infinitesimal external force.
One should notice that adding a force $\bm{f}$ to the equation on the position results in an additional deterministic term in the action, which becomes
\begin{align}
    \mathcal{S}_f &= \mathcal{S} - i \int dt \ \bm{q}\cdot \bm{f}
    \; ,
\end{align}
and one can define the corresponding dynamical partition function $Z_{d,f}$ like above.
Since
\begin{align}
    \left\langle \bm{r}(t)\right\rangle = \frac{1}{Z_{d,f}} \int \mathcal{D}\left[\bm{r},\bm{q}\right] \mathcal{D}\left[\theta,\lambda\right] d\bm{X}_0 \ \bm{r}(t) \ e^{-\mathcal{S}_f},
\end{align}
one can also notice that the response function
\begin{align}
    R_{\bm{r},\bm{r}}(t,t') &= \left. \frac{\delta \left\langle \bm{r}(t) \right\rangle}{\delta \bm{f}(t')}\right|_{\bm{f} = 0},
\end{align}
can be written as
\begin{align}
    R_{\bm{r},\bm{r}}(t,t') &=  - \left.\left\langle  \bm{r}(t)\cdot  \frac{\delta \mathcal{S}_f}{\delta \bm{f}(t')} \right\rangle\right|_{\bm{f} = 0} = i \left\langle  \bm{r}(t)\cdot \bm{q}(t') \right\rangle,
\end{align}
so that the position linear  response function can  be simply obtained as a derivative of $Z_d$, the partition function without an external force:
\begin{align}
    R_{\bm{r},\bm{r}}(t,t') &= i \left.\frac{\delta^2 Z_d\left[\bm{J}_{\bm{r}}, \bm{J}_{\bm{q}}, J_{\theta}, J_{\lambda}\right] }{\delta \bm{J}_{\bm{r}}(t) \delta \bm{J}_{\bm{q}}(t')}\right|_{\bm{J}_{\bm{r}}= \bm{J}_{\bm{q}}= J_{\theta}= J_{\lambda} = 0}.
\end{align}

These observations will be useful to establish the fluctuation-dissipation theorem (FDT) for this system.
In order to check the form of this theorem, which is the conservation law associated to the invariance under time-reversal symmetry, one must first check what the exact nature of that symmetry is.

\subsection[\hspace{1.75cm} Generalized time reversal symmetry] {Generalized time reversal symmetry\label{app:TRS}}

Let us introduce the notations $\bm{r}_t = \bm{r}(t)$ and $d/dt = d_t$ to lighten the expressions in the following calculation. 
For convenience, we set the origin of times in the middle of the considered time interval so as to make the latter symmetric.
As a result, integrals over time are now computed over the symmetric time-interval $\left[ - \mathcal{T}; \mathcal{T}\right]$. We 
further introduce the notation
\begin{align}
    \int_t \;\; \equiv \int\limits_{-\mathcal{T}}^{\mathcal{T}} dt
    \; .
\end{align}
From the microscopic dynamics, we expect that one should, when reversing time, also multiply the spins by $-1$.
We therefore introduce the following time-reversal transformations of all variables, 
\begin{align}
\begin{split}
   \mathbb{T}\!: \;\; &  \bm{r}_t \to \bm{r}_{-t} \; , \\
     & \theta_t \to \theta_{-t} + \pi \; , \\
   &  i\bm{q}_t \to i \bm{q}_{-t} + \beta \left( d_t \bm{r}_{-t} + \bm{v}_0\right)
    \; , \\
    & i \lambda_t \to i \lambda_{-t} + \beta d_t \theta_{-t} \; ,
    \end{split}
    \label{eq:time-reversal}
\end{align}
that we call  $\mathbb{T}$. The angular transformation implies
\begin{equation}
\bm{s}_t \to -\bm s_{-t} \; . 
\end{equation}
We will see below that the reference velocity $\bm{v_0}$ in the dissipative action will have to be changed as 
$\bm{v}_0 \to - \bm{v}_0$.

As one may expect the deterministic and dissipative contributions to the action 
to remain invariant on their own, we treat them separately.
We first focus on the deterministic part, Eq.~\eqref{eq:Sdet}.
Noticing that under $\mathbb{T}$,  the distribution of initial conditions should be replaced 
according to $\ln P_{-\mathcal{T}}(\bm{X}_{\mathcal{-T}}) \to \ln P_{\mathcal{T}}(\bm{X}_{\mathcal{T}}) $, 
we evaluate it at the transformed variables, 
$ \mathcal{S}_{det}[ \mathbb{T} \bm{Y} ] = 
\mathcal{S}_{det}[ \mathbb{T} \bm{r}, \mathbb{T}\bm{q}, \mathbb{T}\theta,\mathbb{T} \lambda]$:
\begin{align}
    \mathcal{S}_{det}[ \mathbb{T}\bm{Y}] 
    &= 
    \int_t\; \;  \left[ \left(i \bm{q}_{-t} + \beta \left(d_t \bm{r}_{-t} + \bm{v}_0\right)\right)
    \cdot
    \left( d_t^2\bm{r}_{-t} - K d_t\bm{s}_{-t}\right)  
    \right.
    \nonumber\\
    & \qquad\quad
    \left.
    +  \left(i \lambda_{-t} + \beta d_t \theta_{-t}\right) 
    \left( d_t^2 \theta_{-t} + K d_t\bm{r}_{-t} \cdot \bm{s}_{\perp,{-t}} \right)\right]  - \ln P_{\mathcal{T}}(\bm{X}_{\mathcal{T}})
    \nonumber \\
    & =  
    \int_t \;\; \left[ i\bm{q}_{-t}\cdot\left( d_t^2\bm{r}_{-t} - K d_t\bm{s}_{-t}\right)  
    +  i\lambda_{-t} \left( d_t^2 \theta_{-t} + K d_t\bm{r}_{-t} \cdot \bm{s}_{\perp,{-t}} \right)\right] \nonumber  \\
    & \qquad
    + \beta 
    \int_t \;\; 
    \left[
    \left(d_t \bm{r}_{-t} + \bm{v}_0\right) \cdot \left( d_t^2\bm{r}_{-t} - K d_t\bm{s}_{-t}\right) 
    + d_t \theta_{-t}\left( d_t^2 \theta_{-t} + K d_t\bm{r}_{-t} \cdot \bm{s}_{\perp,{-t}} \right)  \right]   
    \nonumber\\
    & \;\;\;
    -  \ln P_{\mathcal{T}}(\bm{X}_{\mathcal{T}}) \nonumber\\
    &=  \int_{t} \;\; \left[ i\bm{q}_{t}\cdot\left( d_{t}^2\bm{r}_{t} + K d_{t}\bm{s}_{t}\right)  + i \lambda_{t} \left( d_{t}^2 \theta_{t} - K d_{t}\bm{r}_{t} \cdot \bm{s}_{\perp,{t}} \right)\right] \nonumber  \\
    & \qquad
    + \beta \int_{t} \;\; \left[ \left(- d_{t} \bm{r}_{t} +   \bm{v}_0\right)
    \cdot \left( d_{t}^2\bm{r}_{t} + K d_{t}\bm{s}_{t}\right) - d_{t} \theta_{t}\left( d_{t}^2 \theta_{t} - K d_{t}\bm{r}_{t} \cdot \bm{s}_{\perp,{t}} \right)  \right]  
     \nonumber\\
    & \;\;\; - \ln P_{\mathcal{T}}(\bm{X}_{\mathcal{T}})
    \; . 
    \label{eq:intermediate}
\end{align}
In the last equality we simply changed variables from $-t$ to $t$ in the time-integrals and we exploited the 
symmetry of the time-interval to further use $\int_{-t} = \int_t$.
The first line in the last expression reproduces the integral terms in the original action \eqref{eq:Sdet}. 
We now need to focus on the second line, especially on the integral terms.
Noticing that 
\begin{align}
    d_{t}\bm{s}_{t} = d_t \theta_{t} \bm{s}_{\perp,t}, 
\end{align}
the integral terms can be rewritten as
\begin{align}
    &\beta \int_{t} \;\; \left[ \left(- d_{t} \bm{r}_{t} + \bm{v}_0\right)
    \cdot \left( d_{t}^2\bm{r}_{t} + K d_{t}\bm{s}_{t}\right)
    -
    d_{t} \theta_{t}\left( d_{t}^2 \theta_{t} - K d_{t}\bm{r}_{t} \cdot \bm{s}_{\perp,{t}} \right)  \right]
    \nonumber\\
    & \qquad =
    \beta \int_{t} \;\; \left[
    - \left(d_{t} \bm{r}_{t} - \bm{v}_0\right) \cdot d_{t}^2\bm{r}_{t}
    - d_{t} \theta_{t}  d_{t}^2 \theta_{t}
    + K \bm{v}_0 \cdot d_{t}\bm{s}_{t} \right]
    \; . 
    \label{eq:extra-term}
\end{align}
The canonical distribution of momenta can be transformed into one of velocities: 
\begin{align}
    P(\bm{p}, \bm{\omega}) &= Z^{-1}  \
    e^{
    - \beta \left( \frac{p^2}{2} + \frac{\omega^2}{2} - K \bm{p}\cdot \bm{s} - \bm{v_0}\cdot \bm{p}\right) }
    = 
    {Z'}^{-1} 
    e^{- \beta \left( \frac{v^2}{2} + \frac{\dot{\theta}^2}{2} - \bm{v_0}\cdot \left(\bm{v} + K \bm{s}\right)\right) 
    }
    \nonumber\\
    &
   = {Z''}^{-1} 
   e^{- \beta \left( \frac{\left(\bm{v} - \bm{v}_0\right)^2}{2} + \frac{\dot{\theta}^2}{2} - K \bm{v_0}\cdot \bm{s}\right) 
   }
   =
   P(\bm{v}, \dot\theta)
    \; ,
\end{align}
where we absorbed constant terms into the partition function at each step.
Consequently,
\begin{align}
    \ln P(\bm{v}, \dot\theta) &= - \beta \left( \frac{\left(\bm{v} - \bm{v}_0\right)^2}{2} + \frac{\dot{\theta}^2}{2} - K \bm{v_0}\cdot \bm{s}\right) - \ln Z''
    \equiv \ln P_t
    \; , 
\end{align}
so that the right-hand-side in Eq.~(\ref{eq:extra-term}) can be written as
\begin{align}
    & - \beta \int_{t} \; 
    \left[ \left(d_{t} \bm{r}_{t} - \bm{v}_0\right) \cdot d_{t}^2\bm{r}_{t}   + d_{t} \theta_{t}  d_{t}^2 \theta_{t}
     - K \bm{v}_0 \cdot d_{t}\bm{s}_{t}  \right] = \int_{t} \; d_t \ln P_t 
     \nonumber\\
     & \qquad\qquad\qquad = \ln P_{\mathcal{T}} - \ln P_{-\mathcal{T}}
    \; . \label{eq:TRS_eqdistri}
\end{align}
We therefore see that the first term in this equation cancels the last contribution in \eqref{eq:intermediate} and we are 
left with the original distribution at the initial time $-\mathcal{T}$.
In the end, we find that 
\begin{align}
    \mathcal{S}_{det}[ \mathbb{T}\bm{Y}] 
   &= i \int_{t} \left[ \bm{q}_{t}\cdot\left( d_{t}^2\bm{r}_{t} + K d_{t}\bm{s}_{t}\right)  +  \lambda_{t} \left( d_{t}^2 \theta_{t} - K d_{t}\bm{r}_{t} \cdot \bm{s}_{\perp,{t}} \right)\right] - \ln P_{-\mathcal{T}} \nonumber\\
   & = \mathcal{S}_{det}[\bm{Y}] 
    \; . 
\end{align}
We conclude that the deterministic part of the action is invariant under the tine-reversal transformation introduced.

While here we focus on the case of a single particle to keep the discussion lighter, one could have considered $N$ particles coupled through a repulsive potential $U$ and a ferromagnetic coupling $J$ as per Eq.~(\ref{eq:Hamiltonian}) throughout this calculation.
Since the total potential energy stemming from interactions $V$ is even under time-reversal symmetry, the contribution of these terms would, like in standard equilibrium system~\cite{ArnoulxdePirey2022}, only modify Eq.~(\ref{eq:TRS_eqdistri}) to include terms of the form $\beta d_t V$, and thus produce the full interacting canonical distribution.
As a result, considering the full interacting case does not affect the definition of the time-reversal symmetry.

We now check how the dissipative part of the action behaves under time-reversal.
We need to change $\bm{v}_0 \to - \bm{v}_0$ which appears explicitly in the dissipative action, so that
\begin{align}
     \mathcal{S}_{diss}[ \mathbb{T}\bm{Y}, -\bm{v}_0]  
    &= 
     \int_{t} \;\; 
    \left[ \gamma_t\left( i \bm{q}_{-t} + \beta \left(  d_t \bm{r}_{-t} 
    + \bm{v}_0\right) \right)  
    \cdot  
    \left( d_t\bm{r}_{-t} + \bm{v}_0\right) 
    +
    \gamma_r\left( i \lambda_{-t} + \beta d_t \theta_{-t}  \right) d_t\theta_{-t} \right] 
    \nonumber \\
    & \quad
     - k_BT 
     \int_{t} \;\; \left\{
     \gamma_r\left( i \lambda_{-t} + \beta d_t \theta_{-t}  \right)^2 +
      \gamma_t\left[ i \bm{q}_{-t} - \beta ( d_t \bm{r}_{-t} + \bm{v}_0 ) \right]^2 
      \right\}
     \nonumber\\
    &= 
     i \int_{-t} \left[\gamma_t\bm{q}_{-t}  \cdot  \left( d_t\bm{r}_{-t} + \bm{v}_0\right) + \gamma_r\lambda_{-t} d_t\theta_{-t} \right] 
    \nonumber\\
    & \quad
    +\beta 
    \int_{t} \;\; \left[ \gamma_t\left(  d_t \bm{r}_{-t} + \bm{v}_0\right)^2 +  \gamma_r\left( d_t \theta_{-t}\right)^2 \right]  
    \nonumber \\
    &\quad 
    +  k_BT   \int_{t} \;\;   \left(\gamma_r\lambda_{-t}^2 + \gamma_t\bm{q}_{-t}^2\right)
    - 2 i \int_{t} \left[\gamma_t\bm{q}_{-t}  \cdot  \left( d_t\bm{r}_{-t} + \bm{v}_0\right) + \gamma_r\lambda_{-t} d_t\theta_{-t} \right] 
    \nonumber\\
    & \quad
    - \beta  \int_{t} \;\; \left[ \gamma_r\left(d_t \theta_{-t}\right)^2 + \gamma_t\left( d_t \bm{r}_{-t} + \bm{v}_0\right)^2 \right] 
    \nonumber\\
    &=
     -i \int_{t} \;\; \left[\gamma_t\bm{q}_{-t}  \cdot  \left( d_t\bm{r}_{-t} + \bm{v}_0\right) 
     + \gamma_r \lambda_{-t} d_t\theta_{-t} \right]  
     + 
      k_BT \int_{t} \;\;  \left(\gamma_r\lambda_{-t}^2 + \gamma_t\bm{q}_{-t}^2\right) \nonumber \\ 
    &=  i \int_t \; \left[\gamma_t\bm{q}_{t}  \cdot  \left( d_{t}\bm{r}_{t} - \bm{v}_0\right) + \gamma_r\lambda_{t} d_{t}\theta_{t} \right]   +  k_BT \int_{t} \; \left(\gamma_r\lambda_{t}^2 + \gamma_t\bm{q}_{t}^2\right)
    \nonumber \\
    &=  \mathcal{S}_{diss}[\bm{Y},\bm{v}_0] \; .
\end{align}
The steps of the calculation above are the following: the first equation is obtained by applying the transformation to each term of the action.
Then, to get the second line, we expand all the products of the line above.
We then simplify by regrouping proportional terms to get the third line.
Finally, we change variables in the integral, using $t \to -t$.

The result is that if one reverses time, it is sufficient to reverse the sign of the damping term to recover the same action: in other words, we are here describing the symmetry between the original Langevin equations,
\begin{align}
\begin{split}
    \ddot{\bm{r}} + K \dot{\bm{s}} + \gamma_t \left(\dot{\bm{r}} - \bm{v}_0 \right) - \sqrt{2 \gamma_t k_B T} \bm{\eta} &= \bm{0} \; , \\
    \ddot{\theta} - K \dot{\bm{r}} \cdot \bm{s}_{\perp} + \gamma_r \dot{\theta} (t) - \sqrt{2 \gamma_r k_B T} \xi &= 0
    \; ,
    \end{split}
\end{align}
and the time-reversed ones
\begin{align}
\begin{split}
    \ddot{\bm{r}} + K \dot{\bm{s}} - \gamma_t \left(\dot{\bm{r}} - \bm{v}_0 \right) - \sqrt{2 \gamma_t k_B T} \bm{\eta} &= \bm{0} \; , \\
    \ddot{\theta} - K \dot{\bm{r}} \cdot \bm{s}_{\perp} - \gamma_r \dot{\theta} (t) - \sqrt{2 \gamma_r k_B T} \xi &= 0
    \; .
    \end{split}
\end{align}

The generalized TRS also reflects the fact that the damping term was chosen consistently with the symmetries of the system.
In particular, it could have been tempting to introduce a damping acting not on the velocity, but directly on the momentum, for instance via a term $\gamma_t \bm{p}$ instead of $\gamma_t \bm{v}$.
However, such a damping term does not admit a generalized TRS.
The reason is that there is no way to make the transformation of $\bm{q}_t$ accommodate this sort of damping: any proposition of the form $i \bm{q}_t \to i \bm{q}_{-t} + \beta d_t \bm{p}_{-t}$ would create spurious $K \bm{s}$ terms that would not be compensated by the equilibrium canonical distribution.

\subsection[\hspace{1.75cm} Fluctuation-Dissipation Theorem]{Fluctuation-Dissipation Theorem\label{app:FDT}}

We found that the dynamical action is symmetric under the transformation $\mathbb{T}$ defined in Eq.~\eqref{eq:time-reversal}. 
This symmetry can be translated into conservation laws, in the spirit of Noether's theorem, by writing that correlators between variables are invariant under $\mathbb{T}$.
In the following, we investigate each relevant pair of variables separately.

\subsubsection[\hspace{1.75cm} Position observables]{Position observables.}

One can write that
\begin{align}
    \left\langle \bm{r}_t i \bm{q}_{t'} \right\rangle_{\mathcal{S}} &= 
    \left\langle \mathbb{T}\bm{r}_t \mathbb{T} i \bm{q}_{t'} \right\rangle_{\mathbb{T}\mathcal{S}}=\left\langle \bm{r}_{-t} i \bm{q}_{-t'} \right\rangle_{\mathcal{S}} + 
    \beta \left\langle \bm{r}_{-t} \cdot \left(d_{t'} \bm{r}_{-t'} + \bm{v}_0\right) \right\rangle_{\mathcal{S}}.
\end{align}
Using the definitions of the space-space correlation and response functions,
\begin{align}
\begin{split}
    R_{\bm{r},\bm{r}}(t, t') &\equiv i \left\langle  \bm{r}(t)\cdot \bm{q}(t') \right\rangle \, , \\
    C_{\bm{r},\bm{r}}(t, t') &\equiv \left\langle \bm{r}(t)\bm{r}(t')\right\rangle \, ,
    \end{split}
\end{align}
we find that
\begin{align}
    R_{\bm{r},\bm{r}}(t,t') - R_{\bm{r},\bm{r}}(- t,-t') &= 
    \beta d_{t'} C_{\bm{r},\bm{r}}(-t,-t') + \beta \bm{v}_0 \cdot  \left\langle \bm{r}_{-t}  \right\rangle_{\mathcal{S}}
    \; .
\end{align}
In other words, we find the usual fluctuation-dissipation theorem for spatial correlations and response for $\bm{v}_0 = \bm{0}$, and an extra term that depends on one of the times only that represents the mean drift of the particle due to the velocity imposed by the bath otherwise.
However, this contribution disappears if one defines the correlations of positions in the co-moving frame that moves at $\bm{v}_0$, $\bm{\delta r}(t) = \bm{r}(t) - \bm{v}_0 t$, much like in Ref.~\cite{Gomez-Solano2009}.
Here, since $\langle d_t\bm{r}_t\rangle = \bm{v}_0$, working in the co-moving frame is equivalent to considering the connected correlation function, $C^c_{\bm{a},\bm{b}}(\tau) = C_{\bm{a},\bm{b}}(\tau) - \langle \bm{a}(t)\rangle \cdot \langle \bm{b}(t')\rangle$, so that we actually find
\begin{align}
    R_{\bm{r},\bm{r}}(t,t') - R_{\bm{r},\bm{r}}(- t,-t') &= 
    \beta d_{t'} C^c_{\bm{r},\bm{r}}(-t,-t')
    \; .
\end{align}
In that last expression, we may introduce $\tau = t - t'$ and invoke time-translational invariance in the co-moving frame (where all single-time observables have constant average) to write the FDT in a more usual way,
\begin{align}
    R_{\bm{r},\bm{r}}(\tau) - R_{\bm{r},\bm{r}}(- \tau) &= 
    - \beta d_\tau C^c_{\bm{r},\bm{r}}(\tau).
\end{align}
Furthermore, using causality and assuming $\tau>0$, this expression assumes the simpler form
\begin{align}
    R_{\bm{r},\bm{r}}(\tau) &= 
    - \beta d_\tau C^c_{\bm{r},\bm{r}}(\tau).
\end{align}

We here obscured a technical but important point for the discussion in the main text: the expressions above only hold if the particles being considered are either placed in a bounded space (for instance a box with periodic boundary conditions like those used in simulations throughout this paper), or in contact with a confining potential.
Otherwise, a single particle diffusing in unbounded space generically breaks the FDT~\cite{Cugliandolo1994}, as the distribution of positions cannot become stationary even after infinite times.
Throughout this section, and in the following, we thus implicitly assume that we are working in a finite volume or in a confining potential.

\subsubsection[\hspace{1.75cm} Spin observables]{Spin observables.}
\label{subsec:spin}

One can reproduce the same calculation for spin-spin correlations and linear response. 
The trick is to introduce the equations of motion on the spin vector itself.
Since $\bm{s}$ is a unit vector, one must account for the constraint $|\bm{s}| = 1$, which introduces a centripetal term.
Using $\dot{\bm{s}} = \dot{\theta}\bm{s}_{\perp}$ and $\ddot{\bm{s}} = \ddot{\theta}\bm{s}_{\perp} - \dot{\theta}^2 \bm{s}$, the deterministic Hamiltonian equation of motion on $\theta$ translates to
\begin{align}
    \ddot{\bm{s}} + |\dot{\bm{s}}|^2 \bm{s} - K (\bm{p} \cdot \bm{s}_{\perp})\bm{s}_{\perp} = \bm{0}
    \; ,
\end{align}
where the $|\dot{\bm{s}}|^2 \bm{s}$ term is the centripetal acceleration maintaining the unit-norm constraint, and a projection onto $\bm{s}_\perp$ yields the Hamiltonian equation of motion on $\theta$.
The Langevin equation~(\ref{eq:Langevin-s}) for the spin reads
\begin{align}
	\ddot{\bm{s}} + |\dot{\bm{s}}|^2 \bm{s} - K (\dot{\bm{r}} \cdot \bm{s}_{\perp})\bm{s}_{\perp} + \gamma_r \dot{\bm{s}} - \sqrt{2 \gamma_r k_B T}  \bm{s}_{\perp} \; \xi &= \bm{0}
	\;,
\end{align}
which boils down to the Langevin equation on the angle, Eq.~(\ref{eq:Langevin_omega}), by projecting on $\bm{s}_\perp$.

This equation can be treated exactly like the equation on $\theta$ and one can reconstruct the dynamical action in terms of $\bm{s}$ and its conjugate field $\bm{\lambda}$.
A subtlety, however, is that the noise is now formally multiplicative, as the prefactor of the unit-variance zero-mean scalar 
white noise $\xi$ is now
\begin{align}
    \bm{g}(\bm{s}) = \sqrt{2\gamma_r k_B T} \bm{s}_\perp,
\end{align}
which depends explicitly on the state of the system.
However, this subtlety is immediately dispelled by the fact that the dynamics are underdamped, meaning that the noise depends on $\bm{s}$ but appears in the equation on $\ddot{\bm{s}}$, not the one on $\dot{\bm{s}}$.
More concretely, when writing the dynamics in simplified Hamiltonian form for Cartesian component $a$,
\begin{align}
\begin{split}
    \dot{s}_a &= \omega_a
    \; , \\
    \dot{\omega}_a &= F_a( \bm{s}) - \gamma_r \omega_a + g_a(\bm{s}) \xi
    \; ,
\end{split}
\end{align}
occulting the dependence on $\bm{r}$ and $\bm{p}$ of the deterministic part $\bm{F}$ of the equation for brevity.
One may then introduce discretized dynamics with an It\={o}-Stratonovich interpolation constant $\alpha \in[0;1]$, such that the finite-difference version of the dynamics between times $t$ and $t+dt$ reads
\begin{align}
\begin{split}
    \delta s_a
    & \equiv s_a(t+dt) - s_a(t) = \omega_{a}\left(t + \alpha\,dt\right)\,dt \\
    \delta \omega_{a} & \equiv \omega_a(t+dt) - \omega_a(t) 
    \\
    & = \left[F_a\left(\bm{s}(t+\alpha\,dt)\right) - \gamma_r\,\omega_{a}\left(t + \alpha\,dt\right) + g_a\left(\bm{s}(t+\alpha\,dt)\right)\,\xi_t \right]\,dt
    \; ,
    \end{split}
\end{align}
where the normal random variable $\xi_t\sim N(0,1)$ is one increment of the noise.
Additional  terms may come from the multiplicativeness of the noise appear when 
expanding the right-hand sides of these equations to order $dt$.
The equation on $\delta s_a$ yields no extra term, only $\delta s_a = \omega_a(t)dt + O(dt^{3/2})$.
In the equation on $\delta\omega_a$, additional  terms may only come from the Taylor expansion of $g$,
\begin{align}
    g_a\bigl(\bm{s}(t+\alpha\,dt)\bigr) = g_a\bigl(\bm{s}(t)\bigr) 
    + \alpha \sum_b \frac{\partial g_a}{\partial s_b}\,\delta s_b + \ldots
\end{align}
Introducing $\Delta W_t = \xi_t dt$, which verifies $\langle \Delta W_t\rangle = 0$ and $\langle \Delta W_t^2\rangle = dt$ where the average is over the normal distribution, the noise term verifies
\begin{align}
  & 
    g_a\bigl(\bm{s}(t+\alpha\,dt)\bigr)\;\xi_t\,dt  
    \nonumber\\
    & \qquad 
    = 
    g_a(\bm{s})\,\Delta W_t
    + \alpha \sum_b \frac{\partial g_a}{\partial s_b}\, \omega_{b}\,\Delta W_t\,dt
    + \alpha^2 \sum_b \frac{\partial g_a}{\partial s_b}\, g_b\,(\Delta W_t)^2\,dt
    + \ldots 
    \label{eq:noise_expansion_ito_strato}
\end{align}
Since the dynamics are non-anticipating (the value of the noise at the current step is independent of the value of the state before the update~\cite{Gardiner1983}), one then has
\begin{align}
    \left\langle g_a(\bm{s})\,\Delta W_t\right\rangle =0 \; , \qquad\qquad
    \left\langle \frac{\partial g_a}{\partial s_b}\, \omega_{b}\,\Delta W_n\right\rangle = 0 \; ,
\end{align}
so that the leading order term is the term proportional to $\alpha^2$, which is of order $dt^2$ and is thus discarded.

Thus, this particular choice of multiplicative noise does not yield additional terms in the discretized dynamics, 
and the naïve chain rule for derivatives can be used.
Reproducing the same steps as in the path integral written in terms of $\theta$, the two parts of the action then read (here setting $\gamma_r = \gamma_t = \gamma$ for simplicity),
\begin{align}
\begin{split}
    \mathcal{S}_{det} &= i \int dt \ \left[ \bm{q}\cdot \left( \ddot{\bm{r}} + K \dot{\bm{s}}\right)  + \bm{\lambda} \cdot \left( \ddot{\bm{s}} + |\dot{\bm{s}}|^2 \bm{s} - K (\dot{\bm{r}} \cdot \bm{s}_{\perp})\bm{s}_{\perp} \right)\right] - \ln P_0(\bm{X}_0) \; ,\\
    \mathcal{S}_{diss}&= i \gamma \int dt \ \left[  \bm{q}\cdot  \left(\dot{\bm{r}} - \bm{v}_0\right) + \bm{\lambda} \cdot \dot{\bm{s}} \right]  + \gamma k_B T \int dt \left(\lambda_{\perp}^2 + \bm{q}^2 \right) \; ,
\end{split}
\end{align}
with $\lambda_{\perp} = \bm{\lambda}\cdot \bm{s}_{\perp}$.
Since the projection of the deterministic part of the spin equation of motion onto $\bm{s}$ identically yields $0$ at all times, only $\lambda_\perp$ needs to be considered throughout the action, so that the angular action is recovered,
\begin{align}
\begin{split}
    \mathcal{S}_{det} &= i \int dt \ \left[ \bm{q}\cdot \left( \ddot{\bm{r}} + K \dot{\bm{s}}\right)  + \lambda_\perp \bm{s}_\perp\cdot \left( \ddot{\bm{s}} + |\dot{\bm{s}}|^2 \bm{s} - K (\dot{\bm{r}} \cdot \bm{s}_{\perp})\bm{s}_{\perp} \right)\right] - \ln P_0(\bm{X}_0) \; ,\\
    \mathcal{S}_{diss}&= i \gamma \int dt \ \left[  \bm{q}\cdot  \left(\dot{\bm{r}} - \bm{v}_0\right) + \lambda_\perp \bm{s}_\perp \cdot \dot{\bm{s}} \right]  + \gamma k_B T \int dt \left(\lambda_{\perp}^2 + \bm{q}^2 \right) \;.
\end{split}
\end{align}

As shown above, the MSRJD action written in terms of the spin vector reduces to the angular action with $\lambda_\perp = \bm{\lambda}\cdot\bm{s}_\perp$ replacing $\lambda$.
The spin-spin FDT can therefore be derived directly from the angular TRS, Eq.~\eqref{eq:time-reversal}.
The physical response of $\bm{s}$ to a magnetic field $\bm{h}$ coupled as $\delta\mathcal{H} = - \, \bm{h}\cdot\bm{s}$ in the Hamiltonian is
\begin{align}
    R_{s_a, s_b}(t - t') = i\left\langle s_a(t)\, \lambda(t')\, {s_\perp}_b(t') \right\rangle \; .
\end{align}
This can be seen as follows.
The perturbation enters the angular equation of motion through the Hamiltonian derivative, $-\partial_\theta(-\bm{h}\cdot\bm{s}) = \bm{h}\cdot\bm{s}_\perp$, so that the perturbed equation of motion residual acquires an extra term $- \, \bm{h}\cdot\bm{s}_\perp$.
In the MSRJD action, this adds $-i\lambda(t')\,\bm{h}(t')\cdot\bm{s}_\perp(t')$ to $\mathcal{S}$, and the path-integral weight $e^{-\mathcal{S}}$ becomes $e^{-\mathcal{S}_0 + i\int d\tau\,\lambda(\tau)\,\bm{h}(\tau)\cdot\bm{s}_\perp(\tau)}$.
Taking the functional derivative at $\bm{h} = \bm{0}$ then yields the expression above, where the ${s_\perp}_b$ factor reflects the fact that the field acts as a torque on the constrained spin, not as a Cartesian force on an unconstrained vector.
Under $\mathbb{T}$, both $\bm{s}$ and $\bm{s}_\perp$ flip sign (since $\theta \to \theta + \pi$), and these two sign flips cancel.
Defining
\begin{align}
\begin{split}
    R_{\bm{s},\bm{s}}(t - t') &\equiv \sum_a R_{s_a, s_a}(t - t') \; , \\
    C_{\bm{s},\bm{s}}(t- t') &\equiv \left\langle \bm{s}(t)\cdot\bm{s}(t')\right\rangle \; ,
    \end{split}
\end{align}
and applying the angular TRS as in the position sector, we find
\begin{align}
    R_{\bm{s},\bm{s}}(\tau) - R_{\bm{s},\bm{s}}(- \tau) &= - \beta d_\tau C_{\bm{s},\bm{s}}(\tau)
    \; .
\end{align}
Since the spin degree of freedom has a stationary average in equilibrium, $d_t\langle\bm{s}_t\rangle = 0$, this expression may be rewritten transparently for the connected correlations,
\begin{align}
    R_{\bm{s},\bm{s}}(\tau) - R_{\bm{s},\bm{s}}(- \tau) &= - \beta d_\tau C^c_{\bm{s},\bm{s}}(\tau)
    \; .
\end{align}
Finally, enforcing causality and $\tau > 0$,
\begin{align}
    R_{\bm{s},\bm{s}}(\tau) &= - \beta d_\tau C^c_{\bm{s},\bm{s}}(\tau)
    \; .
\end{align}

\subsubsection[\hspace{1.75cm} Crossed relations]{Crossed relations.}
\label{subsec:crossed}

Finally, one can seek cross-correlation and crossed response functions, defined through
\begin{align}
\begin{split}
    R_{\bm{r},\bm{s}}(t - t') &\equiv i \left\langle  r_a(t)\, \lambda(t')\, (s_\perp)_a(t') \right\rangle \; , \\
    R_{\bm{s},\bm{r}}(t - t') &\equiv i \left\langle  \bm{s}(t)\cdot \bm{q}(t') \right\rangle \; , \\
    C_{\bm{r},\bm{s}}(t- t') &\equiv \left\langle \bm{r}(t)\bm{s}(t')\right\rangle \; .
    \end{split}
\end{align}
The first two functions do not necessarily have to be equal: one describes the response, in space, to a field acting on spins; while the other describes the response of spins when a force acts on positions. 

First off, one can write that 
\begin{align}
    \left\langle \bm{s}_t i \bm{q}_t' \right\rangle_{\mathcal{S}} &= \left\langle \mathbb{T}\bm{s}_t \mathbb{T} i \bm{q}_t' \right\rangle_{\mathbb{T}\mathcal{S}},
\end{align}
so that
\begin{align}
    R_{\bm{s},\bm{r}}(t,t') + R_{\bm{s},\bm{r}}(-t,-t') &= \beta d_{t'} C_{\bm{r},\bm{s}}(-t,-t') - \beta \bm{v}_0 \cdot \left\langle \bm{s}_{-t}\right\rangle_{\mathcal{S}} \, ,
\end{align}
where the action of the bath velocity on the spin field appears explicitly, like in position-position correlations.
The cure is, once again, to work in the co-moving frame and introduce the connected correlation function, yielding
\begin{align}
    R_{\bm{s},\bm{r}}(t,t') + R_{\bm{s},\bm{r}}(-t,-t') &= \beta d_{t'} C^c_{\bm{r},\bm{s}}(-t,-t')
\end{align}
and, using time-translational invariance in the co-moving frame
\begin{align}
    R_{\bm{s},\bm{r}}(\tau) + R_{\bm{s},\bm{r}}(-\tau) &= - \beta d_{\tau} C^c_{\bm{r},\bm{s}}(\tau)
\end{align}
then finally invoking causality,
\begin{align}
    R_{\bm{s},\bm{r}}(\tau) &= - \beta d_{\tau} C^c_{\bm{r},\bm{s}}(\tau).
\end{align}

Likewise, using the definition $R_{\bm{r},\bm{s}}(t-t') = i\langle r_a(t)\,\lambda(t')\,(s_\perp)_a(t')\rangle$ and applying $\mathbb{T}$ to the angular fields (under which $\bm{r}_t \to \bm{r}_{-t}$, $(s_\perp)_a \to -(s_\perp)_a$, and $i\lambda_{t'} \to i\lambda_{-t'} + \beta d_{t'}\theta_{-t'}$), we find
\begin{align}
    R_{\bm{r},\bm{s}}(\tau) + R_{\bm{r},\bm{s}}(-\tau) = - \beta d_\tau C^c_{\bm{r},\bm{s}}(\tau)
    \; .
\end{align}
After invoking causality, finally,
\begin{align}
    R_{\bm{r},\bm{s}}(\tau) = - \beta d_\tau C^c_{\bm{r},\bm{s}}(\tau)
    \; .
\end{align}

\subsubsection[\hspace{1.75cm} Summary]{Summary.}

All in all, the connected FDT equations read, for $\tau > 0$ and using causality:
\begin{align}
    R_{\bm{a},\bm{b}}(\tau) &=
    - \beta\, \partial_\tau C^c_{\bm{a},\bm{b}}(\tau)
\end{align}
with $\bm{a} = \bm{r}$ or $\bm{a} = \bm{s}$ and $\bm{b} = \bm{r}$ or $\bm{b} = \bm{s}$.

The Onsager-Casimir reciprocity $R_{\bm{s},\bm{r}}(\tau) = -R_{\bm{r}(\tau),\bm{s}}$ (equivalently, $\chi_{\bm{r},\bm{s}}(\tau) = -\chi_{\bm{s},\bm{r}}(\tau)$) then follows directly: since $\bm{s}$ is odd under $\mathbb{T}$ while $\bm{r}$ is even, the $\mathbb{T}$-symmetry of the equilibrium state implies $C^c_{\bm{r},\bm{s}}(\tau) = -C^c_{\bm{r},\bm{s}}(-\tau)$, \textit{i.e.} the crossed correlator is an odd function of $\tau$.
Combined with the relation $C^c_{\bm{s},\bm{r}}(\tau) = C^c_{\bm{r},\bm{s}}(-\tau)$ (time-translational invariance), one has $C^c_{\bm{s},\bm{r}}(\tau) = -C^c_{\bm{r},\bm{s}}(\tau)$, so that
\begin{align}
    R_{\bm{s},\bm{r}} = -\beta\,\partial_\tau C^c_{\bm{s},\bm{r}} = \beta\,\partial_\tau C^c_{\bm{r},\bm{s}} = -R_{\bm{r},\bm{s}} \; ,
\end{align}
all evaluated at $\tau$.
Thus, Onsager-Casimir reciprocity is a direct algebraic consequence of all four FDT equations having the standard sign $-\beta\partial_\tau C^c$ and of the crossed correlator being odd under $\mathbb{T}$.

\subsection[\hspace{1.75cm} Onsager-Machlup action construction]{Onsager-Machlup action construction\label{app:OM}}

We derive the Onsager-Machlup (OM) action by integrating out the response fields from the MSRJD action.
The Onsager-Machlup action gives the probability of a trajectory directly, $\mathbb{P}[\text{path}] \propto e^{\mathcal{A}_{OM}}$, without reference to response fields.

In the scalar angle formulation, the Lagrangian equations of motion read
\begin{align}
\begin{split}
    \ddot{\bm{r}} + K \dot{\theta}\, \bm{s}_{\perp} + \gamma_t(\dot{\bm{r}} - \bm{v}_0) &= \sqrt{2\gamma_t k_BT}\, \bm{\eta} \; , \\
    \ddot{\theta} - K \dot{\bm{r}} \cdot \bm{s}_{\perp} + \gamma_r \dot{\theta} &= \sqrt{2\gamma_r k_BT}\, \xi \; ,
    \end{split}
\end{align}
where we used $\bm{p} = \dot{\bm{r}} + K\bm{s}$, and $\omega = \dot{\theta}$.
Crucially, both noise sources are \emph{additive} in this formulation (the multiplicative character of the angular noise is absorbed into the scalar projection).
The MSRJD path-integral weight is $e^{-\mathcal{S}}$ with
\begin{align}
    \mathcal{S} &= i \int dt \left[ \bm{q}\cdot \bm{E}_{\bm{r}} + \lambda\, E_{\theta} \right] +  k_BT \int dt \left(\gamma_r\lambda^2 + \gamma_t\bm{q}^2\right) - \ln P_0(\bm{X}_0) \; ,
\end{align}
where the equation-of-motion ``residuals'' are
\begin{align}
    \bm{E}_{\bm{r}} &\equiv \ddot{\bm{r}} + K \dot{\theta}\, \bm{s}_{\perp} + \gamma_t(\dot{\bm{r}} - \bm{v}_0) \; , 
    \label{eq:Er_def} 
    \\
    E_{\theta} & \equiv \ddot{\theta} - K \dot{\bm{r}} \cdot \bm{s}_{\perp} + \gamma_r \dot{\theta} \; . 
    \label{eq:Etheta_def}
\end{align}

The response fields $\bm{q}$ and $\lambda$ appear quadratically in $k_BT(\gamma_t\bm{q}^2 + \gamma_r\lambda^2)$ and linearly in the coupling to $\bm{E}_{\bm{r}}$ and $E_\theta$, so the path integral over the response fields is Gaussian:
\begin{align}
   & \int \mathcal{D}\bm{q}\, \mathcal{D}\lambda \; e^{-\mathcal{S}} 
    \nonumber\\
    & \qquad = 
    P_0 \prod_{t} \left[\int d^2\bm{q}\, d\lambda \; \exp\left( -\gamma_t k_BT |\bm{q}|^2 - i \bm{q}\cdot\bm{E}_{\bm{r}} - \gamma_r k_BT \lambda^2 - i\lambda E_\theta \right)\right] \; .
\end{align}
Each time-slice integral is a standard Gaussian of the form $\int d\bm{Q}\, e^{-a |\bm{Q}|^2 - i\bm{Q}\cdot\bm{b}} \propto \exp\left(-|\bm{b}|^2/4a\right)$, giving
\begin{align}
    \int \mathcal{D}\bm{q}\, \mathcal{D}\lambda \; e^{-\mathcal{S}} &\propto P_0 \exp\left[ - \int dt\, \left( \frac{|\bm{E}_{\bm{r}}|^2 }{4\gamma_t k_BT} +\frac{E_\theta^2}{4\gamma_r k_BT}\right) \right] \; .
\end{align}

The Onsager-Machlup action is therefore
\begin{align}
    \mathcal{A}_{OM}  = 
    &
    - \frac{1}{4 k_BT} \int dt \left[ \frac{1}{\gamma_t}\left|\ddot{\bm{r}} + K\dot{\theta}\,\bm{s}_\perp + \gamma_t(\dot{\bm{r}} - \bm{v}_0)\right|^2  
    + \frac{1}{\gamma_r}\left(\ddot{\theta} - K\dot{\bm{r}}\cdot\bm{s}_\perp
    + \gamma_r\dot{\theta}\right)^2 \right]
    \nonumber\\
    & + \ln P_0(\bm{X}_0) 
    \; .
    \label{eq:AOM}
\end{align}
The probability of a trajectory conditional on its initial condition is
\begin{align}
    \mathbb{P}\left[\bm{r}(\cdot), \theta(\cdot) \,\middle|\, \bm{X}_0\right] \propto e^{\mathcal{A}_{OM}} \; .
\end{align}
Its physical interpretation is that trajectories are weighted by how far they deviate from the deterministic dynamics, with the deviation measured in units of the noise intensities $\gamma_{t,r} k_BT$.

\subsection[\hspace{1.75cm} Entropy production]{Entropy production\label{app:EP}}

The Onsager-Machlup action lets one compute the entropy production from the ratio of forward and time-reversed path probabilities.
Under the extended time reversal $\mathbb{T}$ (which sends $\bm{r}_t \to \bm{r}_{-t}$, $\theta_t \to \theta_{-t} + \pi$, $\bm{v}_0 \to -\bm{v}_0$), the residuals transform as
\begin{align}
\begin{split}
    \bm{E}_{\bm{r}}^{\mathbb{T}} &= \bm{E}_{\bm{r}} - 2\gamma_t(\dot{\bm{r}} - \bm{v}_0) \; , \\
    E_\theta^{\mathbb{T}} &= E_\theta - 2\gamma_r\dot{\theta} \; .
    \end{split}
\end{align}
The difference of Onsager-Machlup actions is thus
\begin{align}
    \mathcal{A}_{OM}^{\mathbb{T}} - \mathcal{A}_{OM} = 
    & 
    - \frac{1}{4 k_BT} \int dt\left[ \frac{1}{\gamma_t}\left(|\bm{E}_{\bm{r}}^{\mathbb{T}}|^2 - |\bm{E}_{\bm{r}}|^2\right) + \frac{1}{\gamma_r}\left((E_\theta^{\mathbb{T}})^2 - E_\theta^2 \right)\right] 
    \nonumber\\
    & + \ln P_\mathcal{T} - \ln P_{-\mathcal{T}} 
    \; .
\end{align}
Expanding the squares:
\begin{align}
\begin{split}
    |\bm{E}_{\bm{r}}^{\mathbb{T}}|^2 - |\bm{E}_{\bm{r}}|^2 &= -4\gamma_t\, \bm{E}_{\bm{r}}\cdot(\dot{\bm{r}} - \bm{v}_0) + 4\gamma_t^2 |\dot{\bm{r}} - \bm{v}_0|^2 \; , \\
    (E_\theta^{\mathbb{T}})^2 - E_\theta^2 &= -4\gamma_r\, E_\theta\, \dot{\theta} + 4\gamma_r^2 \dot{\theta}^2 \; .
    \end{split}
\end{align}
Substituting Eqs.~\eqref{eq:Er_def}--\eqref{eq:Etheta_def} and simplifying then yields
\begin{align}
   &  \bm{E}_{\bm{r}}\cdot(\dot{\bm{r}} - \bm{v}_0) + E_\theta\, \dot{\theta} - \left(\gamma_t|\dot{\bm{r}} - \bm{v}_0|^2 + \gamma_r\dot{\theta}^2\right) 
   \nonumber\\
   & 
   \qquad\qquad\qquad
   = \frac{d}{dt}\left[\frac{(\dot{\bm{r}} - \bm{v}_0)^2}{2} + \frac{\dot{\theta}^2}{2} - K \bm{v}_0 \cdot \bm{s}\right] \; .
    \label{eq:total_derivative}
\end{align}
The expression in brackets is precisely $-k_BT \ln P_t + \text{const}$, so
\begin{align}
    \frac{1}{k_BT}\int dt\, \frac{d}{dt}\left[\frac{(\dot{\bm{r}} - \bm{v}_0)^2}{2} + \frac{\dot{\theta}^2}{2} - K \bm{v}_0 \cdot \bm{s}\right] = -\left[\ln P_\mathcal{T} - \ln P_{-\mathcal{T}}\right] \; .
\end{align}
Everything cancels out as expected, yielding
\begin{align}
    \mathcal{A}_{OM}^{\mathbb{T}} - \mathcal{A}_{OM} = 0 \; .
    \label{eq:OM_TRS}
\end{align}
The entropy production under the extended time reversal $\mathbb{T}$ vanishes identically, confirming detailed balance.
This provides an independent verification of the time-reversal symmetry, directly at the level of trajectory probabilities rather than through response-field identities.

\subsection[\hspace{1.75cm} Spurious entropy production under incomplete TRS] {Spurious entropy production under incomplete TRS\label{app:SpuriousEP}}

We now derive the spurious entropy production obtained when using the incomplete time reversal instead of the generalized one.
For simplicity, in this section, we set $\gamma_t = \gamma_r = \gamma$.
Under the \emph{incomplete} time reversal $\mathbb{T}_S$, one reverses time but does \emph{not} flip spins or self-propulsion:
\begin{align}
    \mathbb{T}_S: \quad \bm{r}_t \to \bm{r}_{-t}\;, \qquad \theta_t \to \theta_{-t}\;, \qquad \bm{s} \to \bm{s}\;, \qquad \bm{v}_0 \to \bm{v}_0 \;.
\end{align}
The reversed residuals under this transformation are (after changing integration variable $t \to -t$):
\begin{align}
    \hat{E}_{\bm{r}} &= \bm{E}_{\bm{r}} - 2K\dot{\bm{s}} \;, \label{eq:Erhat_def}\\
    \hat{E}_\theta &= E_\theta + 2K\dot{\bm{r}}\cdot\bm{s}_\perp \;. \label{eq:Ethetahat_def}
\end{align}
Using $|\hat{E}_{\bm{r}}|^2 - |\bm{E}_{\bm{r}}|^2 = (\hat{E}_{\bm{r}} - \bm{E}_{\bm{r}})\cdot(\hat{E}_{\bm{r}} + \bm{E}_{\bm{r}})$ and similarly for the angular sector:
\begin{align}
    & |\hat{E}_{\bm{r}}|^2 - |\bm{E}_{\bm{r}}|^2 + \hat{E}_\theta^2 - E_\theta^2
    \nonumber\\
    & \qquad\qquad 
    = -4K\dot{\bm{s}} \cdot \bm{E}_{\bm{r}} + 4K^2\dot{\theta}^2
    + 4K(\dot{\bm{r}}\cdot\bm{s}_\perp)\,E_\theta + 4K^2(\dot{\bm{r}}\cdot\bm{s}_\perp)^2 \;.
\end{align}
Substituting the definitions~\eqref{eq:Er_def}--\eqref{eq:Etheta_def}, the $K^2$ terms cancel exactly,
\begin{align}
    -\dot{\bm{s}}\cdot(K\dot{\bm{s}}) + K^2\dot{\theta}^2 &= -K^2\dot{\theta}^2 + K^2\dot{\theta}^2 = 0 \;, \\
    (\dot{\bm{r}}\cdot\bm{s}_\perp)(-K\dot{\bm{r}}\cdot\bm{s}_\perp) + K^2(\dot{\bm{r}}\cdot\bm{s}_\perp)^2 &= 0 \;,
\end{align}
using $|\dot{\bm{s}}|^2 = \dot{\theta}^2$ and $|\bm{s}_\perp|^2 = 1$.
The remaining terms are:
\begin{align}
\begin{split}
    &-4K\dot{\bm{s}}\cdot\left[\ddot{\bm{r}} + \gamma_t(\dot{\bm{r}} - \bm{v}_0)\right] + 4K(\dot{\bm{r}}\cdot\bm{s}_\perp)\left[\ddot{\theta} + \gamma_r\dot{\theta}\right] \\
    &\quad = 4K\left[-\dot{\theta}\,\bm{s}_\perp\cdot\ddot{\bm{r}} + (\dot{\bm{r}}\cdot\bm{s}_\perp)\,\ddot{\theta}\right] + 4K\gamma\left[-\dot{\theta}\,\bm{s}_\perp\cdot(\dot{\bm{r}} - \bm{v}_0) + \dot{\theta}\,(\dot{\bm{r}}\cdot\bm{s}_\perp)\right] \;.
\end{split}
    \label{eq:stdT_remaining}
\end{align}
The friction terms simplify to $4K\gamma\,\dot{\theta}\,\bm{v}_0\cdot\bm{s}_\perp$, i.e.\ in terms of $\theta$:
\begin{align}
    4K\gamma\,\dot{\theta}\,\bm{v}_0\cdot\bm{s}_\perp = 4K\gamma\,\dot{\theta}\left(-v_{0x}\sin\theta + v_{0y}\cos\theta\right) = -4K\gamma\,\bm{v}_0\cdot\dot{\bm{s}} \;.
\end{align}
Writing $\bm{v}_0\cdot\bm{s} = v_{0x}\cos\theta + v_{0y}\sin\theta$, the friction piece is a total derivative:
\begin{align}
    \gamma\,\dot{\theta}\,\bm{v}_0\cdot\bm{s}_\perp = \gamma\,\frac{d}{dt}(\bm{v}_0\cdot\bm{s}) \;.
\end{align}
The acceleration cross terms, however, are \emph{not} a total derivative.
Writing $\bm{s}_\perp = (-\sin\theta, \cos\theta)$ and using the shorthand $v_\perp \equiv \dot{\bm{r}}\cdot\bm{s}_\perp = -\dot{x}\sin\theta + \dot{y}\cos\theta$ for the velocity component perpendicular to the spin, the difference of Onsager-Machlup actions reads
\begin{align}
    \mathcal{A}_{OM}^{\mathbb{T}_S} - \mathcal{A}_{OM} = \ln\frac{P_\mathcal{T}}{P_{-\mathcal{T}}} - \frac{K}{\gamma k_BT}\int dt\,\left[v_\perp\,\ddot{\theta} - \dot{\theta}\,(\ddot{\bm{r}}\cdot\bm{s}_\perp) + \gamma\,\frac{d}{dt}\!\left(\bm{v}_0\cdot\bm{s}\right) \right] \;.
    \label{eq:EPR_stdT}
\end{align}
All quantities are expressed in terms of $\theta$, $\dot{\theta}$, $\ddot{\theta}$, $\dot{\bm{r}}$, and $\ddot{\bm{r}}$, with
\begin{align}
    v_\perp = -\dot{x}\sin\theta + \dot{y}\cos\theta \;, \qquad
    \ddot{\bm{r}}\cdot\bm{s}_\perp = -\ddot{x}\sin\theta + \ddot{y}\cos\theta \;.
\end{align}

The average of Eq.~\eqref{eq:EPR_stdT} can be evaluated using the equations of motion.
In the Langevin dynamics, $\bm{E}_{\bm{r}} = \sqrt{2\gamma k_BT}\,\bm{\eta}$ and $E_\theta = \sqrt{2\gamma k_BT}\,\xi$, so
\begin{align}
    \ddot{\theta} &= K v_\perp - \gamma\dot{\theta} + \sqrt{2\gamma k_BT}\,\xi \;, \\
    \ddot{\bm{r}}\cdot\bm{s}_\perp &= -K\dot{\theta} - \gamma(v_\perp - \bm{v}_0\cdot\bm{s}_\perp) + \sqrt{2\gamma k_BT}\,\bm{\eta}\cdot\bm{s}_\perp \;.
\end{align}
Substituting:
\begin{align}
\begin{split}
    v_\perp\,\ddot{\theta} - \dot{\theta}\,(\ddot{\bm{r}}\cdot\bm{s}_\perp)
    &= v_\perp\left[K v_\perp - \gamma\dot{\theta}\right] - \dot{\theta}\left[-K\dot{\theta} - \gamma v_\perp + \gamma\,\bm{v}_0\cdot\bm{s}_\perp\right] + \text{noise} \\
    &= K\left[v_\perp^2 + \dot{\theta}^2\right] - \gamma\,\dot{\theta}\,\bm{v}_0\cdot\bm{s}_\perp + \text{noise} \;,
\end{split}
\end{align}
where the $\gamma v_\perp\dot{\theta}$ terms cancel. The noise terms $\propto v_\perp\xi - \dot{\theta}\,\bm{\eta}\cdot\bm{s}_\perp$ have zero mean (white noise is uncorrelated with the adapted processes $v_\perp$ and $\dot{\theta}$).
Adding the friction total derivative from Eq.~(\eqref{eq:EPR_stdT}), the $\bm{v}_0\cdot\bm{s}_\perp$ terms cancel, leaving
\begin{align}
    \left\langle v_\perp\,\ddot{\theta} - \dot{\theta}\,(\ddot{\bm{r}}\cdot\bm{s}_\perp) + \gamma\,\frac{d}{dt}(\bm{v}_0\cdot\bm{s}) \right\rangle = K\left\langle v_\perp^2 + \dot{\theta}^2 \right\rangle \;.
\end{align}
The entropy production under $\mathbb{T}_S$ is usually defined with the opposite sign convention~\cite{Seifert2012} $\sigma_{\mathbb{T}_S} = \mathcal{A}_{OM} - \mathcal{A}_{OM}^{\mathbb{T}_S}$, so the average steady-state entropy production rate is
\begin{align}
    \langle \dot{\sigma}_{\mathbb{T}_S} \rangle = \frac{K^2}{\gamma k_BT}\left\langle v_\perp^2 + \dot{\theta}^2 \right\rangle \;.
    \label{eq:avg_EPR_stdT}
\end{align}
This is manifestly non-negative. At $\bm{v}_0 = 0$, the system is in thermal equilibrium and equipartition gives $\langle v_\perp^2 \rangle = k_BT$ (one transverse translational 
degree of freedom, averaged over the isotropic $\theta$-distribution) and $\langle\dot{\theta}^2\rangle = k_BT$, so
\begin{align}
    \langle \dot{\sigma}_{\mathbb{T}_S} \rangle = \frac{2K^2}{\gamma} \;.
    \label{eq:EPR_stdT_equil}
\end{align}
This is nonzero despite the system being in genuine thermal equilibrium. 
The spurious entropy production rate is proportional to $K^2$ and inversely proportional to $\gamma$.
It vanishes only when $K = 0$, i.e.\ when the spin-velocity coupling is absent and $\mathbb{T}_S$ and $\mathbb{T}$ coincide.

The entropy production~\eqref{eq:avg_EPR_stdT} arises because $\mathbb{T}_S$ does not account for the spin flip required by the Hamiltonian structure: it treats the deterministic spin-velocity coupling $K\dot{\bm{s}}$ as an irreversible current and incorrectly attributes dissipation to it.
Two additional comments are in order.
First, this is a constant entropy production rate, akin to a ``housekeeping'' cost, so that the total entropy production of a trajectory with duration $\tau$ grows linearly with $\tau$.
Second, this is the entropy production rate for a single particle.
If one now assumes that $N$ particles are coupled by Hamiltonian~\eqref{eq:Hamiltonian} and initialized in the equilibrium steady state set by the Gibbs measure, the ferromagnetic coupling is left unchanged whether one flips spins or not.
As a result, the only source of entropy production is still the individual spin velocity couplings, and the generalization of Eq.~\eqref{eq:avg_EPR_stdT} is simply
\begin{align}
    \langle \dot{\sigma}_{N,\mathbb{T}_S} \rangle = \frac{K^2}{\gamma k_BT}\sum\limits_{i=1}^N\left\langle v_{\perp,i}^2 + \dot{\theta}_i^2 \right\rangle \;.
    \label{eq:avg_EPR_spurious_Nbody}
\end{align}
In the many-body case, the spins and velocities are still subject to equipartition relations~\cite{Bore2016,Casiulis2019b}.
The angular velocities follow standard equipartition,
\begin{align}
    \left\langle \dot{\theta}_i^2 \right\rangle &= k_B T
\end{align}
while the velocities are Gaussian with width $k_B T$ around the off-zero mean $\bm{v}_0$.
As a result, it is convenient to decompose the velocity into its mean $\bm{v}_0$ and fluctuations $\bm{u}_i$ so that $\bm{v}_{\perp,i} = \bm{v}_0\cdot\bm{s}_{\perp,i} + \bm{u}_i \cdot\bm{s}_{\perp,i}$.
The fluctuations $\bm{u}_i$, by definition, follow a normal distribution with zero mean and width $k_B T$, and
\begin{align}
    \left\langle v_{\perp,i}^2 \right\rangle &= v_0^2 \left\langle \sin^2 \theta_i\right\rangle + \left\langle \left(\bm{u}_{i} \cdot \bm{s}_{\perp,i}\right)^2 \right\rangle
\end{align}
where $\bm{v}_0$ is here assumed to be along $\hat{\bm{e}}_x$ without any loss of generality.
The last term in that last equation is the average of a projection of an isotropic Gaussian variable along a unit vector: it immediately yields $k_B T$, so that 
\begin{align}
    \left\langle v_{\perp,i}^2 \right\rangle &= v_0^2 \left\langle \sin^2 \theta_i\right\rangle + k_B T.
\end{align}
As a result, in the many-body setting, assuming identical particles,
\begin{align}
    \langle \dot{\sigma}_{N,\mathbb{T}_S} \rangle = N\frac{K^2}{\gamma } + \frac{NK^2 v_0^2}{\gamma k_B T} \left\langle \sin^2\theta \right\rangle .
    \label{eq:avg_EPR_spurious_Nbody_final}
\end{align}
Because the source of spurious entropy production is the one-body spin-velocity coupling term, the many-body EPR is extensive, as it would be in an active system where each particle has its own propulsion engine.
Furthermore, note that the EPR is a function of the ferromagnetic ordering: it is lowest when spins are all aligned or antialigned with $\bm{v}_0$ ($\sin \theta_i = 0$ for all particles), and maximal when they are orthogonal to $\bm{v}_0$, with the isotropic case $\left\langle \sin^2\theta \right\rangle = 1/2$ lying in between.

\section[\hspace{2cm} Angular diffusion constant corrections]{Angular diffusion constant corrections}
\label{app:DR}

In the main text, we present a derivation of a simple value of $D_R$ for $v_0 = 0$.
Here, we discuss the validity of the simplified value, then show the expression used in the main text for $v_0 \neq 0$.

\subsection[\hspace{1.75cm}Self-consistent $D_R$: when it matters] {Self-consistent $D_R$: when it matters}

As shown in the main text, the full expression of the rotational diffusion constant $D_R$ is originally obtained within an implicit equation,
\begin{align}
D_R = \frac{k_BT}{\gamma_r + K^2\gamma_t/(\gamma_t + D_R)^2}
  \;,
  \label{eq:DR_self_consistent}
\end{align}
which defines $D_R$ as the real-valued root of a cubic polynomial.
In the main text, we disregard this full expression and use the leading-order value at small $D_R/\gamma_t$,
\begin{align}
    D_R^{(0)} = \frac{k_BT\,\gamma_t}{\gamma_r\gamma_t + K^2} \; .
    \label{eq:DR_K}
\end{align}
Here, we briefly justify this assumption and discuss its failure modes.

To make the discussion simpler, we assume $\gamma_r = \gamma_t = \gamma$, so that 
\begin{align}
D_R = \frac{k_BT/\gamma}{1 + K^2/(\gamma + D_R)^2}
  \;,
  \label{eq:DR_self_consistent_onegamma}
\end{align}
which may be rewritten in terms of two parameters, the usual equilibrium diffusion constant $D_{r} = k_B T/\gamma$ and $\kappa = K/\gamma$,
\begin{align}
D_R = \frac{D_r}{1 + \kappa^2/(1 + D_R/\gamma)^2}
  \;.
  \label{eq:DR_self_consistent_kappa}
\end{align}
In this work, we keep $k_BT$ and $\gamma$ constant throughout, so $D_r$ is also constant, and 
this equation can be considered to depend on the parameter $\kappa$ only.
In particular, the $D_R/\gamma$ on the right-hand-side can be recursively eliminated by replacing $D_R$ by its expression in the right-hand-side, which yields an expansion-form expression for $D_R$ in terms of $\kappa$,
\begin{align}
    D_R &= \frac{D_r}{1 + \kappa^2/(1 + D_R/\gamma)^2} \\
    &=\frac{D_r}{1 + \kappa^2/\Big[1 + \dfrac{D_r/\gamma}{1+ \kappa^2/(1+D_R/\gamma)^2}\Big]^2} 
    \label{eq:first-order}\\
    &= \ldots
\end{align}
In practice, one can truncate this expansion after a finite number $n$ of recursions, by setting the $D_R/\gamma$ 
on the right-hand-side to $0$, obtaining in this way $D_R^{(n)}$.
The expression in the main text is $D_R^{(0)}$, which is valid under the self-consistent assumption that $D_R/\gamma \ll 1$.
Conversely, $\lim_{n\to \infty} D_R^{(n)}$ recovers the full solution to the implicit equation. 

To assess how bad truncating the expansion can be, it is useful to consider $D_R^{(0)}/\gamma$ as an expansion parameter, which is a proxy for $D_R/\gamma$ -- if $D_R^{(0)}/\gamma$ is small, truncating at the zeroth order by taking $D_R/\gamma \ll 1$ is self-consistent, if not, more terms should be considered. 

Concretely, the zeroth order is 
\begin{align}
    \frac{D_R^{(0)}}{\gamma} = \frac{D_r/\gamma}{1 + \kappa^2} 
\end{align}
and this quantity is guaranteed to be small for $\kappa \gg 1$, so that the form in the main text is valid in this regime.

In the opposite limit, $\kappa \to 0$, the recursive expansion of $D_R$ can be coupled to a Taylor expansion in powers of $\kappa$, \textit{e.g.}
\begin{align}
    D_R &\approx D_r \left[1 - \frac{\kappa^2}{(1 + D_R/\gamma)^2}\right].
\end{align}
It is then clear that truncating at higher orders in $D_R^{(n)}$ will contribute to higher-order terms in $\kappa$.
Thus, the approximation in the main text \textit{also} yields the correct leading-order corrections as $\kappa \to 0$.

As a result, one expects, in practice, the worst regime for $K \sim \gamma$, where neither $D_R^{(0)}/\gamma$ nor $\kappa$ can be considered small.
We briefly check this on the $K$ values used in the main text in Table~\ref{tab:DR_corrections}, where we compare the values predicted by $D_R^{(0)}$, $D_R^{(1)}$ and the full self-consistent $D_R$ to a numerical estimate of the slope of the diffusive regime in simulations at $v_0 = 0$ for $N = 10^3$ trajectories.
We find that the agreement is worst at $K = \gamma$ as expected.

\begin{table}[h]
\centering
\begin{tabular}{c|c|c|c|c|c}
$K$ & $D_R^{(0)}$ & $D_R^{(1)}$ (pert.)
    & $D_R$ (self-cons.) & $D_R$ (sim.) \\
\hline
0  & 1.0000 & 1.0000 & 1.0000 & 0.992 \\
1  & 0.5000 & 0.6923 & 0.683  & 0.578 \\
5  & 0.0385 & 0.0414 & 0.0415 & 0.0414 \\
10 & 0.010 & 0.0101 & 0.0101 & 0.0101 \\
\end{tabular}
\caption{Rotational diffusion constant at $v_0 = 0$ for several values
of~$K$, with $\gamma = k_BT = 1$.  ``$D_R^{(1)}$ (pert.)'' is
the first-order perturbative
correction.  ``Self-cons.'' is the
numerical solution of Eq.~\eqref{eq:DR_self_consistent}.}
\label{tab:DR_corrections}
\end{table}

\subsection[\hspace{1.75cm}$D_R$ with a global velocity]{$D_R$ with a global velocity\label{app:DRwithv0}}

The bath velocity $\bm{v}_0$ enters the momentum equation but not the angular equation directly.
However, it generates an effective torque on the spin through the spin-velocity coupling.
To see this, we define the momentum in the co-moving frame, $\bm{q} = \bm{p} - \bm{v}_0$, which obeys
\begin{align}
    \dot{\bm{q}} = \dot{\bm{p}} - \bm{0} = -\gamma_t(\bm{p} - K\bm{s} - \bm{v}_0) + \sqrt{2\gamma_t k_BT}\,\bm{\eta} = -\gamma_t(\bm{q} - K\bm{s}) + \sqrt{2\gamma_t k_BT}\,\bm{\eta} \; .
\end{align}
Then, we rewrite the angular torque $K\bm{p}\cdot\bm{s}_\perp$ in terms of $\bm{q}$:
\begin{align}
    K\bm{p}\cdot\bm{s}_\perp = K(\bm{q} + \bm{v}_0)\cdot\bm{s}_\perp = K\bm{q}\cdot\bm{s}_\perp + K\bm{v}_0\cdot\bm{s}_\perp \;.
\end{align}
Taking $\bm{v}_0 = v_0\hat{\bm{x}}$ and using $\bm{s}_\perp = (-\sin\theta, \cos\theta)$:
\begin{align}
    \bm{v}_0\cdot\bm{s}_\perp = v_0(-\sin\theta) = -v_0\sin\theta \;.
\end{align}
The angular equation of motion therefore becomes
\begin{align}
    \dot{\omega} = K\bm{q}\cdot\bm{s}_\perp \underbrace{- Kv_0\sin\theta}_{\text{aligning torque}} - \gamma_r\omega + \sqrt{2\gamma_r k_BT}\,\xi \; .
    \label{eq:omega_with_v0}
\end{align}
The term $-Kv_0\sin\theta = -\partial_\theta U_{\rm eff}$ derives from the effective potential
\begin{align}
    U_{\rm eff}(\theta) = -Kv_0\cos\theta \; .
    \label{eq:Ueff}
\end{align}
This potential has the same effect as a magnetic field of strength $h = Kv_0$ on the spin, favouring alignment with $\bm{v}_0$, as noted in previous works~\cite{Bore2016,Casiulis2019b}.

We now establish the diffusive constant in the long-time regime under this potential.
We start by establishing the steady-state equations of the system.
Following the steady-state derivation of the main text to write the rescaled damping of the 
angular variables, in steady state we find
\begin{align}
    \gamma_{\rm eff}\,\dot{\theta} = -Kv_0\sin\theta + \sqrt{2\gamma_{\rm eff} k_BT}\,\xi \; .
    \label{eq:overdamped_theta}
\end{align}
The noise amplitude $\sqrt{2\gamma_{\rm eff} k_BT}$ is fixed by the Einstein relation: in equilibrium the steady-state distribution must be $P(\theta) \propto e^{-U_{\rm eff}/k_BT}$, which requires $D_R = k_BT/\gamma_{\rm eff}$, consistent with Eq.~\eqref{eq:DR_K}.

Equation~\eqref{eq:overdamped_theta} describes a Brownian particle on a circle in the periodic potential $U_{\rm eff}(\theta) = -Kv_0\cos\theta$, with effective friction $\gamma_{\rm eff}$ and bare diffusion constant $D_R = k_BT/\gamma_{\rm eff}$.
The effective long-time diffusion constant for overdamped motion in a periodic potential is an exactly solvable problem~\cite{Festa1978,Risken1989}.

In short, the effective diffusion constant is then obtained from the mean time $\langle T \rangle$ to traverse one period of $2\pi$~\cite{Risken1989},
\begin{align}
    D_R = \frac{4\pi^2}{2\langle T \rangle} \;,
\end{align}
where the time may be re-expressed as
\begin{align}
    D_R(v_0) = \frac{D_R(v_0 = 0)\, 4\pi^2}{\displaystyle\int_0^{2\pi}\!e^{\beta U_{\rm eff}(\theta)}\,d\theta \;\int_0^{2\pi}\!e^{-\beta U_{\rm eff}(\theta)}\,d\theta} \; .
    \label{eq:Deff_periodic}
\end{align}
For $U_{\rm eff}(\theta) = -Kv_0\cos\theta$, the integrals evaluate to the modified Bessel function $I_0$:
\begin{align}
    \int_0^{2\pi} e^{\pm\beta Kv_0\cos\theta}\,d\theta = 2\pi\,I_0(\beta Kv_0) \; .
\end{align}
Substituting into Eq.~\eqref{eq:Deff_periodic},
\begin{align}
    D_R(v_0) = \frac{D_R(v_0 = 0)}{\left[I_0(\beta Kv_0)\right]^2} \; .
    \label{eq:DR_v0}
\end{align}
We refer readers interested in the finer details of the calculation to Chapter 11 of Ref.~\cite{Risken1989}.

We now briefly comment on the asymptotic regimes of $D_R(v_0)$.
For weak alignment, ($\beta Kv_0 \ll 1$), $I_0(x) \approx 1 + x^2/4$, so that
\begin{align}
    D_R \approx D_R^{(0)}\left[1 - \frac{(\beta Kv_0)^2}{2} + \cdots\right],
\end{align}
and the leading correction is quadratic in $v_0$.
For strong alignment, on the other hand ($\beta Kv_0 \gg 1$), $I_0(x) \approx e^x/\sqrt{2\pi x}$, so that
\begin{align}
    D_R \approx 2\pi\beta Kv_0\,D_R^{(0)}\,e^{-2\beta Kv_0} \; ,
\end{align}
which is exponentially suppressed.
This is why for large $Kv_0$ the plateau of confined dynamics of the MSAD lasts longer.

\section[\hspace{2cm} Mean-square displacements and Einstein relations]{Mean-square displacements and Einstein relations \label{app:MSD}}

\subsection[\hspace{1.75cm}Mean-square displacement] {Mean-square displacement}

The mean-square displacement (MSD) decomposes as
$\Delta r^2(t) = \Delta r^2_{\rm fluct}(t) + v_0^2 t^2$,
where $\Delta r^2_{\rm fluct}(t) = \langle [\delta\bm{r}(t) - \delta\bm{r}(0)]^2 \rangle$ with $\delta\bm{r} = \bm{r} - \langle\bm{r}\rangle$.
Thus
\begin{align}
    \frac{d}{dt}\Delta r^2_{\rm fluct}(t) = -2\,\frac{d}{dt} C^c_{\bm{r},\bm{r}}(t) \; ,
\end{align}
and the connected FDT, Eq.~\eqref{eq:Rrr}, immediately gives
\begin{align}
    \frac{d}{dt}\Delta r^2_{\rm fluct}(t) = \frac{2}{\beta}\,R_{\bm{r},\bm{r}}(t) \; .
    \label{eq:MSD_response}
\end{align}
Defining the spatial diffusion constant $D_T = \lim_{t\to\infty} \Delta r^2_{\rm fluct}(t)/(2d\,t)$ with $d=2$ and integrating Eq.~\eqref{eq:MSD_response}:
\begin{align}
    D_T = \lim_{t\to\infty}\frac{1}{d\beta t}\int_0^t d\tau\, R_{\bm{r},\bm{r}}(\tau) \equiv \frac{\mu_T}{\beta} \; ,
    \label{eq:Einstein_T}
\end{align}
where $\mu_T$ is the translational DC mobility per component, which can be seen as the time-averaged response.
In the limit of infinitesimal forcing, which implies linear response, the integrated response is in practice given by the ratio of the displacement due to the forcing to its intensity, so that for a force acting along coordinate $y$, the response along coordinate $x$ is
\begin{align}
    \chi_{x,y}(t) \equiv \frac{\left\langle x(t) \right\rangle_{f_y} - \left\langle x(t)\right\rangle_0}{f_y} = \int\limits_{0}^t dt' \; R_{xy} (t,t') 
    \;.
\end{align}
In the limit of long times, the average velocity relaxes to its steady-state value so that $(\left\langle x(t) \right\rangle_{f_y} - \left\langle x(t)\right\rangle_0)/t \to  v_{x}^\infty (f_y) -  v_x^\infty(0)$ with $v_x^\infty(f) = \lim_{t\to \infty}\left\langle \dot{x} (t) \right\rangle_{f}$.
As a result,
\begin{align}
\mu_T = \frac{1}{d} \sum\limits_{a=1}^d \frac{v_a^\infty (f_a) - v_a^\infty(0)}{f_a}
\; .
\end{align}
As a result, if a force $\bm{F}_{\rm ext}$ is applied along any direction, one will then have the vector equation
\begin{align}
\bm{v}^\infty (\bm{F}_{\rm ext}) - \bm{v}^\infty(0) = \mu_T \bm{F}_{\rm ext}
\; .
\end{align}

To compute $\mu_T$ for our dynamics, notice that the steady-state mean momentum equation in the presence of an external force reads $\gamma_t\langle\bm{p}\rangle = \gamma_t K\langle\bm{s}\rangle + \gamma_t\bm{v}_0 + \bm{F}_{\rm ext}$, or $\langle\dot{\bm{r}}\rangle - \bm{v}_0 = \bm{F}_{\rm ext}/\gamma_t$ 
so that $\mu_T = 1/\gamma_t$.
This Einstein relation holds for all values of $K$ and $\bm{v}_0$.
Intermediate-time MSDs do have weak $K$ and $v_0$ dependences, which we do not discuss further given the agreement with simple Brownian predictions.

\subsection[\hspace{1.75cm}Mean-square angular displacement] {Mean-square angular displacement \label{app:MSAD}}

For 2d spins $\bm{s} = (\cos\theta, \sin\theta)$, where we here treat the angle $\theta$ as unbounded so that $\Delta\theta^2(t) = \langle[\theta(t) - \theta(0)]^2\rangle$ can grow indefinitely.
Since $\langle\omega\rangle = 0$ in the steady state, $\langle\theta(t)\rangle = \theta_0$ and there is no ballistic advection piece.
Writing $\theta(t) - \theta(0) = \int_0^t \omega(t')\,dt'$:
\begin{align}
    \frac{d}{dt}\Delta\theta^2(t) = 2\int_0^t C_{\omega,\omega}(\tau)\,d\tau \; ,
    \label{eq:MSAD_Comega}
\end{align}
where $C_{\omega,\omega}(\tau) = \langle\omega(t)\,\omega(t')\rangle$ is the stationary angular velocity autocorrelation.

The time-reversal transformation $\theta_t \to \theta_{-t} + \pi$, $i\lambda_t \to i\lambda_{-t} + \beta\,\partial_t\theta_{-t}$ yields, for $\tau > 0$:
\begin{align}
    R_{\theta,\theta}(\tau) = -\beta\,\langle\theta(-t)\,\omega(-t')\rangle \; .
\end{align}
Differentiating with respect to $\tau$ gives the angular velocity FDT:
\begin{align}
    R_{\omega,\theta}(\tau) \equiv \partial_\tau R_{\theta,\theta}(\tau) = \beta\,C_{\omega,\omega}(\tau) \; ,
    \label{eq:FDT_omega}
\end{align}
where $R_{\omega,\theta}$ is the response of $\omega$ to an external torque $\Gamma_{\rm ext}$.
Combining Eqs.~\eqref{eq:MSAD_Comega} and~\eqref{eq:FDT_omega}:
\begin{align}
    \frac{d}{dt}\Delta\theta^2(t) = \frac{2}{\beta}\,R_{\theta,\theta}(t) \; ,
    \label{eq:MSAD_response}
\end{align}
the exact analogue of Eq.~\eqref{eq:MSD_response} -- in particular, the time-integral of the response defines the DC angular mobility,
\begin{align}
    \mu_R \equiv \lim_{t\to \infty}\frac{1}{t}\int\limits_{0}^t dt'\,R_{\theta,\theta}(t') \approx\lim_{t\to\infty} R_{\theta,\theta} \; ,
    \label{eq:MSAD_mobility_FDT}
\end{align}
or equivalently the diffusion constant,
\begin{align}
    D_R \equiv \frac{\mu_R}{\beta}
\label{eq:MSAD_diffusion_FDT}
\end{align}
which is computed in the main text.
In the following, we provide derivations for the full MSAD expressions used in the figures of the main text.

\subsection[\hspace{1.75cm}MSAD prediction at zero global velocity] {MSAD prediction at zero global velocity\label{app:MSAD_v0=0}}

At $v_0 = 0$, the MSAD can be computed exactly from the \emph{angular velocity} correlator alone, using
Eq.~\eqref{eq:MSAD_Comega}.
To do so, it is useful to define the transverse and parallel projections of the canonical momentum onto the spin,
\begin{align}
  v_\perp \equiv \bm{p}\cdot\bm{s}_\perp \;, \qquad
  v_\parallel \equiv \bm{p}\cdot\bm{s} \;.
\end{align}
The angular velocity equation at $v_0 = 0$ reads
\begin{align}
  \dot{\omega} &= K\,(\bm{p}\cdot\bm{s}_\perp)
  - \gamma_r\,\omega + \sqrt{2\gamma_r k_BT}\,\xi
  \\
  &= K\,v_\perp - \gamma_r\,\omega + \sqrt{2\gamma_r k_BT}\,\xi.
  \label{eq:omega_eq_derived}
\end{align}

We then write an equation on the transverse velocity.
To do so, we differentiate the projection $v_\perp = \bm{p}\cdot\bm{s}_\perp$, using $\dot{\bm{s}}_\perp = -\omega\,\bm{s}$ (which follows from $\bm{s}_\perp = ( -\sin\theta,\cos\theta)$ and $\dot{\theta} = \omega$),
\begin{align}
  \dot{v}_\perp
  = \dot{\bm{p}}\cdot\bm{s}_\perp + \bm{p}\cdot\dot{\bm{s}}_\perp
  = \dot{\bm{p}}\cdot\bm{s}_\perp - \omega\,v_\parallel \;.
  \label{eq:vperp_chain_rule}
\end{align}
The momentum equation at $v_0 = 0$ gives
$\dot{\bm{p}} = -\gamma_t\bm{p} + \gamma_t K\bm{s} + \text{noise}$.
Projecting onto $\bm{s}_\perp$,
\begin{align}
  \dot{\bm{p}}\cdot\bm{s}_\perp
  = -\gamma_t\,v_\perp + \sqrt{2\gamma_t k_BT}\,\eta_\perp \;,
\end{align}
where $\eta_\perp \equiv \bm{\eta}\cdot\bm{s}_\perp$ is still a standard white noise.
Substituting back into~\eqref{eq:vperp_chain_rule},
\begin{align}
  \dot{v}_\perp = -\gamma_t\,v_\perp
  - \omega\,v_\parallel
  + \sqrt{2\gamma_t k_BT}\,\eta_\perp \;.
  \label{eq:vperp_exact}
\end{align}

In spite of the nonlinear $v_\parallel$ term, the equations for the two-time \emph{correlators} $\langle\omega(\tau)\,\omega(0)\rangle$ and $\langle v_\perp(\tau)\,\omega(0)\rangle$ close into a $2\times 2$ linear system.
The reason is that the $v_\parallel$ term generates a three-point function, that averages to zero
\begin{align}
  \frac{d}{d\tau}\langle v_\perp(\tau)\,\omega(0)\rangle = -\gamma_t\,\langle v_\perp(\tau)\,\omega(0)\rangle - K\,\langle\omega(\tau)\,\omega(0)\rangle - \underbrace{\langle\omega(\tau)\,v_\parallel(\tau)\, \omega(0)\rangle}_{\displaystyle = 0} .
\end{align}
Indeed, at $v_0 = 0$ the system is in thermal equilibrium, where the stationary distribution is the Boltzmann measure proportional to $ e^{-\beta H}$ with $H = \frac{1}{2}|\bm{p} - K\bm{s}|^2 + \frac{1}{2}\omega^2$.
In this measure, the velocity variables $(\omega, v_\perp, v_\parallel)$ are jointly Gaussian with zero mean.
Thus, by Wick's theorem, all odd-order correlators of zero-mean Gaussian variables vanish; in particular, the three-point function $\langle\omega(\tau)\,v_\parallel(\tau)\,\omega(0)\rangle = 0$.

Therefore, the two-time correlators satisfy the linear system generated by the $2\times 2$ system $(v_\perp, \omega)$ alone,
\begin{align}
  \dot{\omega} &= K\,v_\perp - \gamma_r\,\omega + \sqrt{2\gamma_r k_BT}\,\xi(t) \;, \\
  \dot{v}_\perp &= -K\,\omega - \gamma_t\,v_\perp + \sqrt{2\gamma_t k_BT}\,\eta_\perp(t) \;,
\label{eq:omega_vperp_system}
\end{align}
where $\xi$ and $\eta_\perp$ are independent white noises.
These equations are to be understood as generating the correct two-time correlators, not as the exact stochastic dynamics of $v_\perp$ (which includes the additional $v_\parallel$ term~\eqref{eq:vperp_exact}).

The system may be rewritten in vector form as
\begin{align}
    \dot{\bm{X}} = M\bm{X} + D_M \bm{\Xi}
\end{align}
with $\bm{X} = \begin{pmatrix}
    \omega \\
    v_\perp
\end{pmatrix}$, $ \bm{\Xi}$ a vector of $2$ independent unit-variance zero-mean white noises, and $M$ and $D_M$ are the drift matrix and the diffusion matrix, respectively.
The drift matrix is
\begin{align}
  M = \begin{pmatrix} -\gamma_r & K \\ -K & -\gamma_t \end{pmatrix} ,
  \label{eq:drift_matrix_2x2}
\end{align}
and the diffusion matrix $D_M = 2k_BT\,\mathrm{diag}(\gamma_r, \gamma_t)$.
The equilibrium covariance matrix is proportional to identity, $\Sigma_M = k_BT\,I$.
Indeed, both $\omega$ and $v_\perp$ have variance $k_BT$ by equipartition, and the cross-correlation $\langle\omega\,v_\perp\rangle = 0$, as one may verify from the Lyapunov equation
$M\Sigma_M + \Sigma_M M^T + D_M = 0$, which yields 
$k_BT(M+M^T) + D_M = 0$.

The eigenvalues of $M$ are
\begin{align}
  \lambda_\pm = -\bar{\gamma} \pm \sqrt{\delta^2 - K^2} \;, \label{eq:eigenvalues_2x2}
\end{align}
where
\begin{align}
  \bar{\gamma} &= \frac{\gamma_t + \gamma_r}{2} \;, \\
  \delta &= \frac{\gamma_t - \gamma_r}{2} \;.
\end{align}

When $K > |\delta|$, the eigenvalues form a complex-conjugate pair:
\begin{align}
  \lambda_\pm &= -\bar{\gamma} \pm i\Omega \;, \\
  \Omega &\equiv \sqrt{K^2 - \delta^2}. \label{eq:Omega_def}
\end{align}

The two-time correlator of angular velocities is defined as $C_{\omega,\omega}(\tau) = k_BT\,(e^{M\tau})_{11}$.
Writing $M = -\bar{\gamma}\,I + N$ with
\begin{align}
    N = \begin{pmatrix}\delta & K \\ -K &
-\delta\end{pmatrix},
\end{align}
noting that $N^2 = (\delta^2 - K^2)\,I = -\Omega^2 I$ for
$K > |\delta|$, the matrix exponential is
$e^{M\tau} = e^{-\bar{\gamma}\tau}[\cos(\Omega\tau)\,I +
(\sin(\Omega\tau)/\Omega)\,N]$, yielding
\begin{align}
    C_{\omega,\omega}(\tau) = k_BT\,e^{-\bar{\gamma}\tau}
    \left[\cos(\Omega\tau)
    + \frac{\delta}{\Omega}\,\sin(\Omega\tau)\right]
  \;, \qquad \tau \geq 0 \;.
  \label{eq:Cww_oscillatory}
\end{align}
This can be written more compactly as
\begin{align}
    C_{\omega,\omega}(\tau) &= k_BT\,(K/\Omega)\,e^{-\bar{\gamma}\tau}\cos(\Omega\tau - \psi) \\
    \psi &= \arctan\frac{\delta}{\Omega}.
\end{align}
The  correlator \emph{oscillates}: at each half-period $\pi/\Omega$, angular momentum is transferred to and from the transverse velocity $v_\perp$ via the ``gyroscopic'' coupling.  

Integrating~\eqref{eq:Cww_oscillatory}, we find
\begin{align}
  \int_0^\infty C_{\omega,\omega}(\tau)\,d\tau
  = k_BT\,\frac{\bar{\gamma} + \delta}{\bar{\gamma}^2 + \Omega^2}
  = \frac{k_BT\,\gamma_t}{\gamma_t\gamma_r + K^2} = D_R^{(0)} \;,
  \label{eq:GK_oscillatory}
\end{align}
using $\bar{\gamma} + \delta = \gamma_t$ and $\bar{\gamma}^2 + \Omega^2 = \gamma_t\gamma_r + K^2$. 
This confirms consistency with the long-time rotational diffusion constant derived in the main text.

Substituting~\eqref{eq:Cww_oscillatory} into Eq.~\eqref{eq:MSAD_Comega} and evaluating the integral $\Delta\theta^2(t) = 2\int_0^t(t-\tau)\,C_{\omega,\omega}(\tau)\,d\tau$
by computing
\begin{align}
    \int_0^t(t-\tau)\,e^{-\bar{\gamma}\tau}e^{i\Omega\tau}\,d\tau
= (zt - 1 + e^{-zt})/z^2
\end{align}
with $z = \bar{\gamma} - i\Omega$, then taking real and imaginary parts to identify cosine and sine parts, one obtains after simplification:
\begin{align}
  \Delta\theta^2_{\rm osc}(t)
  = \frac{2k_BT}{\gamma_t\gamma_r + K^2}\left[ \gamma_t\,t - \frac{\gamma_t^2 - K^2}{\gamma_t\gamma_r + K^2} + \frac{K}{\Omega}\,e^{-\bar{\gamma}t}\, \cos\left(\Omega\,t + 2\varphi - \psi\right)\right] 
  \; ,
  \label{eq:msad_osc_general}
\end{align}
where the two phase angles are
\begin{align}
  \varphi = \arctan\frac{\Omega}{\bar{\gamma}} \;, \qquad
  \psi = \arctan\frac{\delta}{\Omega} 
  \;.
  \label{eq:phase_angles}
\end{align}

These expressions are the main result of this section, and constitute the model that we use in the main text as a theory line for $v_0 = 0$.
We now discuss a few of its properties.
First, we check the short-time limit of the dynamics.
We verify that $\Delta\theta^2_{\rm osc}(0) = 0$, as the constant term and the $\cos(2\varphi - \psi)$ term cancel exactly.
Indeed,
\begin{align}
    \frac{K}{\Omega}\cos(2\varphi-\psi) = \frac{\gamma_t^2-K^2}{
\gamma_t\gamma_r+K^2}
\end{align}.
Furthermore,
\begin{align}
    \frac{d}{dt}\Delta\theta^2_{\rm osc}(0) = 0
\end{align}
as the linear coefficient $\gamma_t$ cancels against the derivative of the exponential--cosine term.
The second time derivative, however, yields
\begin{align}
    \frac{d^2}{dt^2} \Delta\theta^2_{\rm osc}(0)= 2C_{\omega,\omega}(0) = 2k_BT.
\end{align}
which is consistent with a short-time ballistic regime $\Delta\theta^2 \approx k_BT\,t^2$.
On the other hand, as $t \to \infty$, the exponential term vanishes and $\Delta\theta^2_{\rm osc}(t) \to 2D_R^{(0)}\,t$ with $D_R^{(0)} = k_BT\gamma_t/(\gamma_t\gamma_r + K^2)$, as predicted in the main text.

Finally, for $\gamma_t = \gamma_r \equiv \gamma$: $\bar{\gamma} = \gamma$,
$\delta = 0$, $\Omega = K$, $\psi = 0$,
$\varphi = \arctan(K/\gamma)$, and the result
simplifies to
\begin{align}
  \Delta\theta^2_{\rm osc}(t)\big|_{\gamma_t=\gamma_r=\gamma} = \frac{2k_BT}{\gamma^2 + K^2}\left[ \gamma\,t - \cos(2\phi) + e^{-\gamma t}\cos(Kt + 2\phi) \right] , \label{eq:msad_osc_equal_friction}
\end{align}
where we distinguished the simplified notation $\phi = \arctan(K/\gamma)$.
This is the expression used in practice in the main text.

\subsection[\hspace{1.75cm}Non-zero global velocity: linear approximation of confined dynamics] {Non-zero global velocity: linear approximation of confined dynamics\label{app:MSAD_v0>0}}

In the case $Kv_0 >0$, as noted in~\ref{app:DRwithv0}, the angular dynamics behave as if they felt an external confining potential pinning the spin to $\bm{v}_0$.
As a result, in the case $Kv_0 >0$, we treat the MSAD as a piecewise function.
At long times, the MSAD is diffusive with the diffusion constant predicted in~\ref{app:DRwithv0}.
At short times, instead, we develop a model of confined motion by linearizing the dynamics.
We will then combine the two by taking the max of the prediction of the two models at any time point.

We here assume $\bm{v}_0 = v_0 \hat{\bm{e}}_x$ without any loss of generality.
We then linearize dynamics around the potential minimum $\theta = 0$, where $\sin\theta \approx \theta$ and $\cos\theta \approx 1$.  
Defining the transverse velocity $v_y = p_y - K\sin\theta \approx p_y - K\theta$ and using the kinematic relation $\dot{\theta} = \omega$, the coupled system for $(\theta, \omega, v_y)$ reads (with $\gamma_t = \gamma_r \equiv \gamma$ to make the discussion lighter)
\begin{align}
\begin{split}
  \dot{\theta} &= \omega \;, \\
  \dot{\omega} &= -Kv_0\,\theta + K\,v_y - \gamma\,\omega
                  + \sqrt{2\gamma k_BT}\,\xi(t) \;, \\
  \dot{v}_y &= -K\,\omega - \gamma\,v_y
               + \sqrt{2\gamma k_BT}\,\eta_y(t) \;.
\end{split}
\label{eq:linearized_3x3}
\end{align}

The deterministic part is described by the drift matrix
\begin{align}
  A = \begin{pmatrix} 0 & 1 & 0 \\ -Kv_0 & -\gamma & K \\ 0 & -K & -\gamma \end{pmatrix} ,
  \label{eq:drift_matrix}
\end{align}
and the noise covariance matrix is $D = 2\gamma k_BT\,\mathrm{diag}(0, 1, 1)$.

We write the linearized system~\eqref{eq:linearized_3x3} in compact form as
\begin{align}
  \dot{\bm{q}} = A\,\bm{q} + B\,\bm{\zeta}(t) \;,
  \label{eq:compact_langevin}
\end{align}
where $\bm{q} = (\theta, \omega, v_y)^T$ is the state vector, $A$ is the drift matrix~\eqref{eq:drift_matrix}, $\bm{\zeta}(t) = (\xi(t), \eta_y(t))^T$ is the noise vector with $\langle\zeta_i(t)\zeta_j(t')\rangle = \delta_{ij}\delta(t-t')$,
and
\begin{align}
  B = \sqrt{2\gamma k_BT}
  \begin{pmatrix} 0 & 0 \\ 1 & 0 \\ 0 & 1 \end{pmatrix}
\end{align}
is the noise coupling matrix (note that $D = BB^T$).

The equal-time covariance matrix
$\Sigma_{ij} \equiv \langle q_i\,q_j \rangle$
satisfies the continuous Lyapunov equation
\begin{align}
  A\,\Sigma + \Sigma\,A^T + D = 0 \;.
  \label{eq:lyapunov}
\end{align}
In particular, just like in the $v_0 = 0$ discussion in~\ref{app:MSAD_v0=0}, the Lyapunov equation implies that the covariance matrix $\Sigma$ is \emph{diagonal}.
We thus only check its diagonal elements.
The first one is $\Sigma_{11} = \langle\theta^2\rangle = k_BT/(Kv_0)$, which represents equipartition in the harmonic potential $\frac{1}{2}Kv_0\,\theta^2$).
The second one, $\Sigma_{22} = \langle\omega^2\rangle = k_BT$ is standard equipartition for the angular momentum with unit moment of inertia.

Likewise, $\Sigma_{33} = \langle v_y^2\rangle = k_BT$.
This last statement could look surprising at first glance given the presence of the spin-velocity coupling.
However, as noted in past work~\cite{Casiulis2019b}, the canonical measure of the system can be rewritten in terms of velocities rather than momenta, leading to a distribution of velocities that factorizes,
\begin{align}
    P_v(\bm{v}) = \frac{1}{Z_v}e^{-\beta (\bm{v} - \bm{v}_0)^2}.
\end{align}
It immediately follows that the fluctuations of the velocity are given by standard equipartition (it is still a Gaussian with width $k_B T$), but with a mean $\bm{v}_0$ that is off-zero.
Thus, in particular, for the component of the velocity orthogonal to $\bm{v}_0$, $\langle v_y\rangle = 0$ and $\langle v_y^2\rangle = k_BT$.

The formal solution of Eq.~\eqref{eq:compact_langevin} gives, for $\tau > 0$, the two-time correlation matrix
\begin{align}
  \langle \bm{q}(t+\tau)\,\bm{q}(t)^T \rangle = e^{A\tau}\,\Sigma \;.
  \label{eq:two_time_corr}
\end{align}
This follows from $\bm{q}(t+\tau) = e^{A\tau}\bm{q}(t) + \text{noise integral}$, where the noise integral is uncorrelated with $\bm{q}(t)$.
In component notation, $\langle q_i(t+\tau)\,q_j(t)\rangle = (e^{A\tau}\,\Sigma)_{ij}$.
As a result, the MSAD within the linearized approximation is given by
\begin{align}
  \Delta\theta^2_{\rm lin}(\tau)
  &= \langle[\theta(t+\tau) - \theta(t)]^2\rangle \nonumber\\
  &= \langle\theta(t+\tau)^2\rangle + \langle\theta(t)^2\rangle - 2\langle\theta(t+\tau)\,\theta(t)\rangle \nonumber\\
  &= 2\bigl[\Sigma_{11} - (e^{A\tau}\,\Sigma)_{11}\bigr] \;,
  \label{eq:msad_linearized}
\end{align}
where we used stationarity ($\langle\theta^2\rangle = \Sigma_{11}$ at all times) and Eq.~\eqref{eq:two_time_corr} for the cross-term.

Since $\Sigma = k_BT\,\mathrm{diag}\bigl(1/(Kv_0),\,1,\,1\bigr)$, the product $(e^{A\tau}\Sigma)_{11} = (e^{A\tau})_{11}\,\Sigma_{11}$, and the MSAD reduces to
\begin{align}
  \Delta\theta^2_{\rm lin}(\tau) = \frac{2k_BT}{Kv_0}\Bigl[1 - \bigl(e^{A\tau}\bigr)_{11}\Bigr] \;.
  \label{eq:msad_lin_explicit}
\end{align}
The $(1,1)$ element of the matrix exponential is obtained by Lagrange interpolation (sometimes called Lagrange-Sylvester formula in this context, due to its introduction by Sylvester~\cite{Sylvester1883}), as follows.
Let $s_1, s_2, s_3$ be the three roots of the characteristic polynomial of $A$,
\begin{align}
  s^3 + 2\gamma\,s^2 + (\gamma^2 + K^2 + Kv_0)\,s + Kv_0\gamma = 0 \;.
  \label{eq:char_poly}
\end{align}
Then $(e^{A\tau})_{11} = \sum_i c_i\,e^{s_i\tau}$ with weights
\begin{align}
  c_i = \frac{(s_i + \gamma)^2 + K^2}{(s_i - s_j)(s_i - s_k)}
  \;, \qquad \{i,j,k\} = \{1,2,3\} \;.
  \label{eq:weights}
\end{align}
Substituting into~\eqref{eq:msad_lin_explicit}:
\begin{align}
  \Delta\theta^2_{\rm lin}(\tau)
  = \frac{2k_BT}{Kv_0}\left[
    1 - \sum_{i=1}^{3}
    \frac{(s_i + \gamma)^2 + K^2}{(s_i - s_j)(s_i - s_k)}\,
    e^{s_i\tau}\right] .
  \; .
  \label{eq:msad_lin_closed}
\end{align}

The weights satisfy $\sum_i c_i = 1$, $\sum_i c_i s_i = 0$, and $\sum_i c_i s_i^2 = -Kv_0$, ensuring the correct initial conditions, $\Delta\theta^2_{\rm lin}(0) = 0$, $\frac{d}{dt}\Delta\theta^2_{\rm lin}(0) = 0$, and the ballistic onset $\Delta\theta^2_{\rm lin}(\tau) \approx k_BT\,\tau^2$ at $\tau \to 0$.
In the long-time regime,  $\tau \to \infty$, since all eigenvalues of $A$ have negative real part, $e^{A\tau} \to 0$, and
\begin{align}
  \Delta\theta^2_{\rm lin}(\infty) = 2\,\Sigma_{11} = \frac{2k_BT}{Kv_0} \;,
\end{align}
recovering the equipartition result.

The linearized theory~\eqref{eq:msad_linearized} is exact in the harmonic approximation, but it confines the angle to a single potential well forever. 
In reality, the periodic potential $U_{\rm eff}(\theta) = -Kv_0\cos\theta$ has a finite barrier $\Delta U = 2Kv_0$, and thermal fluctuations allow rare barrier-crossing events that restore diffusion on the Kramers timescale $\tau_K \sim (\gamma_R^{\rm eff})^{-1} e^{2\beta Kv_0}$.
On times $t \gg \tau_K$, the angle performs a random walk of step size $2\pi$ with rate $\sim 1/\tau_K$, giving the effective long-time diffusion constant
$D_R(v_0) = D_R^{(0)}/[I_0(\beta Kv_0)]^2$ (Eq.~\eqref{eq:DR_v0}).
On these long timescales, the details of the intra-well dynamics are averaged out: the angle effectively behaves as a free rotor with a renormalized diffusion constant~$D_R(v_0)$.
We therefore model the long-time MSAD by a diffusive regime at $D_R(v_0)$:
\begin{align}
  \Delta\theta^2_{\rm Kramers}(t)
    = 2D_R(v_0) t.
  \label{eq:msad_kramers}
\end{align}

The full MSAD theory used in the numerical plots is the pointwise maximum of the two branches:
\begin{align}
  \Delta\theta^2_{\rm theory}(t) = \max\bigl(
    \Delta\theta^2_{\rm lin}(t),\;
    \Delta\theta^2_{\rm Kramers}(t)
  \bigr) \;.
  \label{eq:msad_composite}
\end{align}
The linearized branch dominates at short and intermediate times (ballistic $\to$ plateau), while the Kramers branch takes over at long times when barrier hopping restores diffusion.
The crossover between the two branches occurs near the Kramers time $\tau_K$, where the linearized plateau $2k_BT/(Kv_0)$ is overtaken by the growing Kramers curve $2D_R(v_0)\,t$.

It is challenging to model analytically the actual shape of the crossover between the linearized theory and the Kramers regime. Hence, we have chosen to represent the crossover through a max function.
One could in principle choose a different shape (\textit{e.g.} an exponential crossover with a characteristic time, but such a choice would introduce superfluous complexity in the model).
Furthermore, note that the terminal diffusion constant is asymptotically exact, so that the precise choice of crossover shape is negligible at long times.

\section[\hspace{2cm} Overdamped dynamics with a tachostat]{Overdamped dynamics with a tachostat\label{app:overdamped}}

Finally, we briefly focus on the limit of overdamped dynamics, Eqs.~\eqref{eq:Langevin_r_overdamped},~\eqref{eq:Langevin_theta_overdamped}, which has been the focus of recent work~\cite{Chen2026}, but was only considered for $v_0 = 0$.

Without loss of generality, we assume that $\bm{v}_0 = v_0 \hat{\bm{e}}_x$, since it is the only term that breaks rotational invariance, and write the overdamped equations explicitly in terms of Cartesian coordinates,
\begin{align}
  \gamma_t v_x &= \gamma_t v_0 + K\omega\sin\theta + \sqrt{2\gamma_t k_BT}\,\eta_x \;,
  \label{eq:od_vx} \\
  \gamma_t v_y &= -K\omega\cos\theta + \sqrt{2\gamma_t k_BT}\,\eta_y \;,
  \label{eq:od_vy} \\
  \gamma_r\omega &= K(v_y\cos\theta - v_x\sin\theta) + \sqrt{2\gamma_r k_BT}\,\xi \;.
  \label{eq:od_omega}
\end{align}
Substituting Eqs.~\eqref{eq:od_vx} and~\eqref{eq:od_vy}
into Eq.~\eqref{eq:od_omega} to eliminate $v_x$ and~$v_y$, the angular dynamics follow
\begin{align}
  \gamma_{\rm eff}\,\omega = & 
    -Kv_0\sin\theta
    + \frac{K}{\gamma_t}\bigl(-\sqrt{2\gamma_t k_BT}\,\eta_x\sin\theta
    + \sqrt{2\gamma_t k_BT}\,\eta_y\cos\theta\bigr)
    \nonumber\\
    & \; 
    + \sqrt{2\gamma_r k_BT}\,\xi \; ,
  \label{eq:od_omega_closed}
\end{align}
with the effective rotational friction we already obtained in the context of underdamped dynamics,
\begin{align}
  \gamma_{\rm eff} = \gamma_r + \frac{K^2}{\gamma_t}
  \; .
  \label{eq:gamma_eff_od}
\end{align}
Written in this form, the three noise sources combine into a single effective angular noise with the diffusion coefficient $D_R = k_BT/\gamma_{\rm eff}$, so that in particular the angular dynamics obey the Fokker-Planck equation
\begin{align}
    \partial_t P(\theta,t) = \partial_\theta\!\left[\frac{Kv_0\gamma_t\sin\theta}{\Delta}\,P(\theta,t) + D_R\,\partial_\theta P(\theta,t)\right] ,
\end{align}
whose stationary solution is the von~Mises distribution
\begin{align}
  P_{\rm st}(\theta)
    = \frac{e^{\beta K v_0 \cos\theta}}{2\pi I_0(\beta K v_0 )} \;.
  \label{eq:von_Mises}
\end{align}
This is the Boltzmann distribution for the potential $U(\theta) = -Kv_0\cos\theta$, confirming that this choice of overdamped limit preserves an equilibrium structure.

As noted in Ref.~\cite{Chen2026}, the overdamped dynamics can be written in the compact friction-matrix form,
\begin{align}
  \underbrace{\begin{pmatrix}
    \gamma_t\mathbf{I}_2 & K\hat{\bm{s}}_\perp \\
    -K\hat{\bm{s}}_\perp^T & \gamma_r
  \end{pmatrix}}_{\displaystyle \mathbf{A}(\theta)}
  \begin{pmatrix}
    \dot{\bm{r}} \\ \dot{\theta}
  \end{pmatrix}
  =
  \begin{pmatrix}
    \bm{F} \\ \tau
  \end{pmatrix}
  + \boldsymbol{\zeta}(t) \;,
  \label{eq:friction_matrix}
\end{align}
where $\bm{F}$ and $\tau$ are the force and torque (here $\bm{F} = \gamma_t v_0\hat{\bm{x}}$, $\tau = 0$ for the free single particle), and $\langle\boldsymbol{\zeta}(t)\boldsymbol{\zeta}(t')^T\rangle = 2k_BT\,\mathrm{diag}(\gamma_t,\gamma_t,\gamma_r)\,\delta(t-t')$.
Adopting the notation $\Delta = \gamma_t\gamma_r + K^2$ from Ref.~\cite{Chen2026}, inverting the friction matrix gives the mobility $\mathbf{M} = \mathbf{A}^{-1}$:
\begin{align}
  \mathbf{M}(\theta) =
  \begin{pmatrix}
    \mathbf{M}_{rr} & \mathbf{M}_{r\theta} \\[4pt]
    \mathbf{M}_{\theta r} & M_{\theta\theta}
  \end{pmatrix}
  =
  \begin{pmatrix}
    \dfrac{1}{\gamma_t}\!\left[\mathbf{I}_2
      - \dfrac{K^2}{\Delta}\,\hat{\bm{s}}_\perp\hat{\bm{s}}_\perp^T\right]
    & -\dfrac{K}{\Delta}\,\hat{\bm{s}}_\perp \\[10pt]
    \dfrac{K}{\Delta}\,\hat{\bm{s}}_\perp^T
    & \dfrac{\gamma_t}{\Delta}
  \end{pmatrix}
  \label{eq:full_mobility}
\end{align}
where $\bm{I}_2$ is the identity matrix.
The translational block decomposes as
\begin{align}
  \mathbf{M}_{rr} = \frac{1}{\gamma_t}\,\hat{\bm{s}}\hat{\bm{s}}^T
    + \frac{\gamma_r}{\Delta}\,
      \hat{\bm{s}}_\perp\hat{\bm{s}}_\perp^T \;.
  \label{eq:Mrr}
\end{align}
Mobility along~$\hat{\bm{s}}$ takes the bare value $1/\gamma_t$; perpendicular to~$\hat{\bm{s}}$ it is reduced to $\gamma_r/\Delta$.
The off-diagonal blocks $\mathbf{M}_{r\theta} = -(K/\Delta)\hat{\bm{s}}_\perp$ and $\mathbf{M}_{\theta r} = (K/\Delta)\hat{\bm{s}}_\perp^T$ are the cross-mobilities that interconvert torques and velocities.

With these notations, the diffusion matrix of the three-variable Fokker-Planck equation for $(x,y,\theta)$ is $D_{ij} = k_BT\,(\mathbf{M}\,\mathrm{diag}(\gamma_t,\gamma_t,\gamma_r)\,\mathbf{M}^T)_{ij}$.
Using the expression of the mobility matrix,
\begin{align}
  D_{x\theta} = D_{y\theta} = 0 \;.
  \label{eq:cross_diff_zero}
\end{align}
so that the angular and position parts of the Fokker-Planck equation decouple, $P_{\rm st}(x,y,\theta) \propto P_{\rm st}(\theta)$.
The translational diffusion tensor is given by
\begin{align}
  \bm{D}_T(\theta) &= k_BT\,\mathbf{M}_{rr} = \frac{k_BT}{\gamma_t}\,\hat{\bm{s}}\hat{\bm{s}}^T + \frac{k_BT\gamma_r}{\gamma_t \gamma_r + K^2}\,\hat{\bm{s}}_\perp\hat{\bm{s}}_\perp^T \;, \label{eq:DT}
\end{align}
This writing explicitly defines instantaneous diffusion constants parallel and perpendicular to the spin, $D_\parallel = k_B T/\gamma_t$ and $D_\perp = k_B T/\gamma_{\rm eff}$.
This anisotropic diffusion is reminiscent of what is observed in living systems, see \textit{e.g.} Refs.~\cite{Liverpool2003,Kurzthaler2016,Zhang2023}.

Inverting the friction matrix and using this notation, the dynamics thus reduce to
\begin{align}
  \dot{x} &= v_0\!\left[1 - \frac{K^2}{\gamma_t\gamma_r + K^2}\sin^2\!\theta\right] + \frac{K\sin\theta}{\gamma_t\gamma_r + K^2}\,\zeta_\theta + \frac{1}{\gamma_t}\!\left[1-\frac{K^2\sin^2\!\theta}{\gamma_t\gamma_r + K^2}\right] \zeta_x 
  \nonumber\\
  & \;\;\; + \frac{K^2\sin\theta\cos\theta}{\gamma_t(\gamma_t\gamma_r + K^2)}\,\zeta_y \;,
    \label{eq:xdot_mob} \\
  \dot{y} &= \frac{K^2 v_0\sin\theta\cos\theta}{\gamma_t\gamma_r + K^2} - \frac{K\cos\theta}{\gamma_t\gamma_r + K^2}\,\zeta_\theta + \frac{K^2\sin\theta\cos\theta}{\gamma_t(\gamma_t\gamma_r + K^2)}\,\zeta_x 
  \nonumber\\
  & \;\;\; 
  + \frac{1}{\gamma_t}\!\left[1-\frac{K^2\cos^2\!\theta}{\gamma_t\gamma_r + K^2}\right]
      \zeta_y \;,
    \label{eq:ydot_mob} \\
  \dot{\theta} &= -\frac{Kv_0\gamma_t\sin\theta}{\gamma_t\gamma_r + K^2} + \frac{\gamma_t}{\gamma_t\gamma_r + K^2}\,\zeta_\theta + \frac{K}{\gamma_t\gamma_r + K^2}\!\left(-\sin\theta\,\zeta_x + \cos\theta\,\zeta_y\right),
    \label{eq:thetadot_mob}
\end{align}
where $\zeta_x$, $\zeta_y$, $\zeta_\theta$ are the independent white noises with $\langle\zeta_\alpha(t)\zeta_\beta(t')\rangle = 2k_BT\,\mathrm{diag}(\gamma_t,\gamma_t,\gamma_r)_{\alpha\beta} \,\delta(t-t')$.

In particular, this cumbersome expression reveals a non-trivial effect in the overdamped dynamics with $v_0 >0$: the mean velocity along $x$ is renormalized by the spin-velocity coupling.
To see it, consider the deterministic parts of the position equations that we shall call $v_x$ and $v_y$, and average them over the stationary measure of the angles $P_{\rm st}(\theta)$, Eq.~\eqref{eq:von_Mises}.
We find
\begin{align}
    \left\langle v_x \right\rangle_{\rm st} &\equiv \int\limits_{-\pi}^\pi d\theta P_{\rm st} (\theta) v_x(\theta) = v_0\left[1 - \frac{K^2}{\gamma_t \gamma_r +K^2}\, \frac{I_1(\beta Kv_0)}{\beta Kv_0\,I_0(\beta Kv_0)}\right] 
    \; , \\
    \left\langle v_y \right\rangle_{\rm st} &\equiv \int\limits_{-\pi}^\pi d\theta P_{\rm st} (\theta) v_y(\theta) = 0
    \; .
\end{align}
The correction to the mean velocity vanishes when $v_0 =0$ or $K = 0$ (so that the system does not spontaneously start moving without a tachostat or a spin-velocity coupling).
In the limit $\beta K v_0 \to \infty$, the correction also vanishes, as the spin is locked onto the velocity of the tachostat.
In the opposite limit, $\beta K v_0 \to 0$, 
\begin{align}
    \left\langle v_x \right\rangle_{\rm st} \underset{\beta K v_0 \to 0}{\sim} v_0\left[1 - \frac{1}{2}\frac{K^2}{\gamma_t \gamma_r + K^2}\right],
\end{align}
which is maximally reduced when $K^2 \gg \gamma_t\gamma_r$, leading to $\left\langle v_x \right\rangle_{\rm st} = v_0/2$.
The reduction of the velocity does not break the equilibrium structure, as $\theta$ still obeys the correct stationary distribution and ESSR (with the effective damping $\gamma_{\rm eff}$).
In overdamped dynamics, it is the reduced velocity times time that should be removed from the position 
to calculate the displacements and recover the FDT.

\section[\hspace{2cm} Numerical methods]{Numerical methods\label{app:NumericalMethods}}

\subsection[\hspace{1.75cm}Langevin simulations] {Langevin simulations}

We perform single-particle Langevin simulations of both free and driven dynamics.
In free simulations, we numerically integrate the Hamiltonian equations of motion, Eqs.~\eqref{eq:Langevin_p},~\eqref{eq:Langevin_omega},~\eqref{eq:Langevin_r},~\eqref{eq:Langevin_theta} under the It\={o}-convention discretization
\begin{align}
    \bm{r}(t + dt) &= \bm{r}(t) + \bm{v}_{det}(t) dt, \\
    \theta(t + dt) &= \theta(t) + \omega_{det}(t) dt, \\
    \bm{p}(t+dt)   &= \bm{p}(t) + \bm{F}_{det}(t) dt + \bm{F}_{stoch}(t)\sqrt{dt}, \\
    \omega(t+dt)   &= \omega(t) + \Gamma_{det}(t) dt + \Gamma_{stoch}(t) \sqrt{dt},
\end{align}
where the deterministic parts of the equations are computed using a Runge-Kutta scheme of order 4 to reproduce the scheme of Refs.~\cite{Casiulis2019,Casiulis2019b,Casiulis2019c}.
As for the stochastic part of the dynamics, we use
\begin{align}
\bm{F}_{stoch}(t) &= \sqrt{2 \gamma_t k_B T } \bm{\eta}(t), \\
\omega_{stoch}(t) &= \sqrt{2 \gamma_r k_BT} \xi(t),
\end{align}
with $\eta_x, \eta_y, \xi$ independent random numbers drawn from a unit-variance, zero-mean normal distribution.
Throughout the simulations, we set $\gamma_t = \gamma_r = \gamma$ for simplicity, and we set the units of time and space through $t \mapsto \gamma t$ and $\bm{r} \mapsto \sqrt{\beta} \gamma \bm{r} $ (or, equivalently, by setting $\gamma = 1$ and $k_B T = 1$).
The remaining dimensionless parameters for free dynamics are then $ \sqrt{\beta} v_0$, $\sqrt{\beta} K$, as well as the dimensionless discretization time-step $\gamma dt$.
Throughout simulations, we set $\gamma dt = 10^{-3}$, and $\bm{v}_0 = v_0 \hat{\bm{e}}_x$.

Driven simulations are performed by adding additional terms to the deterministic part of the dynamics.
Force driving is performed by adding an external force $\bm{f}_{\text{ext}}$ to the momentum equation, Eq.~\eqref{eq:Langevin_p}, torque driving by adding an external torque $\Gamma_{\text{ext}}$ to the angular momentum equation, Eq.~\eqref{eq:Langevin_omega}, and field driving by adding an external field $\bm{h}_{\text{ext}}$ via a term $\bm{h}_{\text{ext}}\cdot \bm{s}_{\perp}$ to the angular momentum equation.
In order for linear response theory to hold for these external components, they have to verify $A_{\text{ext}}/\gamma \ll \sqrt{2 k_B T}$ with $A$ representing the modulus of any of these external components.
In practice, we set $\beta f_{\text{ext}}/\gamma = 0.1$, $\beta h_{\text{ext}}/\gamma = 0.2$, $\beta \Gamma_{\text{ext}}/\gamma = 0.2$.
Without loss of generality, we furthermore set the force and field to be positively along $\hat{\bm{e}}_x$ for most checks, save for the Onsager-Casimir reciprocity, where we set the force to positively be along $\hat{\bm{e}}_y$.

Throughout the paper, we initialize the system using the equilibrium distribution for a free particle, given in the main text.
The dynamics are then run for a dimensionless time $\gamma t_{\text{therm}} = 100$ before measurements are performed, and before external forces, torques, or fields are introduced.

\subsection[\hspace{1.75cm}Common random numbers method] {Common random numbers method\label{app:CRN}}

Throughout the main text, we present curves that were obtained using the common random numbers method to reduce the finite-sample effects on the statistics.
In short, the common random number methods is a standard variance reduction method~\cite{Kahn1953} used when trying to compare the outcome of two different stochastic strategies.
It consists in setting identical seeds to the pseudo-random number generators in both strategies, so that the observed comparative observables (\textit{e.g.} the difference across strategies in the the mean of an observable) display a much smaller finite-sample variance, as the overall variance is reduced by the covariance between noise sources.

Concretely, we endowe each simulated particle by an identity, that is uniquely associated to its index $i$, with $1\leq i \leq N$.
This particle $i$ then undergoes the exact same noise history $(\eta_{x,i}, \eta_{y,i},\xi_{i})$ across simulation designs, whether it undergoes free dynamics, driven dynamics under a constant force, driven dynamics under a constant torque, etc.

This is particularly important for \textit{short-time} driven dynamics (\textit{e.g.} MSD values in the case of an external force).
Indeed, in that case, the quantities of interest are of the form
\begin{align}
    \chi \approx \frac{\langle \Delta x \rangle_{f_{\text{ext}}} - \langle \Delta x \rangle_{\text{free}}}{f_{\text{ext}}}
\end{align}
which is exactly the average of a difference between two quantities obtained from two different stochastic models.
The issue in that case is that the signal to noise ratio of the measurement in the numerator at time $t$, when using naïve sampling, scales like
\begin{align}
    SNR(t) &= \frac{f_{\text{ext}} t/\gamma}{\sqrt{2 D t / N}} \propto \frac{f_{\text{ext}} \sqrt{N t}}{\gamma \sqrt{D}}.
\end{align}
In other words, at any finite $N$, there is a time below which signal will be drowned in noise.
To make matters worse, note that $f_{\text{ext}}$ cannot be increased too much, or the system will leave the linear response regime.
Last but not least, recall that the angular diffusion constant gets exponentially suppressed by $K v_0$ in the limit of large $K v_0$, so that noise dominates until rather large times when using a naïve sampling approach.

This choice does \textit{not} make results such as the Onsager-Casimir reciprocity trivial, it simply tightens the confidence interval we find for agreement between curves when comparing fluctuations and responses with each other at a given finite $N$, thus saving us the trouble of instead making $N$ orders of magnitude larger.
More formally, the common random number method cancels out the diffusive noise contribution $\sqrt{2D t/N}$ by making noise history deterministic, and replaces it by a noise coming from the feedback of non-linear couplings at the next order, $O(f_{\text{ext}}^2)$, so that now
\begin{align}
    SNR \propto \frac{t}{\gamma f_{\text{ext}} G(t)}
\end{align}
with $G(t)$ some function that is dynamics dependent.
In other words, as long as the dynamics of $x$ are not just linearly following $f_{\text{ext}}$ at the deterministic level, the CRN output is still not trivial.

\addcontentsline{toc}{section}{References} 
\bibliographystyle{iopart-num}
\bibliography{PostDoc-StefanoMartiniani}

\end{document}